\documentclass[10pt,aps,pra,twocolumn,superscriptaddress,nofootinbib,floatfix,longbibliography]{revtex4-2}

\usepackage[utf8]{inputenc}  

\usepackage{amsmath}
\usepackage{amsthm}
\usepackage{amssymb}
\usepackage{bm}

\usepackage{graphicx}
\usepackage[table,dvipsnames]{xcolor}
\usepackage{tikz}

\usepackage{booktabs}
\usepackage{tabularx}
\usepackage{multirow}
\usepackage{siunitx}
\usepackage{ragged2e}
\usepackage{adjustbox}

\usepackage{enumitem}
\usepackage{setspace}
\usepackage[normalem]{ulem}

\usepackage{eurosym}
\usepackage{pythonhighlight}

\usepackage[colorlinks, citecolor=blue, linkcolor=blue, urlcolor=blue, linktocpage]{hyperref}

\usepackage{orcidlink}

\setitemize{noitemsep,topsep=0pt,parsep=0pt,partopsep=0pt,leftmargin=*}
\newcommand*{\bra}[1]{\ensuremath{\langle #1 \vert}}
\newcommand*{\ket}[1]{\ensuremath{\vert #1 \rangle}}

\newcommand{\braket}[2]{\langle #1 | #2 \rangle}

\newcommand{\norm}[1]{|| #1 ||}

\usepackage{mdframed}
\usepackage{enumitem}

\usepackage[capitalise]{cleveref}

\begin{document}

\title{Continuous-time quantum walk-based ans\"atze on neutral atom hardware}

\author{Edric Matwiejew  \orcidlink{0000-0002-2480-1633}}
\affiliation{Pawsey Supercomputing Research Centre, Perth, WA, 6152, Australia}

\author{Jonathan Wurtz \orcidlink{0000-0001-7237-0789}}
\affiliation{QuEra Computing Inc., 1284 Soldiers Field Road, Boston, MA, 02135, USA}

\author{Jing Chen \orcidlink{0000-0003-0538-689X}}
\affiliation{QuEra Computing Inc., 1284 Soldiers Field Road, Boston, MA, 02135, USA}

\author{Pascal Jahan Elahi \orcidlink{0000-0002-6154-7224
}}
\affiliation{Pawsey Supercomputing Research Centre, Perth, WA, 6152, Australia}

\author{Tommaso Macri \orcidlink{0000-0002-7778-8014
}}
\affiliation{QuEra Computing Inc., 1284 Soldiers Field Road, Boston, MA, 02135, USA}

\author{Ugo Varetto
\orcidlink{0000-0002-7696-0345}}
\affiliation{Pawsey Supercomputing Research Centre, Perth, WA, 6152, Australia}

\date{\today}
\begin{abstract}
Continuous-time quantum walks offer provable speedups for certain computational problems, yet translating these advantages to near-term hardware remains challenging. We realize variational ans\"atze based on continuous-time quantum walks on an analog neutral-atom processor. For unentangled targets, we derive closed-form expressions for near-optimal control parameters that transfer directly to hardware with minimal calibration. On QuEra's Aquila processor we observe the super-quadratic convergence characteristic of efficient quantum walk algorithms, visible at low circuit depth, with theory predicting stronger speedups as hardware improves. For entangled targets, specifically symmetric superpositions in the Rydberg-blockaded subspace, we introduce an optimization protocol exploiting spectral properties of the walk dynamics. The required evolution time scales inversely with the spectral gap, offering an advantage over adiabatic protocols, whose evolution time scales as the inverse square of the spectral gap. We verify this scaling behavior on Aquila and confirm that the prepared states are coherent superpositions via quench dynamics. Our results establish a practical pathway from abstract quantum walk algorithms to analog quantum processors, demonstrating that the dynamics underlying their potential for super-quadratic quantum speedup are accessible on current devices.
\end{abstract}

\maketitle

\section{Introduction}
Quantum search and optimization are key applications for near-term quantum computers~\cite{Preskill18}. Algorithms for unstructured or spatial search and the Quantum Approximate Optimization Algorithm (QAOA) exemplify the unique properties of quantum computing: superposition to explore exponentially many computational pathways in quantum parallel, entanglement to encode a cost landscape, and interference to amplify desired states~\cite{farhi2014quantum,Ambainis2007}. These algorithms typically utilize an ansatz\footnote{A parameterised family of circuits used to prepare candidate states, with the parameters chosen (typically by a classical outer loop) to maximize the probability of sampling desirable solutions.} comprised of an interleaved sequence of unitaries. At each ansatz iteration, the first diagonal unitary phase-encodes the cost or validity of each solution state, and the second \emph{mixing unitary} drives amplitude transfer between these states. This layered structure can be naturally understood through the framework of continuous-time quantum walks (CTQWs), where coherent dynamics of quantum states traversing a graph underpin the convergence behavior of the algorithm.

CTQWs provide a natural framework for quantum algorithm design, with coherent dynamics over a graph underlying methods for spatial search and optimization~\cite{childsuniversal2009,childs_exponential_2003}. In particular, CTQW-based search algorithms achieve the optimal quadratic improvement in query complexity for unstructured search~\cite{Ambainis2007}, and operate in a non-adiabatic regime that can saturate fundamental quantum speed limits~\cite{Deffner2017,christandl2004perfect}.

These algorithms can all be expressed as a \emph{phase-walk ansatz}, an alternating sequence of two interleaved unitaries. With each \emph{ansatz iteration} a phase separator applies rotations that encode the problem at hand, and a \emph{mixing unitary} carries out a CTQW that drives amplitude transfer between states. While the query-complexity bound for unstructured search is quadratic, the phase-encoding layer can carry additional information about global problem structure. When symmetries of the \emph{mixing graph} (whose edges define allowed transitions) reflect the structure of the solution space, compounding interference across symmetric pathways can coherently focus amplitude on target states, yielding super-quadratic convergence~\cite{matwiejew_quantum_2023,bennett2024non}. Mixing graphs that enable such ``shortcut'' dynamics are especially significant for near-term quantum devices, where practical advantage requires sufficient amplification within time-limited coherence windows.

Continuous-time quantum walks have been demonstrated in the position basis across a variety of platforms, including walks over line graphs using photonic waveguides~\cite{peruzzo2010quantum}, neutral atoms~\cite{preiss2015strongly} and trapped ions~\cite{tamuraquantum2020}, as well as two-dimensional grids using superconducting qubit arrays~\cite{yan2019strongly, gong2021quantum} and tweezer-programmable strontium-atom lattices~\cite{young_tweezer-programmable_2022}. Beyond spatial walks, variational circuits that implement CTQWs in the computational basis have been demonstrated with photonic waveguides~\cite{quexperimental2024}, and CTQW-like dynamics have been observed in quantum many-body scar dynamics in Rydberg chains \cite{Choi2019} on neutral atom systems \cite{Bernien2017}. Full phase-walk ans\"{a}tze with alternating phase-encoding and mixing unitaries have been implemented for QAOA on superconducting processors~\cite{Harrigan2021} and trapped ions~\cite{Shaydulin2024}.

Despite this progress, each approach addresses only part of the phase-walk framework. Position-basis implementations demonstrate coherent quantum dynamics but cannot leverage the full computational-basis Hilbert space and, as the walk graph is fixed by hardware geometry, are not easily tailored to match the structure of a target problem. Gate-based QAOA implementations on superconducting and trapped-ion processors do include phase encoding. However, the mixing unitary is typically limited to a transverse-field operator on an unconstrained Hilbert space, requiring either penalty terms or post-selection to enforce feasibility constraints-- both of which incur overhead that grows with system size~\cite{Hadfield_2019}. Recent QAOA studies have reported evidence of scaling advantages over classical algorithms, but these claims derive entirely from noiseless classical simulations, with hardware demonstrations showing only Grover-like amplification at low circuit depth~\cite{boulebnane2024,Shaydulin2024}.

Analog neutral-atom platforms provide an ideal environment for the exploration of phase-walk ans\"atze. The blockade interaction, whereby strong Rydberg-state repulsion suppresses simultaneous excitation of nearby atoms, enables efficient implementation of constrained CTQWs whose walk graphs possess non-trivial topology with direct relevance to constrained combinatorial optimization~\cite{Hadfield_2019,marsh_combinatorial_2020,Fuchs2024}. The analog evolution has the potential to natively implement continuous-time walk dynamics, while global and local detuning provide flexible means to apply site-dependent phase shifts encoding the cost function. This combination yields a hardware-efficient realization of the full phase-walk ansatz in which both constraint enforcement and the mixing unitary are ``free'' in terms of gate overhead.

In this work, we report the implementation of continuous-time quantum walk (CTQW)-based variational ans\"{a}tze on analog-mode neutral-atom hardware, using QuEra's Aquila processor~\cite{wurtz2023aquila}. Unlike prior Rydberg-atom demonstrations that sample low-energy configurations or observe emergent many-body dynamics~\cite{Ebadi_2022,Bernien2017}, we target the preparation of specific, known benchmark states and systematically quantify convergence as a function of ansatz depth. This provides a controlled setting that isolates the intrinsic performance of CTQW-based dynamics from problem-dependent complexity.

We focus on two classes of target states: computational-basis product states, and ``bracelet'' states--equal-weight superpositions over all basis states related by the dihedral symmetries of the walk graph. The former relies on dynamics that are analogous to the preparation of an \emph{unknown} unentangled state in search or optimization. The latter generalize Dicke states, the symmetric entangled states widely used in quantum metrology and constrained optimization~\cite{Bartschi2019,Hadfield_2019}, to an independent-set subspace. While Dicke states span a subspace that grows only linearly with system size, bracelet states live in an exponentially large symmetric sector, making their preparation a more demanding test of coherent control. Furthermore, their preparation probes whether the non-adiabatic dynamics for preparation of product states in constrained Hilbert spaces~\cite{Lukin2024shortcut} via quantum many-body scar dynamics~\cite{Bernien2017,Choi2019} extend to the preparation of similarly constrained entangled targets. 

For product-state preparation, we derive closed-form expressions for near-optimal walk parameters, providing analytically guided initial points via a method that is generalizable to walk graphs defined by independent-set subspaces. For bracelet states, we introduce an optimization protocol based on the spectral properties of the walk dynamics~\cite{Krovi2007}. By comparing ideal CTQW predictions, noiseless Rydberg emulation, and measurement from noisy hardware, we identify the depth at which super-quadratic amplification persists and the crossover beyond which noise suppresses convergence, providing a practical figure of merit for NISQ implementations of CTQW-based algorithms. Finally, we employ quench dynamics under the walk Hamiltonian to probe coherence of the prepared bracelet states, providing empirical evidence that the observed populations arise from coherent superposition rather than an incoherent mixture.

The remainder of this work is as follows. In \cref{sec:ctqw}, we introduce the phase-walk ans\"{a}tze, performance metrics used in later analysis, the representative target states, and the variational optimization strategies. In \cref{sec:rydberg_atoms}, we describe the mapping of the ans\"{a}tze to the Rydberg Hamiltonian on Aquila. Results of state preparation on hardware are compared with the scaling predicted by ideal CTQW dynamics and noiseless Rydberg emulation in \cref{sec:state_pre_results}. In \cref{sec:quenches}, we apply quenches on the prepared states as an indicator of coherent state preparation. Finally, in \cref{sec:conclusion}, we summarize our conclusions. Overall, our results establish constrained-subspace CTQWs as a framework for entangled state preparation on analog-mode neutral-atom processors, and demonstrate that CTQW-based ans\"{a}tze for known states can exhibit the super-quadratic convergence characteristic of efficient quantum-walk protocols on currently available hardware.

\section{Phase-walk ans\"{a}tze}
\label{sec:ctqw}

\begin{figure}
    \centering
    \includegraphics[width=0.9\linewidth]{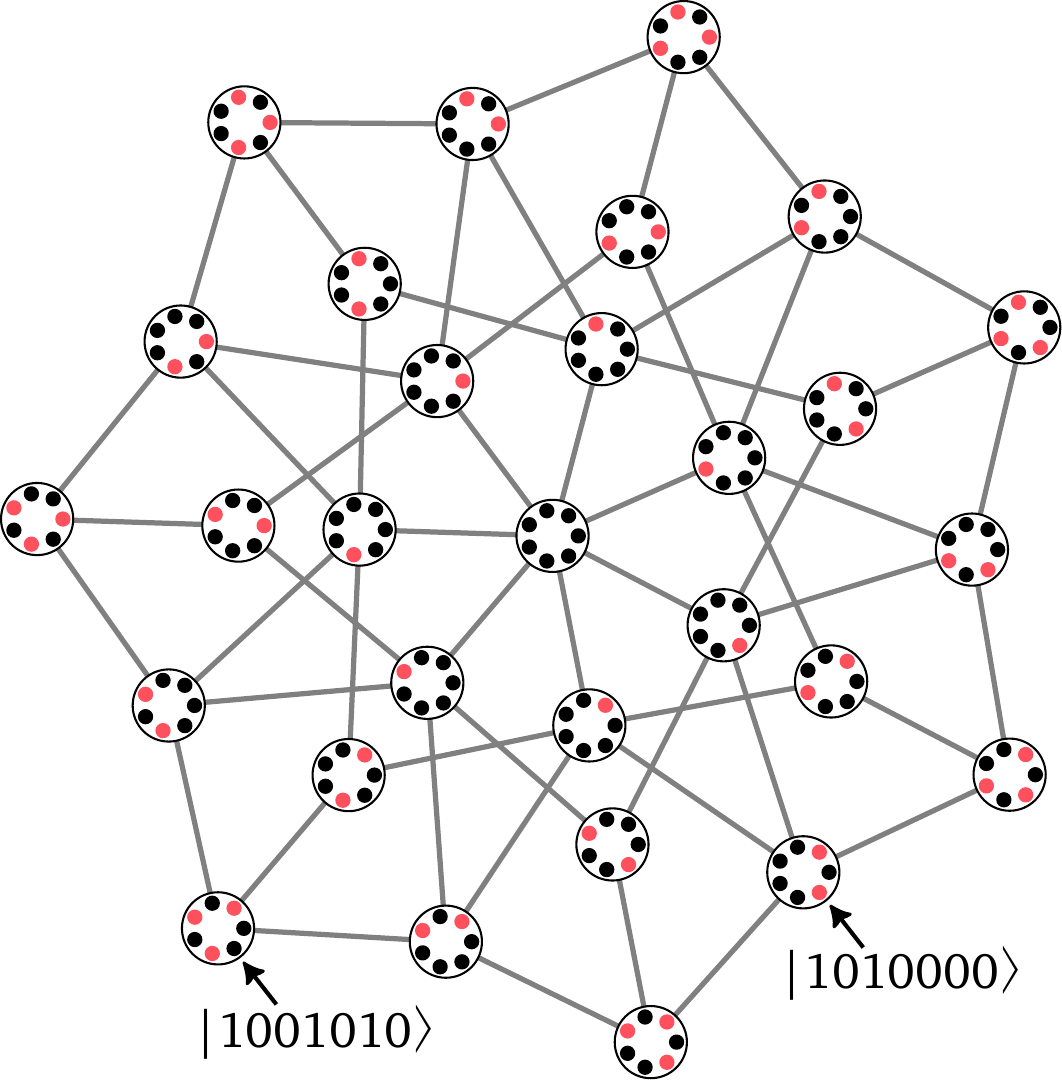}
    \caption{An example walk graph for independent set constrained subspace walks. Each vertex is an independent set in a ring of 7 vertices labeled by bitstring $z$ and representing basis vector $|z\rangle$, with red dots indicating inclusion of that physical vertex in the set. There is an edge between vertices iff the Hamming distance of the bitstring label is 1. }
    \label{fig:hamiltonian}
\end{figure}

\subsection{Quantum Walk-based Ans\"atze}

A Continuous-Time Quantum Walk (CTQW)~\cite{venegasandracaquantum2012} is the coherent evolution of a quantum state under the unitary time-evolution operator
\begin{equation}
\hat{U}_W(\tau) = \exp \bigl(-\mathrm{i}\,\tau\,\hat{\mathcal{G}}\bigr),
\end{equation}
where the generator
\begin{equation}
\hat{\mathcal{G}} = \sum_{(a,b) \in \mathcal{E}} w_{ab}\ket{a}\bra{b},
\end{equation}
is obtained from the (possibly weighted) adjacency matrix of a quantum walk graph $\mathcal G=(\mathcal V,\mathcal E)$ with vertex set $\mathcal V$, edge set $\mathcal E$, and real, symmetric weights $w_{ab}=w_{ba}$. Each vertex is matched to a basis state of a quantum state $a\leftrightarrow |a\rangle$, and the terms of the graph adjacency matrix are matched to the generator.

In the context of algorithms for search and optimization, CTQWs are often paired with a diagonal unitary that adds phase to the graph vertices weighted by $\phi_a=\gamma c_a$, defined by a (variational) scaling parameter $\gamma$ and the generator

\begin{equation}
    \hat C = \sum_{a\in \mathcal V}c_a|a\rangle\langle a|.
\end{equation}

Together the two unitaries form a parameterized \emph{phase-walk} ansatz for which a common task is to prepare some target or resource state by alternating between the walk (or ``mixer'') generator $\hat{\mathcal G}$ and phase-separator generator $\hat C$ applied to some fiducial initial state $|\psi_0\rangle$

\begin{equation}
\label{eq:phase_walk_ansatz}
    |\psi\rangle = \prod_{q=1}^pe^{-i\tau_q \hat{\mathcal G}}e^{-i\gamma_q \hat C}|\psi_0\rangle, 
\end{equation}
for $2p$ variational parameters $\{\vec \gamma,\vec\tau\}$ -- as exemplified by the quantum approximate optimization algorithm \cite{farhi2014quantum}.

A common representation of basis states is binary bitstrings $a\in \mathcal Z=\{0,1\}^n$, where each bit $z_i$ represents the state of qubit $i$. In this work, we define $\ket{\psi_0}$ as an initial walk from the all-zeros state,
\begin{equation}
    \label{eq:psi_0}
    \ket{\psi_0} \equiv\ket{\tau_0}=e^{-\text{i}\tau_0 \hat{\mathcal G}}|0\rangle.
\end{equation}

The phase-separator generator can be written as some function $c_a \equiv f(z)$, typically a low-order polynomial to conform to the locality constraints of quantum hardware. Here, we choose a first-order function $f(z)=\sum_i c_i z_i$ so that the phase-separator generator can be efficiently written in terms of Pauli matrices

\begin{equation}
\hat C = \frac{1}{2}\sum_{i=1}^{N} c_i\,\bigl(\mathbb I - \hat{\sigma}_z^{(i)}\bigr).
\end{equation}

Similarly, the walk graph is often formed by edges between vertices whose labels differ by a Hamming distance of one or two~\cite{farhi2014quantum,Hadfield_2019,matwiejew_quantum_2023,bennett2024non}:
\begin{equation}
d_h(z^a,z^b) \equiv \lVert z^a - z^b \rVert_1,
\end{equation}
Hamming-distance-based constructions yield graphs that are efficiently implementable and typically highly symmetric (generally vertex-transitive)~\cite{matwiejew_quantum_2023,bennett2024non}. If edges connect labels with $d_h=1$ over all $2^N$ possible bitstrings, then the walk graph is the $N$-dimensional hypercube. A Hamming-distance-1 graph walk generator can be efficiently written as $\hat{\mathcal G}=\sum_i\hat\sigma_x^{(i)}$, and small Hamming distances correspond to low-depth polynomial functions over Pauli operators.

\subsection{Walks in constrained subspaces}

Frameworks for optimization with constraints restrict the vertices to a subset of valid solutions and construct the walk generator so that it preserves this reduced search space as an invariant subspace~\cite{marsh_combinatorial_2020,bennett2024non,Hadfield_2019}. The resulting CTQW is often non-separable and inherits the symmetries of the constraint, while breaking others. As we detail in \cref{sec:rydberg_atoms}, such walks are the focus of this work as neutral-atom platforms offer unique opportunities for their efficient implementation. We constrain the full space of length-$N$ bitstrings according to an underlying \emph{constraint graph} $G=(V,E)$ on $N$ vertices. The walk vertices $\mathcal V$ are then independent-set configurations of $G$,
\begin{equation}
\mathcal V=\Bigl\{\, z\in\{0,1\}^N \ \Big|\ z_i+z_j\le 1 \ \ \forall (i,j)\in E(G) \Bigr\},
\end{equation}
with edges between vertices in $\mathcal V$ having $d_h=1$. The resulting walk graph is a subgraph of the hypercube, and the walk generator can be written efficiently as
\begin{equation}
\hat{\mathcal G}=\sum_{i=1}^N \hat{\mathcal P}\,\hat{\sigma}_x^{(i)}\,\hat{\mathcal P},
\end{equation}
where $\hat{\sigma}_x^{(i)}$ is the Pauli-$X$ operator on qubit $i$, and $\hat{\mathcal P}$ is the projector onto $\mathcal V$ (equivalently, the independent-set subspace of the constraint graph $G$).

In particular, we choose the constraint graph $G$ to be the cycle graph on $N$ vertices (i.e., a ring with nearest-neighbor edges). The resulting walk graph is a \emph{Lucas cube}~\cite{munarinilucas2001}, shown in~\cref{fig:hamiltonian}. We make this choice because its structural properties are relatively well understood, supporting efficient enumeration and generation of $\mathcal V$, and reduced-dimensional representations of $\mathcal G$, at large $N$ (which is not the case for all independent-set subspaces~\cite{Valiant1979}), while remaining suitably complex to produce non-trivial dynamics. Specifically, the number of vertices of the Lucas cube is exponentially smaller than that of the hypercube at the same $N$, yet still grows exponentially as $|\Lambda_N|\sim \varphi^N$, where $\varphi=(1+\sqrt5)/2$ and, as it preserves the dihedral symmetry of $G$, it retains a relatively small but structurally rich dihedrally symmetric sector that grows like $B_N^{(D)}\sim \varphi^N/(2N)$~\cite{ashrafi2016vertex}.

\subsection{State preparation}

A key application of quantum walks is to prepare defined target states $|\psi\rangle$. Commonly, this is some unknown but well-defined state, such as a product state that maximizes the value of some low-order polynomial objective $C(z)$, such as in QAOA \cite{farhi2014quantum}. Similarly, the states could be defined concretely as a resource state with an efficient description, such as the GHZ state $|\psi\rangle = (|00\cdots00\rangle + |11\cdots11\rangle)/\sqrt{2}$. Given the extensive literature on QAOA and its variants to solve combinatorial objectives $C(z)$, this work focuses instead on preparing concretely defined states and characterizing the capacity of variational quantum walks to prepare these target states on blockaded subspaces \cite{Gonzales2025}.

Given some defined target state $|\psi\rangle$, a variational optimizer chooses $p(N+1)$ parameters $\{\vec \gamma,\tau\}$ of the variational ansatz state $|\gamma,\tau\rangle$ to maximize the overlap with the target state
\begin{equation}
    \texttt{MAX}_{\gamma,\tau}:\; \big|\langle \gamma,\tau | \psi\rangle \big|^2.
\end{equation}
Or, similarly, minimize the statistical divergence between the target distribution $P(z)=|\langle z|\psi\rangle|^2$ and the variational distribution $Q(z\,|\,\gamma,\tau)=|\langle z|\gamma,\tau\rangle|^2$.

\subsection{Target states}
\label{sec:target_states}

For the constraint graph of the $N$-vertex ring, we chose two classes of target states. The first class are product states, e.g.~$\ket{\psi}=\ket{0010101}$ specified by a bitstring $z$. While preparing such states for a system of non-interacting qubits is trivial (see \cref{sec:optimization}), preparing these \emph{known} states using walk-based dynamics provides a benchmark for the typical goal of variational ans\"atze: preparing \emph{unknown}, low-entanglement states corresponding to low-cost solutions~\cite{farhi2014quantum,blekos2023review,Hadfield_2019}. Furthermore, because the mixer allows only constraint-respecting transitions, it is not \emph{a priori} clear that arbitrary product states can be efficiently prepared with high fidelity using an ansatz parameterised solely by walk times and locally tuned phase-separator values $\gamma_i$. We focus on two representative product-state targets. The first, denoted $z^{\rm half}$, is the state $\ket{0^{N-2h}(01)^h}$ with $h=\lfloor N/2\rfloor$, corresponding to roughly half the maximum Hamming weight for the ring graph. The second, denoted $z^{\rm MIS}$, is a maximum independent set of the ring: the alternating pattern $(01)^{N/2}$ for even $N$, or $0(10)^{(N-1)/2}$ for odd $N$.

The second class consists of superpositions over the symmetric subspace. An $N$-vertex ring has translational and reflective symmetries described by the dihedral group $D_N$ (i.e., the set of all discrete rotations and reflections). Because the constraint graph $G$ is defined on the ring, these operations relabel vertices without changing which pairs are adjacent (which are referred to in graph theory as \emph{automorphisms} of $G$). We therefore consider the corresponding symmetry-invariant subspace, spanned by equal-weight superpositions of all computational-basis states labeled by bitstrings related by cyclic rotations or reflections,
\begin{equation}
\ket{[z]} = \frac{1}{\sqrt{|[z]|}} \sum_{u \in [z]} \ket{u},
\end{equation}
where $[z]=\{g\!\cdot\!z:\, g\in D_N\}$ is the dihedral orbit of a representative bitstring $z$ \footnote{The first instance in which the symmetric subspace contains states whose entire orbit cannot all be generated by rotations alone occurs at $N=9$, where
\begin{align*}
&\big|\big[010010001\big]\big\rangle=
\,\tfrac{1}{\sqrt {18}}\big(|010010001\rangle+|010010100\rangle+|010001010\rangle+\\&|010001001\rangle+|010100100\rangle+|010100010\rangle+|101001000\rangle+\\&|101000100\rangle+|100101000\rangle+|100100010\rangle+|100010100\rangle+\\&|100010010\rangle+|001010001\rangle+|001010010\rangle+|001001010\rangle+\\&001000101\rangle+|000101001\rangle+|000100101\rangle\big).
\end{align*}
}.

We refer to these states as \emph{bracelet states}, borrowing from the combinatorial term for equivalence classes of bitstrings under cyclic rotations and reflections~\cite{sawada_generating_2001}. The preparation of such states is an informative benchmark, as it requires coherent many-body interference to establish uniform amplitudes across a dihedral orbit of up to $2N$ basis states. Furthermore, bracelet states have potential applications as initial states for constrained optimization~\cite{Hadfield_2019,Bartschi2020} and quantum metrology~\cite{Bartschi2019}.

\subsection{Ans\"{a}tze performance}
\label{sec:performance_metrics}

Phase-walk ans\"{a}tze are heuristic methods for preparing approximate solutions to combinatorial problems. Outside a few special cases, formal guarantees are limited, so benchmarking is primarily empirical and problem-dependent~\cite{farhi2014quantum,marsh_combinatorial_2020,wurtzmaxcut2021}. In this context, an informative metric is the \emph{amplification} of desirable solutions relative to a uniform baseline over the accessible subspace
\begin{equation}
\label{eq:amplification}
A(|\mathcal V|, p) = \frac{|\mathcal V|\, P(z^*, p)}{|z^*|},
\end{equation}
where $P(z^*, p)$ is the probability of preparing the target $\ket{z^*}$ at depth $p$~\cite{matwiejew_quantum_2023,bennett2024non} and $|z^*|$ is the cardinality of the target set of basis states. Effective ans\"{a}tze are expected to closely follow a power-law growth in $|\mathcal V|$,
\begin{equation}
\label{eq:amp_power_law}
A(|\mathcal V|, p) = c\, |\mathcal V|^{\alpha},
\end{equation}
with the exponent $\alpha$ as the key performance indicator~\cite{brassard2000quantum,ambainis2003quantum}. Because the single-shot success scales as $|\mathcal V|^{\alpha-1}$, $\alpha$ induces an effective polynomial speedup order $n$,
\begin{equation}
\label{eq:polynomial_speedup}
n = \frac{1}{1-\alpha},
\end{equation}
which can be understood as a proxy for query-complexity scaling (e.g., at $\alpha=\tfrac12$ the scaling is Grover-like).

For a particular phase-walk ansatze to offer the possibility of quantum-speedup given the overheads of optimization, sampling, and classical post-processing, it should ideally demonstrate two key features. First, at fixed depth $p$, the amplification factor $A$ should grow with problem size as a high-order polynomial in $|\mathcal V|$. Second, at fixed $|\mathcal V|$, the amplification should itself increase polynomially with circuit depth $p$. In both instances, orders $n>2$ indicate the leveraging of global phase-structure to accelerate convergence~\cite{matwiejew_quantum_2023,zalkagrovers1999}.

In the following, we probe these criteria through the preparation of known target states, seeking a reliable protocol for CTQW implementation and an empirical view of potential quantum speedup possible within the degrees of freedom of our ans\"{a}tze when applied to non-trivial problem instances.

\subsection{Variational optimization}\label{sec:optimization}

\subsubsection{Product states}

If the constraint graph $G$ is the null graph, then the walk Hamiltonian is the $N$-dimensional hypercube and the time evolution remains fully separable. With $p = 1$, the state is
\begin{equation}
\ket{\psi}
=\bigotimes_{k=0}^{N-1}e^{-i\tau_{1}\sigma_x^{k}}e^{-i\gamma c_k \sigma_z^{k}}e^{-i\tau_{0}\sigma_x^{k}}\ket{0}.
\end{equation}
Setting the $X$-rotation angles to $(\tau_0,\tau_1)\equiv\{\pi/4,\,3\pi/4\} \pmod{2\pi}$ (in any order) and choosing $\gamma=\pm\pi \pmod{2\pi}$ and $c_k=z_k^*$ enables the preparation of any computational-basis state\footnote{Geometrically, the single-qubit gates perform an $X\text{-}Z\text{-}X$ rotation sequence on the Bloch sphere of each qubit. Starting from $+z$ (i.e., $\ket{0}$), an $X$-rotation by $\tau_0 = \pi/4$ moves the Bloch vector to $-y$. Then, if $c_k = 1$, the intermediate $Z$-rotation moves the vector to $-y$. Finally, an $X$-rotation by $\tau_1 = 3\pi/4$ drives vectors on $+y$ back to $+z$ (i.e., $\ket{0}$) and those on $-y$ to $-z$ (i.e., $\ket{1}$).}. However, when a nontrivial connected constraint graph is introduced, each $\sigma_x^k$ becomes a neighbor-conditioned multi-qubit gate, so the evolution is no longer separable.  Nevertheless, the walk ansatz can still prepare the exact state. Given a phase separator
\begin{equation}
\label{eq:product_state_phasor}
\hat U_{z^*}(\gamma)\ket z = \exp\left(-i{\gamma\sum_{i\,:\,z_i^*=1} z_i}\right)\ket z
\end{equation}
as a sum of Pauli-z terms that are non-zero in the target bitstring, three protocols are possible. The first is a Trotterized adiabatic protocol that uses perturbatively small parameters $\gamma,\tau\ll1$, exact in the $p\to\infty$ limit \cite{Wurtz2022}. The second is a Trotterized mask protocol with $\gamma,\tau\ll1$ but $\gamma\gg\tau$ to ``pin'' the zero qubits in the zero state with a large effective Z field. The third is a fast-forward limit \cite{GuryOdelin2019} with $\gamma=\pi$ and $\tau_0 + p\tau_1=\tfrac{\pi}{2}$, for which we give the following argument. First, the walk Hamiltonian can be split by $U_{z^*}$ into two components:
\begin{equation}
\mathcal G = \mathcal G_+ + \mathcal G_-,
\end{equation}
where $\mathcal G_-$ flips $h_{z^*}(z)$ and $\mathcal G_+$ preserves it. Then,
\begin{equation}
  \mathcal G_\pm = \tfrac12\left(\mathcal G \pm U_{z^*}\mathcal G U_{z^*}\right),
\end{equation}
Here the $+1$ eigenspace of $U_{z^*}$, which contains the ground state $\ket{0}$ and $\ket{z^*}$, is the \emph{positive} subspace and the $-1$ eigenspace 
the \emph{negative} subspace.

The interleaved $U_{z^*}$ effectively toggles a rotation between the negative and positive subspaces, 
\begin{equation}
  R_{\pm}(\tau)=e^{-i\tau(\mathcal G_+ \pm \mathcal G_-)},
\end{equation}
which together produce constructive interference on the positive subspace. To illustrate this effect, consider the even-depth ($2p^\prime = p$) ansatz,
\begin{equation}
  \ket{\tau_0,\tau_1} = \big[ R_{-}(\tau_1)R_{+}(\tau_1)\big]^{p'}\ket{\psi_0},
  \label{eq:ansatz}
\end{equation}
where $\ket{\psi_0}$ is defined as in \cref{eq:psi_0}. By BCH expansion  to second-order, a single pair is,
\begin{equation}
  R_{-}(\tau_1)R_{+}(\tau_1) = \exp\Big[-2i\tau_1\mathcal G_+ - \tau_1^2[\mathcal G_+, \mathcal G_-] + O(\tau_1^3)\Big].
  \label{eq:BCH_product_state}
\end{equation}
The linear $\mathcal G_-$ terms cancel, so to leading order the dynamics are generated by $\mathcal G_+$ with the effective time $p'\tau_1$. 

Projection of $\exp(-i 2 \tau_1 \mathcal G_+)$ onto the positive subspace yields an effective evolution over states
\begin{equation}
\ket{v_j} = \frac{1}{\sqrt{|S_j|}} \sum_{z \in S_j} \ket{z},
\end{equation}
where $S_j$ is the set of bitstrings in the positive subspace with Hamming weight $j$, and the $\ket{v_j}$ are the corresponding uniform ``shell'' states of fixed Hamming weight (so $\ket{v_0} = \ket{0}$ and $\ket{v_k} = \ket{z^*}$). The effective Hamiltonian $J$ is the SU(2) spin-$k/2$ chain Hamiltonian (permutation-symmetric noninteracting spins $1/2$) expressed in the symmetric Dicke basis, so that in this subspace it reduces to a one-dimensional ladder $\{\ket{v_j}\}_{j=0}^k$, which has non-zero entries only on the first off-diagonals:
\begin{equation}
  J_{j,j+1}=J_{j+1,j} = \sqrt{(k-j)(j+1)},
\end{equation}
since each state in $S_j$ has $k-j$ forward neighbors in $S_{j+1}$ and each state in $S_{j+1}$ has $j+1$ backward neighbors in $S_j$. Because $J$ only couples neighboring Hamming-weight shells $j\leftrightarrow j+1$, the first non-vanishing contribution connecting $\ket{v_0}$ to $\ket{v_k}$ appears at order $k$. The lowest-order coupling between $\ket{0}$ and $\ket{z^*}$ then occurs at order $k$ with strength,
\begin{equation}
  J_{0\rightarrow z^*}=\left( \prod_{j = 0}^{k-1} J_{j,j+1} \right)^{1/k}.
\end{equation}
Moreover, since $J$ inherits the bipartite structure of $\mathcal G$, the transition amplitude $\bra{v_k}e^{-iT_{\rm eff}J}\ket{v_0}$ has a fixed parity in $T_{\rm eff}$ determined by $k$,
and,
\begin{equation}
  \bra{v_k}e^{-iT_{\rm eff} J}\ket{v_0} \approx
  \begin{cases}
    -i \sin \big(T_{\rm eff} J_{0 \rightarrow z^*}\big), & k \ \text{odd}, \\
    1 - \cos \big(T_{\rm eff} J_{0 \rightarrow z^*}\big), & k \ \text{even},
  \end{cases}
\end{equation}
which, in either case, gives
\[
  \mathrm{Prob}(\ket{z^*})\propto\sin^{2}\!\big(T_{\rm eff} J_{0 \rightarrow z^*}\big),
\]
maximized at $T_{\rm eff}=\tfrac{\pi}{2 J_{0 \rightarrow z^*}}$, with leakage proportional to $\norm{\tau_{1}^2[\mathcal{G}_{+}, \mathcal{G}_{-}]}$.

As $T_{\rm eff}$ is inversely proportional to $J_{0\rightarrow z^*}$, at low $k$ and $p'$ we seek an initial state that reduces the second-order leakage. Let $\ket{w_1}=\tfrac{1}{\beta_-}\mathcal{G}_-\ket{v_0}$, with $\beta_- = \norm{\mathcal G_- \ket{v_0}}$, and $\beta_+ = \norm{\mathcal G_{+}\ket{v_0}}$. The state prepared by an initial walk over $\mathcal G$ is
\begin{equation}
  \ket{\psi_0} = \ket{v_0} - i \tau_0 \left( \beta_+ \ket{v_1} + \beta_- \ket{w_1} \right) + \mathcal O (\tau_0^2).
\end{equation}
To first order, the amplitude driving the chain is reduced by
\begin{equation}
  \cos \phi = \frac{1}{\sqrt{1 +  (\frac{\beta_-}{\beta_+} \tau_0)^2}},
\end{equation}
Consequently, the coupling $J_{0 \rightarrow z^*}$ is attenuated to the effective coupling $J_{\rm eff}(\tau_0) = J_{0 \rightarrow z^*} \cos(\phi)$,
with
\begin{equation}
\label{eq:product_state_effective_time}
  T_{\rm eff} = \tau_0 + p' \, \tau_1 = \frac{\pi}{2 J_{\rm eff}(\tau_0)}.
\end{equation}
Minimization of the second-order leakage leads to a unique positive-valued solution for $\tau_0$ and $\tau_1$. The second-order term $-\tau_1^2[\mathcal G_+, \mathcal G_-]$ returns amplitude from $\ket{w_1}$ to the positive subspace proportional to $\tau_1^2 \, \kappa_{\rm ret} \, \sin\phi$ and leaks amplitude from $\ket{v_1}$ proportional to $\tau_1^2\kappa_{\rm leak}\cos\phi$, where $\kappa_{\rm ret}$ and $\kappa_{\rm leak}$ are graph-dependent constants. These channels are balanced when
\begin{equation}
  \frac{\kappa_{\rm leak}}{\kappa_{\rm ret}} \, \tan\phi \approx 1.
\end{equation}
As
\begin{equation}
  \tan\phi = \frac{\beta_- \, \tau_0}{\sqrt{1 + \tau_0^2\beta_+^2}},
\end{equation}
to first-order the optimal leakage-reducing $\tau_0$ is
\begin{equation}
\label{eq:tau_0_star}
  \tau_0^*= \frac{\kappa}{\sqrt{\beta_- - \kappa^2 \, \beta_+}},
\end{equation}
where $\kappa = \kappa_{\rm leak}/\kappa_{\rm ret}$, and
\begin{equation}
\label{eq:tau_1_star}
  \tau_1^*=\frac{1}{p'}\left[ \frac{\pi}{2 J_{\rm eff}(\tau_0^*)} - \tau_0^* \right].
\end{equation}
For odd $p$, \cref{eq:tau_1_star} still holds, but the absence of a full $+/-$ pair leaves a linear $\mathcal G_-$ contribution. The resulting leakage scales as $\mathcal O\big(\tau_1-\tau_0\big)$, rather than $\mathcal O(\tau_1^2)$ as in the even (paired) case. In practice, this can be mitigated by further attenuating $J_{\rm eff}$ via a larger $\tau_0$. More generally, while higher-order terms contribute significantly if $\tau_0,\tau_1 \not\ll 1$, near $(\tau_0^*,\tau_1^*)$, quadratic curvature from second-order terms still dominates both the linear odd-$p$ residual leakage and any higher-order corrections. The result is a locally smooth and convex objective function -- with local optimization initiated at $(\tau^*_0, \tau^*_1)$ reliably converging to the unique optimum\footnote{For the unconstrained hypercube, $[\mathcal G_+, \mathcal G_-]=0$. Consequently $R_{-}(\tau_1)R_{+}(\tau_1) = \exp(-2 i \tau_1 \, \mathcal G_+)$ exactly, and perfect transfer is trivial at $p'=1$ with $\tau_0 = 0$.}.

\begin{figure*}[t]
    \centering
    \includegraphics[width=0.9\linewidth]{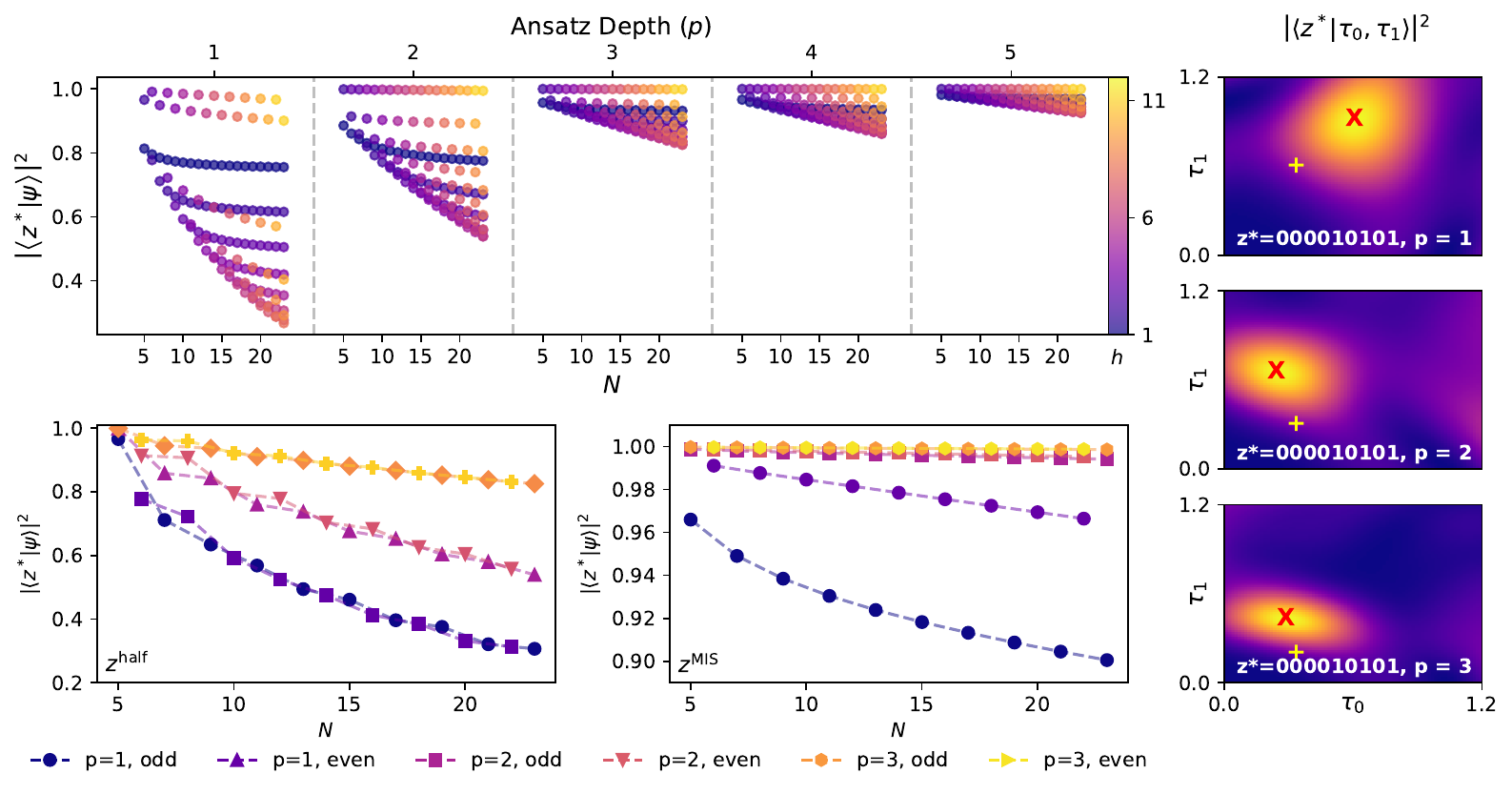}
\caption{Success probability of preparing product states $\ket{z^*} = \ket{0^{N - 2h}(01)^h}$, where $h = 1$ to $\lfloor N/2 \rfloor$, using optimized walk times $\tau_0$ and $\tau_1$, for system sizes $N = 5$ to $23$. \textbf{Top-left} shows success probability across ansatz depths $p = 1$ to $5$ colored by Hamming weight ($h$). \textbf{Bottom-left} highlights the reference states $z^{\rm half}$ and $z^{\rm MIS}$ (see \cref{sec:target_states}). \textbf{Right} shows success probability as a function of $\tau_0$ and $\tau_1$ for the $N=9$ instance of $z^{\rm half}$ at depths $p = 1$, $2$, and $3$. The optimal points (given in \cref{tab:product}) are marked with a red 'X', and theoretical predictions from \cref{eq:tau_0_star,eq:tau_1_star} are indicated by a yellow '+'.}
    \label{fig:ansatz_1}
\end{figure*}

\Cref{fig:ansatz_1} shows the result of running Nelder-Mead from the analytic starting point $(\tau_0^*,\tau_1^*)$ for system sizes $N=5$ to $23$ at depths $p=1$ to $5$.  In every case the solver converges to an optimum in a single basin.  For the largest system $N=23$, the worst-case success probability over all bit-string targets at $p=1$ is only 0.267 (occurring at Hamming weight $h=8$), but this minimum probability rises to 0.925 by $p=5$.  Conversely, the $h=11$ state achieves a probability of 0.901 at $p=1$ and essentially unity at $p=5$. At fixed $N$ the probability profile as a function of Hamming weight is well described by a skewed quadratic, with its lowest point near $h\approx\lfloor N/4\rfloor$ and monotonic increase towards both the all-zero and maximum Hamming weight. The $(\tau_0,\tau_1)$ for the optimum depicted in the right panel are listed in \cref{tab:product}. Each of the three depicted instances show $(\tau_0^*,\tau_1^*)$ within the locally convex region of the optima. Over $p=1$ to $3$ the effective walk time $T_{\rm{eff}}$, as given by \cref{eq:product_state_effective_time}, approaches $\pi/2$.

\subsubsection{Bracelet states}
\label{sec:optimization_bracelet}

\begin{figure}[t!]
    \centering
    \includegraphics[width=0.90\columnwidth]{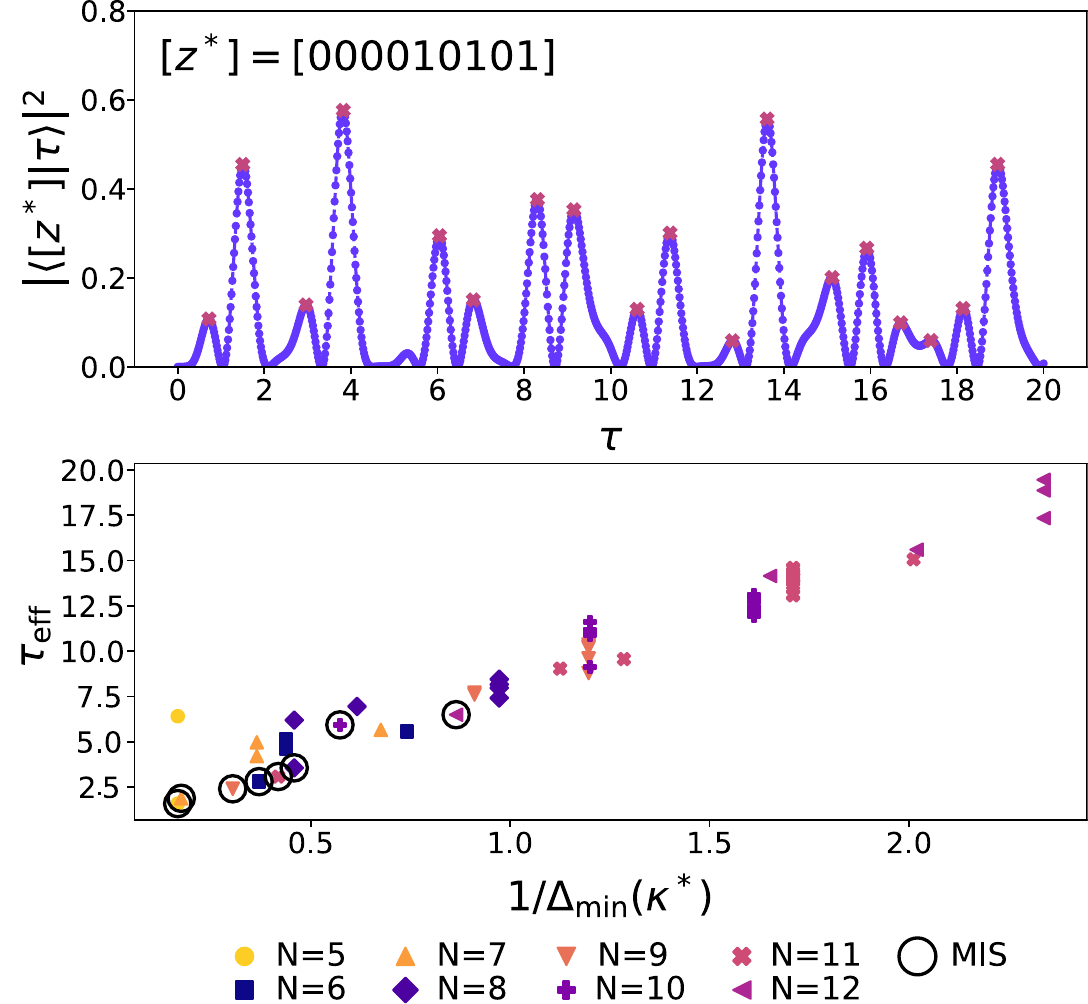}
    \caption{Bracelet state preparation. \textbf{Top}: Population of the $N=9$ target bracelet state $\ket{[z^{\rm half}]}$ during a CTQW from $\ket{0}$, as a function of $\tau$, sampled at $\tau_j = j\,\Delta\tau$ with $\Delta\tau = 0.02$ over $\tau \in [0,20]$. Times at which the population reaches a local maximum exceeding the threshold $1/|D_N|$ (denoted $\tau^{(m)}_{\rm peak}$) are marked with pink `X's; these serve as fixed walk times in the variational optimization of the phases $\bm\gamma$. \textbf{Bottom}: Accumulated walk time $\tau_{\mathrm{eff}}$ required to reach a success probability $\ge 0.98$, plotted against the inverse resolvable spectral minimum $1/\Delta_{\min}(\kappa^*)$ (see \cref{eq:resolvable_spectral_minimum,eq:bracelet_tau_eff}), for all dihedrally distinct targets $\ket{[z^*]}$ with $N = 5,\dots,12$. We fix $\kappa^* = 7.2 \pm 0.4$ from a stability plateau, identified by maximizing $\mathrm{mean}(r) - 2\,\mathrm{std}(r)$ over sliding $\kappa$ windows (where $r$ is the fit residual) and taking the window center. Maximum-independent-set (MIS) targets are circled in black.}
    \label{fig:ansatz_2}
\end{figure}

The optimization strategy for preparation of bracelet states is guided by two practical considerations, as expanded in \cref{sec:rydberg_atoms}. First, we aim to minimize the total evolution time and, second, utilize the hardware-native realization of the phase separator
\begin{equation}
\label{eq:bracelet_state_phasor_unitary}
    \hat U_h(\gamma) = e^{-i \gamma \hat n},
\end{equation}
where $\hat n = \sum_{i = 1}^N\ket{1}_i\bra{1}_i$ applies a uniform phase to all computational basis states with the same Hamming weight.

The initial stage sweeps the walk generator alone,
\begin{equation}
    \ket{\psi(\tau)} = e^{-i \tau \mathcal G} \ket{0}
\end{equation}
at discrete times $\tau_j=j\Delta\tau$ for $0 < \tau_j \leq \tau_{\rm{max}}$. At each sample we measure the population of the target bracelet subspace $\left|\langle [z^*] | \psi(\tau_j) \rangle\right|^2$ and locate the local maxima $\tau^{(m)}_{\rm{peak}}$ higher than $1/|D_N|$, as demonstrated in the top panel of \cref{fig:ansatz_2}. 

Starting from the second identified peak, we fix the total walk duration to $\tau_{\rm{tot}}=\tau^{(m)}_{\rm{peak}}$ and set the ansatz depth and walk times to,
\begin{equation}
\label{eq:bracelet_p_tau}
    p = \left\lfloor \frac{\tau_{\rm{tot}}}{\tau_{\rm{min}}^{\rm{hw}}} \right\rfloor - 2, \qquad \tau = \frac{\tau_{\rm{tot}}}{p + 1}
\end{equation}
where $\tau_{\rm{min}}^{\rm{hw}}$ is the minimum allowed walk time ($\sim0.4$) and the $-2$ guards against quantization error arising through hardware constraints so the walk times are fixed and only the phase separator $\vec \gamma = \{\gamma_k\}_{k=1}^p$ remain variational. As the resulting landscape is non-convex, we optimize these with the derivative-free COBYLA optimizer initialized at $\gamma_k=0$, stopping the scan over peaks when there is a decrease in success probability or the range of identified peaks is exhausted.

We discard the first peak because the state is concentrated in the $C_N$-symmetric (translation-invariant) subspace. Since the phase separator depends on Hamming weight, it acts as $e^{i\gamma h}\mathbb I$ on each weight-$h$ sector and therefore cannot distinguish distinct $D_N$-symmetric components. The total walk time $\tau_{\rm{tot}}$ must be long enough to accumulate relative phase between $D_N$-symmetric components at fixed $h$ before optimization over $\vec \gamma$ can appreciably improve the success probability.

More generally, the required $\tau_{\rm eff}$ for high-fidelity preparation of $\ket{[z^*]}$ is explainable with a frequency-resolution model. Let $\{(\lambda_r,|r\rangle)\}_{r=1}^d$ be the eigenpairs of the walk generator $\mathcal G$ restricted to the $D_N$-invariant subspace~\cite{Krovi2007}. Because the only interleaved control is the phase separator, resolvability of the target state is governed by spectral gaps $\lambda_{rs}=|\lambda_r-\lambda_s|$ between eigenmodes $r, s \in \{1, \dots, d\}$ that both have appreciable support on the same Hamming-weight sector as the target~\cite{yosi2021}. Consequently, control under variation of $\vec \gamma$ requires order-unity accumulated phase, $\lambda_{rs}\,\tau_{\mathrm{eff}}=\Theta(1)$.

We encode this threshold by introducing the dimensionless constant $\kappa=\Theta(1)$ and enforcing $\lambda_{rs} \tau_{\rm eff} \ge \kappa$. Among resolvable pairs, we then define the resolvable spectral minimum,
\begin{equation}
\label{eq:resolvable_spectral_minimum} 
\Delta_{\text{min}}(\kappa)
=\min_{\lambda_{rs}\ge \kappa/\tau_{\mathrm{eff}}}\lambda_{rs},
\end{equation}
and the model predicts that this low-frequency component sets the timescale, 
\begin{equation}
\label{eq:bracelet_tau_eff} 
\tau_{\mathrm{eff}}\propto \frac{1}{\Delta_{\text{min}}(\kappa)}.
\end{equation}
To identify which gaps are associated with the target Hamming-weight, $h^*=h([z^*])$,  we expand each eigenmode $\ket{r}$ in the bracelet basis $\{\ket{[z]}\}$ and compute its weight in the target sector 
\begin{equation}
u_r=\sum_{\,h([z])=h^*}\big|\braket{[z] | r}\big|^2\in[0,1],
\end{equation}
with $\Delta_{\min}(\kappa)$ chosen from the eigenmodes with non-zero weight. 

The bottom panel of \cref{fig:ansatz_2} shows the accumulated walk time $\tau_{\rm{eff}}$ required to reach high-fidelity preparation against $1/\Delta_{\rm min}(\kappa^*)$ for all dihedrally distinct bracelet targets $\ket{[z^*]}$ at $N=5$ to $12$. These numerical results are consistent with the scaling relationship predicted by the frequency-resolution model. Across all targets, $\tau_{\rm{eff}}$ scales linearly with $1/\Delta_{\rm min}(\kappa^*)$ for a single fixed $\kappa^*$, consistent with \cref{eq:bracelet_tau_eff}. This $1/\Delta_{\rm min}$ scaling stands in contrast to adiabatic state preparation, where the required evolution time scales as $1/\Delta_{\rm min}^2$~\cite{Albash_2018}. The improvement arises because the phase-walk ansatz operates in a non-adiabatic regime, creating a superposition of eigenmodes whose relative phases evolve and interfere constructively on the target rather than adiabatically tracking a single instantaneous ground eigenstate.

Within each $N$, targets that are a maximal independent set consistently exhibit the smallest $\tau_{\rm eff}$ and the largest $\Delta_{\min}$. This is consistent with prior work on \emph{scarred} quantum walks \cite{desaules2022hypergrid}, which places these states on or near a small, low-degree hypercube-like subgraph weakly coupled to the rest of the graph -- with dynamics confined to this subgraph producing sparser, quasi-regular spectra with larger internal gaps, resulting in faster preparation.

\section{Rydberg atom arrays}
\label{sec:rydberg_atoms}

Neutral atom quantum computers are a promising modality for various applications, from gate-based computing \cite{Saffman2016,Morgado2021},  fault-tolerant algorithms \cite{Bluvstein2023,reichardt2025,rodriguez2024,bluvstein2025}, and Hamiltonian simulation \cite{Shaw2024,evered2025,Gonz_lez_Cuadra_2025}. In a neutral atom computer, individual atoms, such as Rubidium 87, are trapped in optical lattices with the qubit encoded into the electronic states of valence electrons \cite{Evered2023}. Quantum operations are implemented by manipulating the electronic state of the atom using finely tuned lasers. Entanglement is mediated by the Rydberg state, a highly excited orbital that strongly interacts via a van der Waals interaction with nearby atoms in the Rydberg state.

In this work, we focus on \emph{analog mode} neutral atom computers, and specifically that of Aquila, QuEra Computing's cloud analog Hamiltonian simulator \cite{wurtz2023aquila}. Here, the qubit is encoded into a ground state $\ket{g}=\ket{5S\frac{1}{2}}$ and Rydberg state $\ket{r}=\ket{70S\frac{1}{2}}$ of Rb87. Due to adjacent Rydberg states constantly interacting, an analog mode computer implements the time-dependent dynamics of an Ising-like Hamiltonian
\begin{multline}
    H(t) = \frac{\Omega(t)}{2}\sum_i e^{i\phi(t)}|g_i\rangle\langle r_i|
    + e^{-i\phi(t)}|r_i\rangle\langle g_i|\\ 
     - \sum_i\big(\Delta(t) + w_i\delta(t)\big)|r_i\rangle\langle r_i|
    + \sum_{ij}V_{ij}|r_ir_j\rangle\langle r_ir_j|.
\end{multline}
The first term is a Rabi drive, which coherently drives each atom between the ground and Rydberg state. The second term is the detuning, which is a Pauli-Z-like term that applies an energy penalty to the Rydberg state. The detuning is separated into a global detuning $\Delta(t)$ that acts on every qubit uniformly, and a local detuning $w_i\delta(t)$ that acts on individual atoms $i$ with weights $w_i$ and time-dependent value $\delta(t)$. The third term is the interaction $V_{ij} = C_6/|\vec x_i - \vec x_j|^6$ between atoms at positions $\vec x_i$ and $\vec x_j$ respectively. In systems such as Aquila used in this work, the typical Rabi frequency is $\sim$2.5MHz and the characteristic inter-atomic distance is $\sim8\mu$m, where up to 256 atoms can be arbitrarily positioned in 2d space.

The goal of hardware implementation is to abstract away the physical Hamiltonian so that it optimally matches the generator of continuous time quantum walks
\begin{equation}
    \mathcal T e^{-i\hbar \int_0^T H(t) dt} \approx \mathcal T e^{-i\int_0^\tau G(\tau)d\tau},
\end{equation}
where $G(t)=X(t)\hat{\mathcal G} + Z_i(t) \hat C$ is the generator of quantum walks in the independent set subspace spanned by the projector $\mathcal P$; $X(t)$ and $Z(t)$ are piecewise constant pulses that recreate the alternating ansatz. The states can be identified with bitstrings by matching $|g\rangle \mapsto |0\rangle$ and $|r\rangle\mapsto |1\rangle$. The Rabi drive can be matched to the $\sigma_x$ term by rescaling time $t = 2\tau / \Omega$.

\begin{figure}
    \centering
    \includegraphics[width=1.0\linewidth]{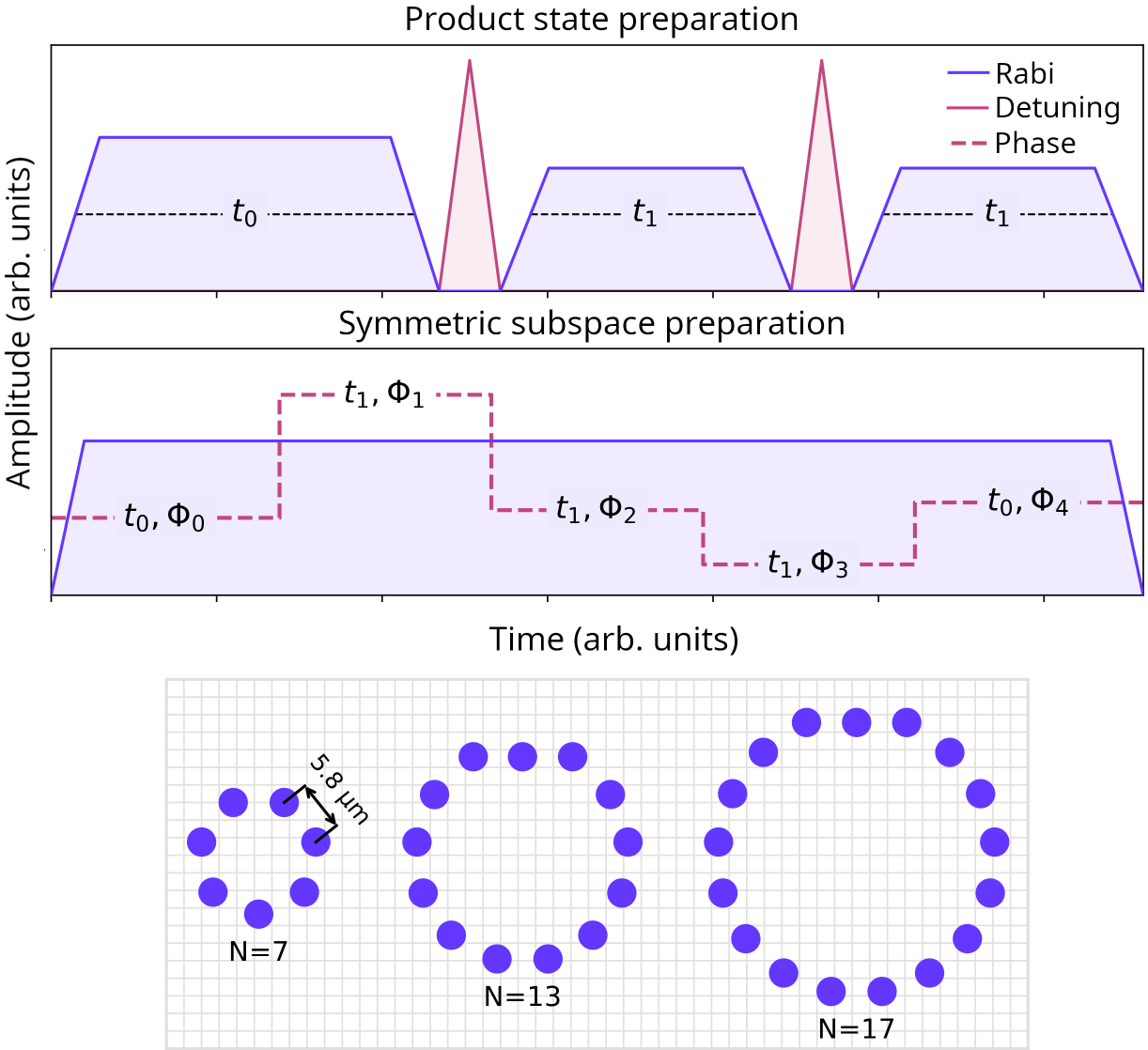}
    \caption{Representative Analog programs to implement quantum walk state preparation. \textbf{Top} plots the analog waveforms for product state preparation. Purple is the waveform for the Rabi frequency, which implements the quantum walk Hamiltonian where the total area of each trapezoid implements a walk of unitless time $\tau_i$. Red is the waveform for the local detuning, which implements the phase-separator $Z$ term on a subset of sites; each triangle has an integrated area of $\pi/2$. The Rabi term is turned off during the phase accumulation to avoid any non-commutative terms. \textbf{Middle} plots the analog waveforms for bracelet state preparation. The red dashed line is the Rabi phase, which is U(1) gauge equivalent to the phase separator via phase jumps. \textbf{Bottom} plots the atom positions for 7, 13, and 17 atoms, with a background grid spacing of $2~\mu$m. Observe that the top of each circle and bottom of $N=13$ and $17$ are flattened to conform to the row constraints of Aquila~\cite{wurtz2023aquila}.}
    \label{fig:analog_program}
\end{figure}

A phase jump of the Rabi drive $\phi(t) = -d\phi\Theta(t-t_0)$ can implement a global phase jump $Z_i(t)= d\phi \delta(t-t_0)$ like that used in state preparation. Local detuning can be used to implement local phase jumps by quickly turning off the Rabi drive, then implementing a fast pulse of local detuning $w_i\delta(t)$ on the target atoms such that $d\phi_i = w_i\int_0^T \delta(t)dt$. On Aquila, this is most naturally done with a 100~ns triangle pulse with an area of the maximum phase jump, as shown in \cref{fig:analog_program}.

The projection to the independent set subspace can be implemented with a key feature of the Rydberg atom Hamiltonian: the \emph{Rydberg blockade}. Due to the strong sixth power of the interaction with the distance between atoms, two atoms close together with a distance less than the \emph{blockade radius} $|\vec x_i-\vec x_j|<r_b$ will have an energy of the doubly-excited Rydberg state that is much higher than any other scale in the problem. These doubly-excited states can then be integrated out of the dynamics, resulting in an effective subspace of all excitations in the space of independent sets \cite{Bernien2017,Ebadi_2022}. The graph connectivity is defined by a unit disk graph where the radius is analogous to the blockade radius of the Hamiltonian.

The blockade radius is defined by the Rabi frequency, detuning, and $C_6$ coefficient. The dynamic blockade radius, as used for encoding blockaded dynamics, is $r_{d}\equiv(C_6/\Omega)^{1/6}$. The static blockade radius, as used for encoding the maximum independent sets into Ising ground states \cite{Ebadi_2022}, is $r_s=(C_6/\Delta)^{1/6}$. For Aquila, $\Omega$ has a maximum of $15.8$ rad/$\mu$s and $C_6=5420503$ $ \mu m^6$rad/$\mu$s, so that the dynamic blockade radius is $r_d\approx 8.367~\mu$m.

To preserve the independent set condition, adjacent atoms must be close together to ensure that they are well within the blockade radius. However, to preserve independent set dynamics, nearby atoms outside of the unit disk radius must be far enough apart to ensure that the $1/R^6$ interaction tail does not spuriously add unwanted interactions. These two factors can be balanced by considering that there is no singular unit disk radius; instead, there is a range of radii that generate the same unit disk graph. The minimum unit disk radius $r_{min}$ is the maximum distance between any two vertices connected by an edge; the maximum unit disk radius $r_{max}$ is the minimum distance between any two vertices not connected by an edge; as an example, consider \cref{fig:rmin_rmax}. To best preserve dynamics, the energy associated with the blockade radius should be much larger (smaller) than the minimum (maximum) distance between atoms
\begin{equation}
    \frac{C_6}{r_\text{min}^6}\gg \frac{C_6}{r_\text{b}^6} \gg \frac{C_6}{r_\text{max}^6}.
\end{equation}
More specifically, the perturbative Schrieffer-Wolff corrections arising from finite blockade energies must be much smaller than the energy-scale of the Rabi drive. This condition can be maximally satisfied by choosing the blockade radius to be $r_\text{b}=\eta \,\sqrt{r_\text{min} r_\text{max}}$ where $\eta\approx0.936$ for the atom geometries explored in this work. For more details, see \cref{sec:SW} and the supplemental of \cite{Bluvstein2021}.

The space of independent sets of an $N$-vertex nearest-neighbor ring can be represented by an $N$-atom ring of atoms in a 2d atom array. Given the ring where each atom is a distance $D$ from the origin, $r_\text{min}=2D\sin(\pi/N)$ and $r_\text{max}=2D\sin(2\pi/N)$. The distance is
\begin{equation}
    \label{eq:atom_distance}
    D=\frac{r_d}{2\eta\sqrt{\sin(\pi/N)\sin(2\pi/N)}}\approx r_d\times\frac{1}{\sqrt{2}}\times\frac{N}{2\pi}.
\end{equation}

In aggregate, it is possible to physically reproduce the state prepared by the phase-walk ansatz (see \cref{eq:phase_walk_ansatz}) by implementing a piecewise-constant, time-dependent evolution with respect to a walk generator
\begin{equation}
    \ket{\psi}
    = \mathcal T \exp\!\bigg\{-\,\mathrm{i}\!\int_0^T \big[X(\lambda)\,\hat{\mathcal G} + Z(\lambda)\,\hat C\big]\, d\lambda\bigg\}\ket{\psi_0},
\end{equation}
where $X(\lambda)$ and $Z(\lambda)$ are non-overlapping, piecewise-constant waveforms whose fragments have area $\tau_q$ or $\gamma_q$,
with an equivalent generator of the Rydberg-atom Hamiltonian for a ring of blockaded atoms by matching the independent-set subspace to the Rydberg-blockade subspace and setting the ring radius as above based on the dynamic blockade radius, using the average Rabi-drive amplitude. The global detuning is set to zero throughout. Global phase shifts are implemented via rotating-frame phase shifts of the Rabi drive \footnote{Note that global phase shifts can be done equivalently through a time-integrated global detuning, but this requires a longer physical evolution time that yields a lower fidelity result.}. Local phase shifts are implemented via the local detuning term by choosing a waveform $\delta(t)$ such that $\int_0^Tw_i\delta(t)dt=\phi_i$ in the physical interval $[0,T]$. Global $\sigma_x$ accumulation is implemented by choosing a waveform $\Omega(t)$ such that $\int_0^T \Omega(t)dt=2\tau$ over the minimized physical interval $[0,T]$. The blockade radius and thus the distance between each atom is set by the average of $\Omega(t)$ and thus varies slightly depending on the ansatz. Waveforms and positions are minimally relaxed to conform to slew rate and atom placement constraints inherent to Aquila \cite{wurtz2023aquila}. An example \texttt{bloqade} program implementing a single walk step is shown in \cref{python:bloqade}, and a representative program is shown in \cref{fig:analog_program}.

\begin{table}
\begin{python}
import numpy as np
import bloqade.analog as ba # QuEra's SDK

# --- Input parameters ---
N = 20 # Set the number of atoms
tau = 2*np.pi*1.2345 # Unitless evolution time
Omega = 2*np.pi*2.5 # Aquila's Rabi frequency
# ---

# Evolution time in usec
evolution_time = tau / Omega
# Dynamic blockade radius in um
d = (5_420_503/Omega)**(1/6)
D = d/(2*np.sqrt(
        np.sin(np.pi/N)*np.sin(2*np.pi/N)))

D *= 1.0 # Optionally add a variational
# fudge factor to optimize the blockade radius.

theta = np.linspace(0,2*np.pi,N+1)[0:N]
positions = np.array([D*np.sin(theta),
                      D*np.cos(theta)]).T
positions = [tuple(q) for q in positions]
positions = ba.atom_arrangement.\\
    ListOfLocations(positions)
rabi_drive =  ba.constant(
              duration=evolution_time,
              value=Omega)
detuning = ba.constant(
              duration=evolution_time,
              value=0.0)
program = ba.rydberg_h(
              atoms_positions=positions,
              amplitude=rabi_drive,
              detuning=detuning)
# Run using emulation
data = program.bloqade.python().run(100)
# Run using Aquila via BraKet
data2 = program.braket.aquila().run_async(100)
\end{python}
\caption{A code snippet outlining a simple \texttt{bloqade.analog} program, which implements a single quantum walk step for a nearest neighbor blockaded ring. The parameters at the top define the program: \texttt{N} defines the number of atoms in the ring; \texttt{phi} sets the total (unitless) evolution time of the quantum walk; \texttt{Omega} defines the Rabi frequency as set by the capabilities of Aquila. Note that this program does not fit Aquila's hardware constraints, such as row spacing and rise time, which have been excluded for simplicity.}\label{python:bloqade}
\end{table}

\section{Results: Preparation of product and bracelet states}
\label{sec:state_pre_results}

We present results for the preparation of the product states $z^{\rm half}$ and $z^{\rm MIS}$, together with their corresponding bracelet states $[z^{\rm half}]$ and $[z^{\rm MIS}]$, as described in \cref{sec:target_states}, using the two phase-walk ans\"{a}tze introduced in \cref{sec:optimization}. These are realized in three scenarios: the ``perfect'' CTQW dynamics, its approximation by the Rydberg Hamiltonian in noiseless emulation, and implementation on Aquila, sharing the same variational parameters across each case. Our comparison focuses on the success probability and amplification relative to a uniform baseline, contrasting the observed amplification as a function of both the cardinality of the invariant (blockaded) subspace $|\mathcal V|$ and the ansatz depth $p$, following the power-law scaling relationships introduced in \cref{sec:performance_metrics}. Particular attention is given to the extent to which the scaling behavior predicted by the ideal CTQW is qualitatively reproduced on noisy hardware.

\begin{figure*}
    \centering
    \includegraphics[width=\linewidth]{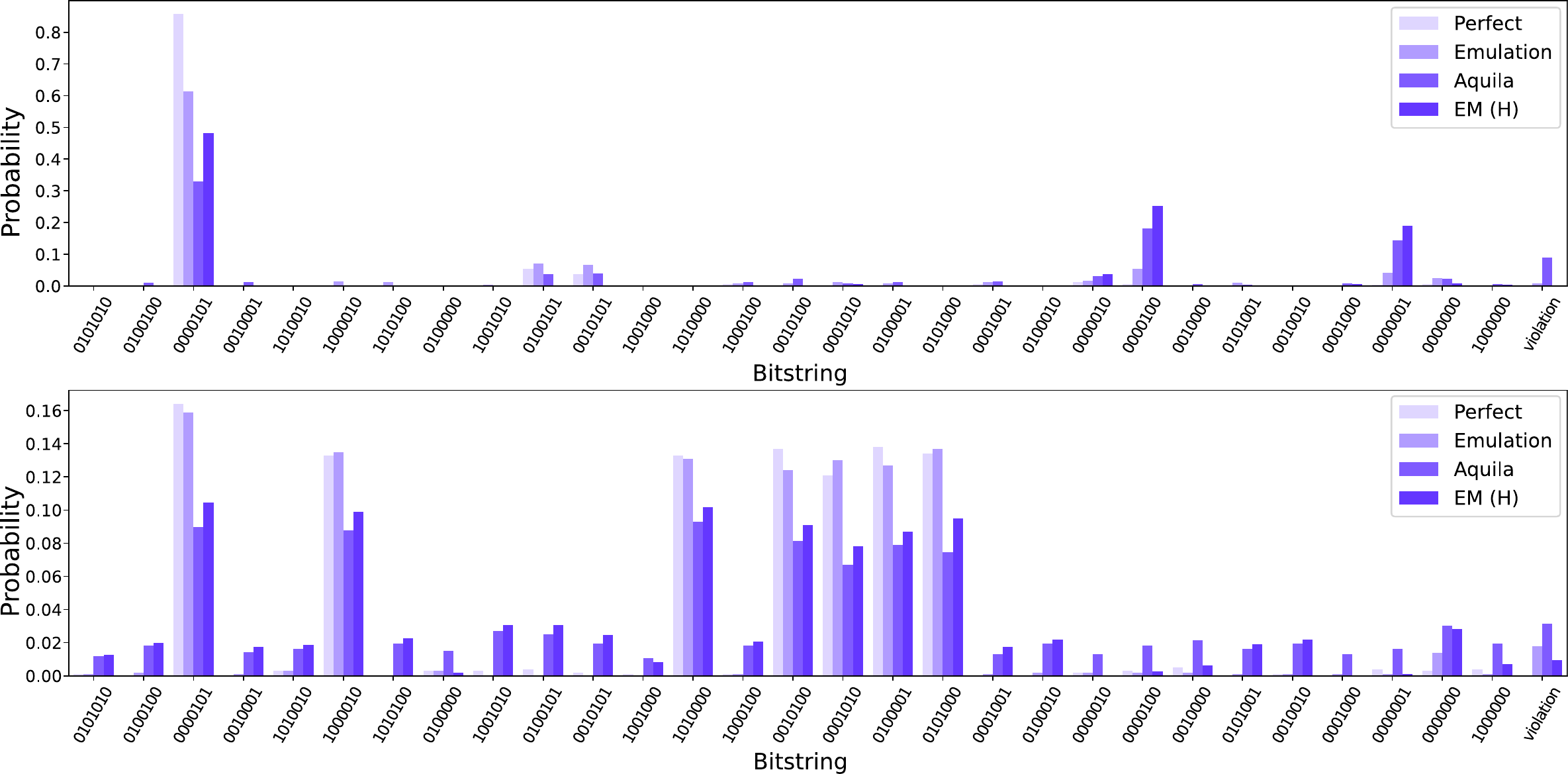}

    \vspace{0.5em}
    
    \small
    \setlength{\tabcolsep}{4pt}
    \renewcommand{\arraystretch}{1.1}
    \begin{tabular}{l c c c c c c}
    \toprule
    State & Depth ($p$) & Na\"ive Count & Perfect & Emulation & EM (H) & EM (E) \\
    \midrule
    $0000101$   & 2  & 305/925 & 0.857 & 0.613 & $0.47\,[0.43, 0.50]$ & $0.63\,[0.58, 0.68]$ \\
    $[0000101]$ & 19 & 529/924 & 0.960 & 0.943 & $0.66\,[0.62, 0.69]$ & $0.93\,[0.90, 0.95]$ \\
    \bottomrule
    \end{tabular}
    \caption{Probability distributions for preparation of the $z^{\rm half}$ product state and the corresponding bracelet state at $N=7$, obtained from numerical simulation of the ideal CTQW dynamics (``Perfect''), noiseless emulation of the Rydberg Hamiltonian (``Emulation''), the raw probability distribution from 1000 shots on Aquila, and the reconstructed distribution (``EM'') that accounts for measurement error. \textbf{Top} shows the product state distribution. \textbf{Middle} shows the bracelet state distribution. The table below summarizes the corresponding success probabilities: the \textbf{State} column labels the target state; \textbf{Ansatz Depth} is the number of walk layers; the \textbf{Na\"{i}ve count} column gives the fraction of bitstrings matching the target directly from raw counts; the \textbf{Perfect} column reports the preparation probability under ideal CTQW dynamics; the \textbf{Emulation} column shows the same quantity using noiseless Rydberg-atom emulation. The \textbf{EM (H)} column gives the success probability from Aquila hardware data after mitigation of measurement errors, while \textbf{EM (E)} reports the corresponding value from 1000 emulated shots convolved with the noisy measurement channel. Agreement between EM (E) and the Emulation column validates that the EM procedure correctly recovers the underlying distribution. Data for all target states are shown in \cref{tab:product,tab:bracelet} of \cref{app:table_details}.}
    \label{tab:stateprep}
\end{figure*}

The unitless walk times $\tau$ are mapped to the minimum feasible physical hardware time given a maximum Rabi frequency of $\Omega = 15.8$ rad/$\mu$s. Evolution under the walk generator is implemented with a hardware-minimum rise and fall time of $0.05~ \mu$s~\cite{wurtz2023aquila}. Consequently, there are four regimes of Rabi drive pulse length. For walk times $\tau < 0.40$, the pulse is triangular with a constant duration of $0.10~\mu$s. Between $0.40 \leq \tau < 0.59$ the pulse is also triangular, with a duration of $\tfrac{4\tau}{15.8}\,\mu$s, increasing from $0.10$ to $0.15~\mu$s. Between $0.59 \leq \tau \leq 0.79$, the optimal pulse is a reduced-amplitude trapezoid of duration $0.15~\mu$s. Beyond $\tau = 0.79$, the pulse duration grows linearly as $0.127\,\tau + 0.05~\mu$s.

The blockaded subspace is encoded into a ring of atoms arranged in a ring, with nearest-neighbor distances given by \cref{eq:atom_distance}. 
Atom positions are rounded to the nearest $0.1~\mu$m, and the tops and bottoms of the rings are flattened to conform to the $2~\mu$m row spacing constraints of Aquila \cite{wurtz2023aquila}. Some example atom positions are shown in \cref{fig:analog_program}. All data from Aquila was acquired using QuEra's exclusive access mode during the month of August 2025, and all shots were post-selected on fully filled arrays.

There is added complexity when using Aquila due to a measurement error of $\sim 7\%$ in misidentification of the Rydberg state as ground and, when using local detuning, an additional $\sim 10\%$ error in misidentification of the ground state, which renders naive frequentist bitstring counting unreliable for large bitstrings. To address this, we employ a Bayesian postprocessing method based on expectation maximization (EM, see~\cref{sec:bayes}), which reconstructs the pre-measurement probability distribution by modeling readout as an asymmetric bit-flip channel, with the probability of the target state taken from the reconstructed distribution unless otherwise noted. Uncertainties in the estimated probabilities are quantified using a nonparametric bootstrap to obtain 95\% confidence intervals.  In \cref{sec:product_state_results}, we fit the amplification $A$ to \cref{eq:amp_power_law} via weighted nonlinear least squares, taking the 95\% CI for $\alpha$ from the fit covariance under the constraint $\alpha < 1$. The intervals are propagated through $n = 1/(1-\alpha)$, with $n\to\infty$ (reported as ``$\ge$'') when the upper bound reaches one.

\begin{figure*}
    \centering
    \includegraphics[width=\linewidth]{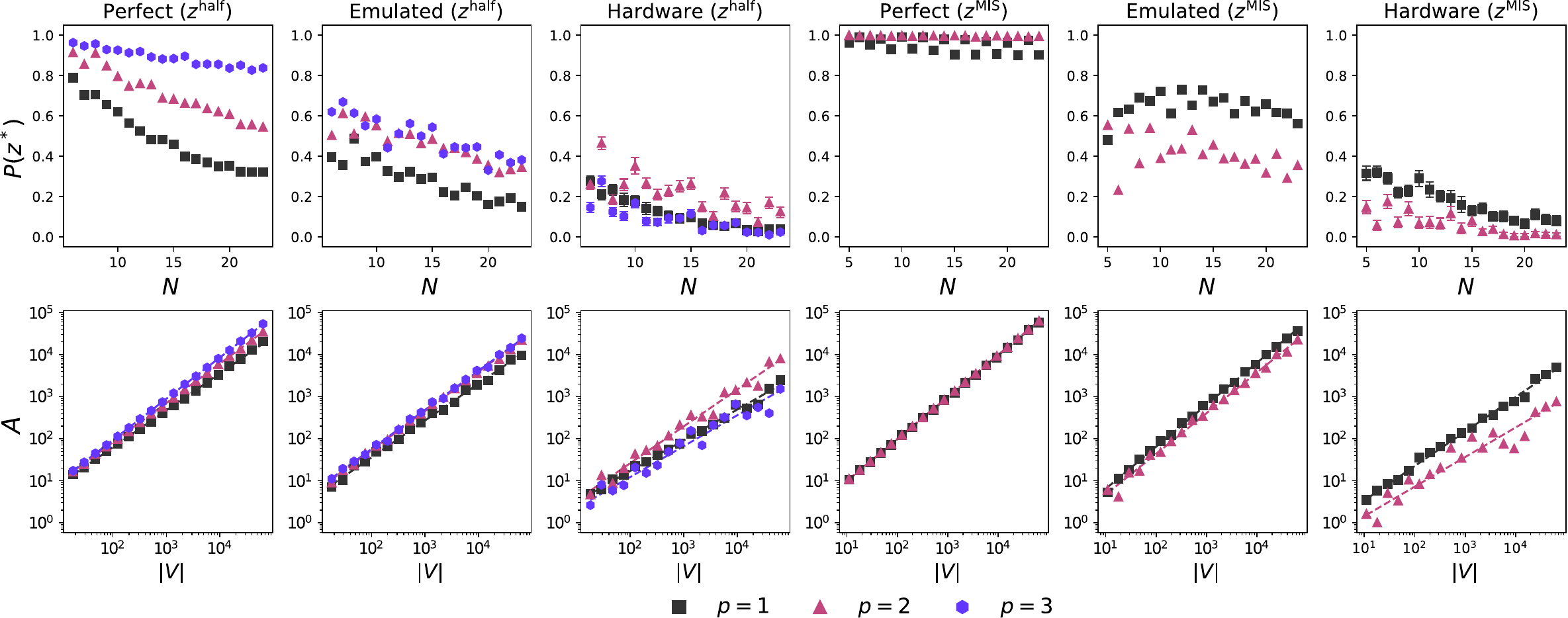}
    \caption{Comparison of success probabilities and amplification factors for $z^{\rm half}$ and $z^{\rm MIS}$ across CTQW (Perfect), noiseless Rydberg emulation, and Aquila hardware. \textbf{Top} shows success probabilities $P(z^*)$, with $z^{\rm half}$ on the \textbf{left} and $z^{\rm MIS}$ on the \textbf{right}. \textbf{Bottom} shows amplification factors $A = |\mathcal V|\,P(z^*)$. Results are shown at depths $p=1,2,3$ for $z^{\rm half}$ and $p=1,2$ for $z^{\rm MIS}$. Amplification plots include power-law fits to \cref{eq:amp_power_law} (dashed lines); fit parameters are reported in \cref{tab:power_law_fits}.}
    \label{fig:4_product_state_all_platforms}
\end{figure*}
\begin{figure}
    \centering
    \includegraphics[width=0.85\linewidth]{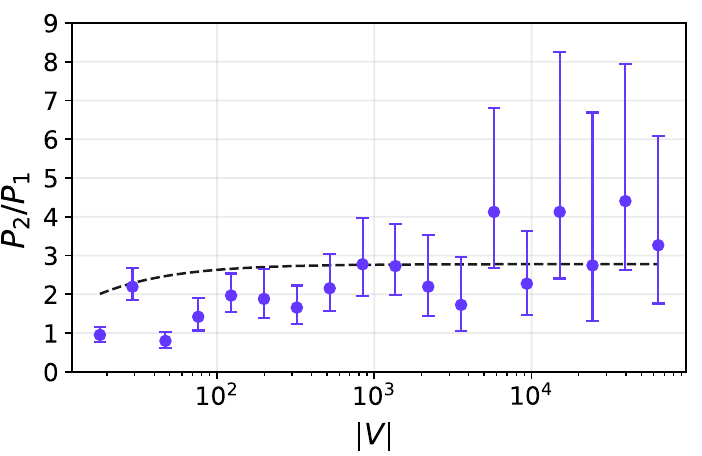}
    \caption{Ratio of success probabilities for preparation of $z^{\rm half}$ at depth $p=2$ ($P_2$) relative to depth $p=1$ ($P_1$) on Aquila. The black dashed line shows the corresponding ratio for a noiseless restricted depth Grover's search over a database of size $|\mathcal V|$ with one marked element.}
    \label{fig:z_half_product_depth_ratio}
\end{figure}

\Cref{tab:stateprep} presents the unprocessed probability distributions for the $z^{\rm half}$ product and bracelet states at $N=7$ across the three considered scenarios, together with the success probabilities obtained from $1000$ shots in noiseless emulation convolved with the asymmetric bit-flip channel, and from the same number of shots on Aquila. Here, and for all other considered target states (see \cref{tab:product,tab:bracelet}), the reconstructed distributions obtained from the scrambled emulation results agree with the true values within the 95\% confidence intervals.

\subsection{Product state preparation}
\label{sec:product_state_results}

The walk parameters for the target $z^{\rm half}$ and $z^{\rm MIS}$ product states are obtained through ideal simulation of a quantum walk generator, that is, a walk without Rydberg interactions, blockade violations, and finite rise times. They are then directly transferred, with no further optimization, to the time-dependent Rydberg Hamiltonian as described above. The local phase separator (see \cref{eq:product_state_phasor}) is encoded into a minimum-duration triangle pulse with a $50\,$ns rise and fall time commensurate with the constraints of Aquila \cite{wurtz2023aquila}, and a peak value of $\Delta_\text{max} = \phi/(50\,\text{ns}) \leq 62~\text{rad}~/\mu\text{s}$. 

In \cref{fig:4_product_state_all_platforms}, the $z^{\rm half}$ targets show higher scaling exponents for both the CTQW reference and noiseless emulation. Via \cref{eq:polynomial_speedup}, the CTQW fits correspond to effective polynomial speedup orders $n = \{8.17 \,[7.80\!-\!8.58],\; 15.3 \,[14.6\!-\!16.2],\; 53.1 \,[48.3\!-\!59.0]\}$, with emulation following closely at $p=1$ and $2$ with $n = \{8.15 \,[7.57\!-\!8.82],\; 15.1 \,[13.8\!-\!16.7]\}$, but diverging at $p=3$ with $n = 14.4 \,[13.2\!-\!15.7]$. On hardware, the same qualitative behavior is observed, albeit diminished: the scaling exponent increases from $p=1$ ($n = 3.70 \,[3.21\!-\!4.35]$) to $p=2$ ($n = 8.07 \,[6.69\!-\!10.1]$) but declines at depth $p=3$ ($n = 3.76 \,[3.15\!-\!4.66]$). 

\Cref{fig:z_half_product_depth_ratio} shows the ratio of success probabilities at $p=2$ relative to $p=1$. Although variability and uncertainty preclude a full fitting-based analysis, the increase in success probability appears to follow a generally rising trend with $N$ which, due to the proportional increase in amplification with $|\mathcal V|$, exceeds the convergence gain predicted by an ideal noiseless restricted-depth Grover search~\cite{zalkagrovers1999} for all even $N \geq 16$ and for odd $N=23$.

For $z^{\rm MIS}$ targets, the CTQW fits remain tightly clustered around unity across all depths. At $p=1$, the effective order is $n = 247 \,[200\!-\!322]$, and at $p=2$ within the limits of numerical precision, the amplification scales directly proportional to $|\mathcal V|$ over the range of considered $N$. This can be explained by noting that in $\mathcal G$ (here a ``Lucas'' cube~\cite{munarinilucas2001}), the subspaces defined by odd and even qubits each induce a hypercube of dimension $\lfloor N/2\rfloor$, containing the antipodal pair $0^N \leftrightarrow 1^N$~\cite{desaules2022hypergrid}, for which the hypercube graph is known to exhibit perfect state transfer~\cite{christandl2004perfect}. Emulated Rydberg dynamics show similarly strong scaling at depth $p=1$, with $n \geq 261 (\alpha \to 1)$, while depth $p=2$ yields a reduced effective order of $n = 30.5 \,[25.9\!-\!37.3]$. On hardware, amplification follows $n = 5.56 \,[4.79\!-\!6.63]$ at $p=1$, dropping to $n = 3.37 \,[2.75\!-\!4.34]$ at $p=2$.

Overall, these results are consistent with the super-linear amplification scaling in $|\mathcal V|$ expected of an efficient phase-walk ansatz, and to a lesser extent with increasing depth. The phase separator relies on local detuning, which carries a shot-to-shot per-site coherent error of roughly 10\%, leading to deviations from the intended $\pm\pi$ phase shifts that compound across iterations. This is likely the primary reason for the breakdown of increased convergence at higher $p$, in addition to the effects of incoherent errors from longer program times. It is therefore notable that the $z^{\rm half}$ targets still show improved convergence at $p=2$. For the $z^{\rm MIS}$ targets, the CTQW dynamics predict convergence near unity across the range of $N$. Consequently, the decrease in amplification from $p=1$ to $p=2$ in both emulation and on-hardware Rydberg dynamics is consistent with saturation of CTQW-based convergence at $p=1$, where further iterations offer little additional amplification and are outweighed by the loss in success probability from accumulated phase errors.

\subsection{Bracelet state preparation}

In bracelet state preparation, direct parameter transfer from the ideal CTQW-based ansatz to its approximation via the Rydberg Hamiltonian is unreliable due to the effect of van der Waals interactions over $\tau_{\rm{eff}}$, which is significantly longer than the $\tau$ used in product state preparation. These interactions introduce an additional positive phase through Rydberg-Rydberg couplings, contributing coherent phase error to the walk evolution and breaking the $-\gamma \equiv \gamma$ equivalence that holds for the ideal CTQW. We mitigate this effect through joint optimization of $\tau_{\rm{eff}}$ and $\bm\gamma$, taking as the objective the mean success probability
$$
\frac{1}{2}\left(P_{\rm CTQW}([z^*]) + P_{\rm Ryd}([z^*])\right),
$$
to identify parameter sets that achieve high convergence simultaneously in both cases, with the global phase separator (see \cref{eq:bracelet_state_phasor_unitary}) encoded into a piecewise constant phase profile of the Rabi drive. 

\begin{figure}
    \centering
    \includegraphics[width=0.8\columnwidth]{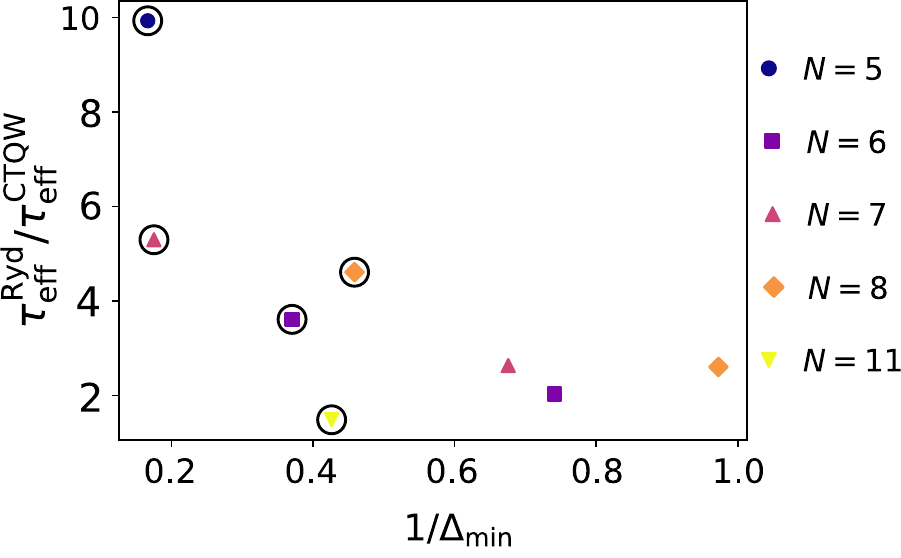}
    \caption{Comparison between the ideal CTQW-based bracelet-state preparation ansatz and its implementation via the Rydberg Hamiltonian for states $\left[z^{\rm half}\right]$ and $\left[ z^{\rm MIS}\right]$ (circled in black) that achieved $\frac{1}{2}\left(P_{\rm CTQW}([z^*]) + P_{\rm Ryd}([z^*])\right) \geq 0.8$ under joint parameter optimization. The horizontal axis shows the CTQW-predicted inverse resolvable spectral minimum $1/\Delta_{\min}$ (also shown in \cref{fig:ansatz_2}), while the vertical axis shows the ratio of optimal walk times found through the procedure described in \cref{sec:optimization_bracelet}.}
    \label{fig:bracelet_time_ratio}
\end{figure}

Compared to the CTQW-derived parameters, the jointly optimal parameters systematically require longer walk times. In \cref{fig:bracelet_time_ratio}, the ratio $\tau_{\rm{eff}}^{\rm{Ryd}}/\tau_{\rm eff}^{\rm CTQW}$ reveals that target states with smaller spectral gaps require the largest inflation in $\tau_{\rm eff}$, up to approximately $10\times$. More resolvable targets cluster around a factor of $1.5$ to $2.5$. This behavior indicates that van-der-Waals-induced phases compress the effective gap, requiring longer walk times to resolve the target.

\begin{figure}
    \centering
    \includegraphics[width=\columnwidth]{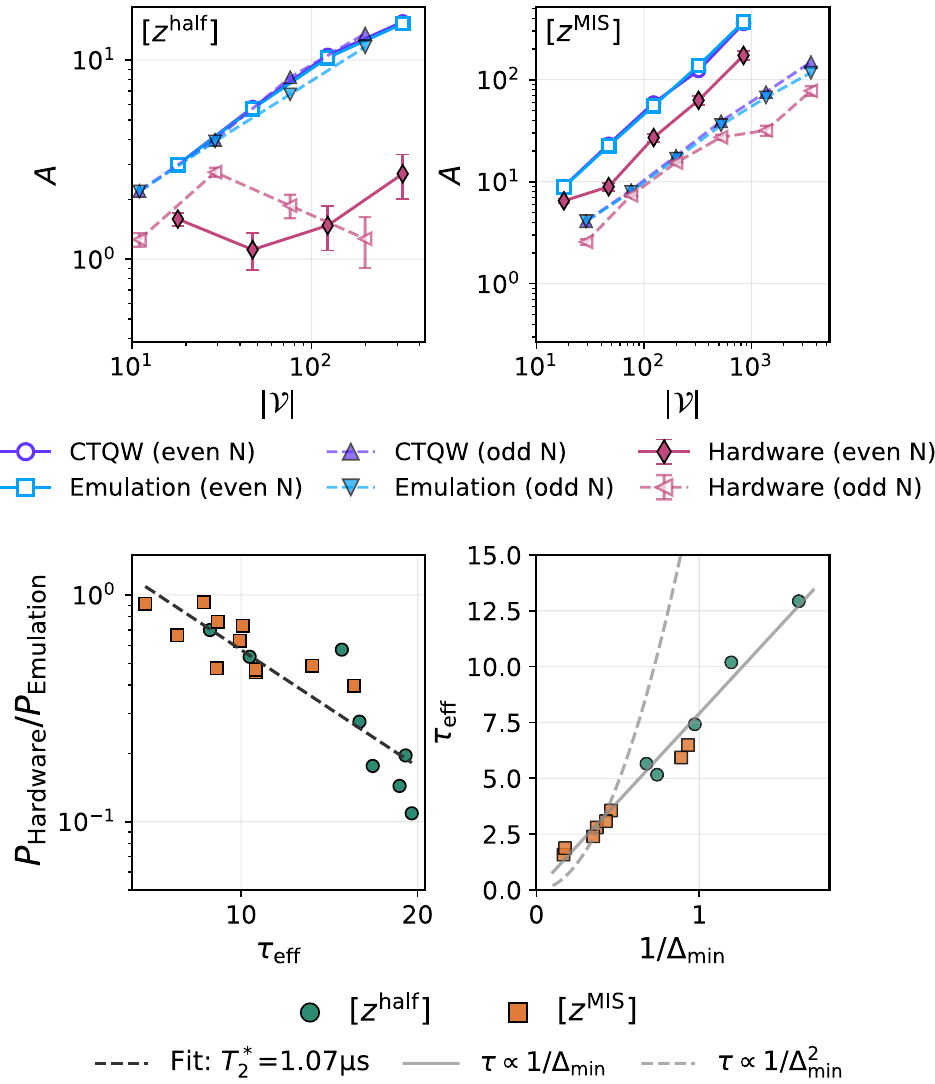}
    \caption{Scaling analysis for bracelet state preparation. \textbf{Top} shows amplification (\cref{eq:amplification}) versus graph size, with $\left[z^{\rm half}\right]$ on the \textbf{left} and $\left[z^{\rm MIS}\right]$ on the \textbf{right}, comparing CTQW, noiseless emulation, and Aquila hardware; data are split by even $N$ (solid lines) and odd $N$ (dashed lines). \textbf{Bottom-left} shows the fidelity ratio $P_{\rm Hardware}/P_{\rm Emulation}$ versus $\tau_{\rm eff}$, decaying exponentially in physical time $t = 2\tau_{\rm eff}/\Omega$, consistent with an effective dephasing time $T_2^* \approx 1.07\,\mu$s. \textbf{Bottom-right} shows $\tau_{\rm eff}$ versus $1/\Delta_{\rm min}$, where $\Delta_{\rm min}$ is the spectral gap.}
    \label{fig:bracelet_scaling}
\end{figure}

\Cref{fig:bracelet_scaling} presents the amplification and fidelity scaling for bracelet state preparation, comparing ideal CTQW dynamics, noiseless Rydberg emulation, and Aquila hardware. Unlike the product state demonstrations, which sweep ansatz depth at fixed $\tau_{\rm eff} \approx \pi/2$, bracelet state preparation is performed at a single optimized $(\tau_{\rm eff}, p)$ pair for each target, with $\tau_{\rm eff}$ ranging from $\sim\!5$ to $\sim\!20$ across targets (see \cref{tab:bracelet} of \cref{app:table_details}). The top panels show amplification (\cref{eq:amplification}) as a function of graph size $|\mathcal{V}|$, with data split by the parity of $N$. For $\left[z^{\rm MIS}\right]$ (top right), power-law fits (\cref{eq:amp_power_law}) to the hardware data yield $\alpha = 0.84 \pm 0.07$ for even-$N$ instances and $\alpha = 0.69 \pm 0.08$ for odd-$N$, both indicating super-quadratic ($\alpha > 0.5$) scaling. The even-$N$ instances correspond to a two-fold degenerate orbit ($|[z]|=2$) and exhibit near-linear scaling across all three cases, while odd-$N$ instances span an $N$-fold superposition and show somewhat reduced but still clear power-law scaling. For $\left[z^{\rm half}\right]$ (top left), noiseless emulation follows the CTQW prediction, but hardware results show no clear power-law trend ($\alpha = 0.18 \pm 0.18$), most likely due to the combined effect of longer walk times and greater ansatz depth.

As shown in the bottom panels of \cref{fig:bracelet_scaling}, $\left[z^{\rm half}\right]$ targets require systematically longer walk times than $\left[z^{\rm MIS}\right]$ targets, which is consistent with the $\left[z^{\rm MIS}\right]$ targets having larger spectral gaps (bottom right). The relative fidelity $P_{\rm Hardware}/P_{\rm Emulation}$ (bottom-left) decays approximately exponentially with the cumulative walk time and, fitting to the form $\exp(-t/T_2^*)$ where $t = 2\tau_{\rm eff}/\Omega$, yields an effective dephasing time $T_2^* \approx 1.07\,\mu$s. Together, these observations indicate that $\left[z^{\rm MIS}\right]$ maintains power-law scaling because its larger spectral gaps permit preparation within the coherence window, while $\left[z^{\rm half}\right]$ targets require walk times that exceed $T_2^*$ and consequently lose their scaling advantage.

The fitted $T_2^*$ is notably shorter than the single-qubit Rabi coherence time of $\sim\!5\,\mu$s~\cite{wurtz2023aquila}. This reduction likely reflects the accumulation of phase errors across the many Rabi phase jumps. Because the ansatz depth scales with the required walk time (see \cref{eq:bracelet_p_tau}), longer $\tau_{\rm eff}$ entails proportionally more phase jumps ($p = 10$--$48$ across the targets studied here). Future work could explore whether fidelity on hardware might be improved by using fewer, longer walk segments. 

Despite these limitations, the hardware results are consistent with the non-adiabatic $\tau_{\rm eff} \propto 1/\Delta_{\rm min}$ scaling predicted by the frequency-resolution model (\cref{eq:bracelet_tau_eff}). As shown in the bottom-right panel of \cref{fig:bracelet_scaling}, the optimized walk times cluster around the $\tau \propto 1/\Delta_{\rm min}$ reference line rather than the $\tau \propto 1/\Delta_{\rm min}^2$ scaling that would arise from adiabatic preparation~\cite{Albash_2018}. While the van-der-Waals-induced phase compression increases the proportionality constant relative to the ideal CTQW (\cref{fig:bracelet_time_ratio}), the scaling relationship is preserved, confirming that the phase-walk ansatz successfully implements the non-adiabatic ``shortcut'' on current hardware.

\section{Results: Coherent quenches of bracelet states}
\label{sec:quenches}

While the projected probability distribution $P(z)=|\langle z|\psi\rangle|^2$ is a reasonable proxy to the fidelity of a state, it is incomplete. Notably, measurements in the Z basis ignore all phase information in the wavefunction, which makes the translationally averaged incoherent mixture
\begin{equation}\label{eq:incoh_rho}
    \rho_{\mathrm{inc}}([z]) = \frac{1}{|[z]|}\sum_{u\in [z]} \ket{u}\!\bra{u},
\end{equation}
and the coherent bracelet state
\begin{equation}\label{eq:coherent_rho}
\rho_{\mathrm{coh}}([z]) = \ket{[z]}\!\bra{[z]}
= \frac{1}{|[z]|}\,\sum_{u,v\in [z]} \ket{u}\!\bra{v}.
\end{equation}
have identical populations ($|\langle z|\rho_{\rm inc}|z\rangle|^2=|\langle z|\rho_{\rm coh}|z\rangle|^2$), even though there are nontrivial off-diagonal matrix elements in $\rho_{\rm coh}$ and (if prepared exactly) none in $\rho_{\rm inc}$. 

We implement quenches from $\ket{[z^{\rm half}]}$ and, for comparison, from the representative product state $\ket{z^{\rm half}}$, using the resulting dynamics as an indicator of bracelet-state preparation, as the post-quench evolution of the coherent bracelet state diverges from that of its incoherent counterpart. The two states are prepared as shown in \cref{tab:stateprep} and the protocol extended by adding an extra Rabi pulse of duration $t=2\tau/\Omega$.

\begin{figure}
    \centering
    \includegraphics[width=0.9\linewidth]{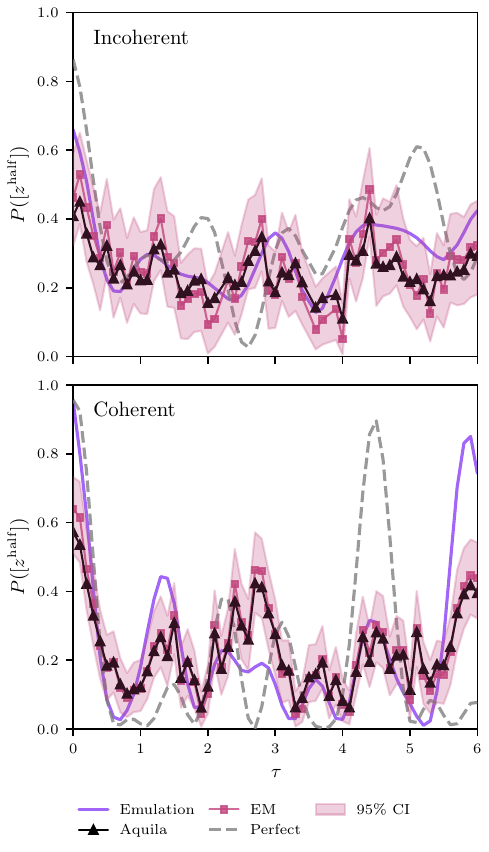}
    \caption{Quenches for target state $\ket{ z^{\rm half} } = \ket{[000101]}$, plotting the probability of measuring $\ket{z^{\rm half}}$ in the dihedrally symmetric subspace of $\mathcal G$. \textbf{Top} shows incoherent product state preparation. \textbf{Bottom} shows bracelet state preparation. Aquila data averaged over $100$ shots per $\tau$.}
    \label{fig:quenches}
\end{figure}

\Cref{fig:quenches} illustrates that, at short evolution times, the dynamics of these two states are equivalent, but at longer times, interference effects cause the distributions to diverge. The coherent and incoherent perfect walk dynamics (gray dashed) show the earliest divergence and clearly distinguish the coherent and incoherent states. Emulated neutral atom dynamics (purple) follow the perfect walk dynamics, as expected from parameter matching the virtual and physical systems. However, at longer times the similarity diverges due to perturbative blockade violations and $R^{-6}$ Rydberg interactions outside of the blockade radius (see \cref{sec:SW}). This suggests that, while for short times, such as state preparation, the two dynamics approximately match. For longer evolution times, the dynamic blockade radius approximation begins to break down, and the Rydberg atom Hamiltonian can no longer accurately represent CTQW on independent set subspaces.

We find that results from Aquila qualitatively track the dynamics of the noiseless emulation. The most prominent difference in the observed dynamics being the higher population at $\tau \approx 5.8$ in the $\ket{[z^{\rm half}]}$ quench (lower panel), which is consistent with the predicted dynamics of the coherent bracelet state. However, there is notable divergence from the noiseless prediction. Most clearly, in the quench from $\ket{[z^{\rm half}]}$ between $\tau\approx1.9$ and $3.2$, two closely spaced peaks merge into a single broader maximum, consistent with the accumulation of dephasing error. For the incoherent quench, the 95\% CI overlaps the ideal curve for most $\tau$, but it has a much broader interval relative to the coherent case -- which could result from use of local detuning during the initial preparation stage (as discussed in \cref{sec:product_state_results}). Overall, although the on-hardware timescales ($\le 2~\mu$s) lie within the Rabi-drive coherence time ($\approx 5~\mu$s), the deviations are not inconsistent with accumulated incoherent error over this window. We leave a deeper investigation of the specific contributing factors to future work.

\section{Conclusion}
\label{sec:conclusion}
This work has presented an implementation and characterization of continuous-time quantum walk (CTQW) based ans\"{a}tze on neutral atom hardware. By transferring the phase-walk structure from abstract models to execution under the Rydberg Hamiltonian, we established a direct link between ideal CTQW dynamics and their realization on a noisy intermediate-scale quantum (NISQ) device. Using known target states as controlled testbeds, we probed how constructive interference, amplification, and entanglement emerge in constrained Hilbert spaces subject to realistic hardware limitations. Variational optimization and error-mitigation were carried out using Pawsey's ``Setonix'' and ``Ella'' supercomputing clusters, while hardware execution was performed on QuEra Computing's cloud-accessible, analog-mode neutral-atom processor Aquila.

Unlike earlier approaches~\cite{farhi2014quantum} that used quantum walks as variational routines for preparing unknown states, here we target nontrivial but efficiently describable states \cite{Gonzales2025} chosen as benchmarks for assessing interference and amplification on NISQ hardware. In particular, we focus on bracelet states that respect the dihedral symmetry of an $N$-vertex ring, and product states whose preparation nonetheless requires nontrivial dynamics through the constrained Hilbert space. For the bracelet states, entanglement emerges from the non-separable structure of the independent-set Hilbert space itself. We find that this constrained space, together with restricted local controls, is sufficient to support variational ans\"{a}tze capable of preparing such structured target states.

We demonstrate a key use of quantum processors as analogue simulators of abstract models: the dynamics of CTQWs on independent-set graphs can be mapped directly onto the nonequilibrium dynamics of the Rydberg Hamiltonian. The independent-set constraint is naturally enforced by the Rydberg blockade, enabling a hardware representation of nonequilibrium dynamics on constrained subspaces that has previously only been explored in the context of quantum scar dynamics~\cite{Bernien2017}. Despite substantial measurement errors, we find that Aquila can reproduce the same target states as the abstract CTQW system through direct parameter transfer with little to no further optimization.

For product-state targets, hardware circuits reproduce the hallmark amplification through constructive interference of efficient CTQW-inspired protocols. The $z^{\rm MIS}$ states benefit from perfect state transfer on an embedded hypercube subgraph, with ideal CTQW dynamics predicting near-unity amplification even at a single ansatz iteration. The $z^{\rm half}$ states provide a more representative benchmark. Here, our results suggest a conservative theoretical upper-bound on the achievable polynomial-order speedup in query complexity of $n \approx 8$ at ansatz depth one, rising to $n \approx 15$ at depth two and $n > 50$ at depth three. On Aquila, we observe $n \approx 4$ at depth one and $n \approx 8$ at depth two, demonstrating that the fast-forward dynamics characteristic of efficient CTQW protocols are already accessible at low circuit depth. This opens the possibility of realizing super-quadratic amplification in algorithmic applications on current hardware.

For bracelet-state targets, our results extend the non-adiabatic shortcut demonstrated for product-state MIS preparation via sweep-quench-sweep protocols~\cite{Lukin2024shortcut} to the preparation of symmetric entangled states, successfully applying a CTQW-based framework that predicts an advantage over adiabatic protocols. While the influence of van der Waals interactions in the Rydberg dynamics precluded the direct transfer of CTQW-derived parameters, we found that it was nevertheless possible to identify parameters that are jointly optimal for both the ideal CTQW and the Rydberg evolution, with  resulting walk times retaining the $\tau_{\rm eff} \propto 1/\Delta_{\rm min}$ scaling predicted by the non-adiabatic protocol.

While this work is an initial step in implementing CTQW on neutral atom hardware, there are many future directions to explore. Instead of directly optimizing to some efficiently describable target state, future work could instead use the quantum walks framework to prepare resource states for hybrid optimization \cite{Wurtz2024} or machine learning tasks \cite{Cerezo2022}. Furthermore, this work focused on the walk graph formed by independent sets of an $N$-vertex ring; future work could implement different walk graphs induced by other independent set graph constraints or Hamiltonian engineering techniques \cite{Choi2020} to explore quantum walks on tailored subspaces.

\section*{Acknowledgements}
This work was supported by resources provided by the Pawsey Supercomputing Research Centre's Setonix Supercomputer~\cite{setonixcomputer}, with funding from the Australian Government and the Government of Western Australia. This work and the Pawsey Supercomputing Research Centre's Quantum Supercomputing Innovation Hub were made possible by a grant from the Australian Government through the National Collaborative Research Infrastructure Strategy (NCRIS). We thank Sheng-Tao Wang and the rest of the theory team at QuEra Computing for their support. PJE and EM thank the entire Quantum Supercomputing Innovation Hub team at Pawsey and QuEra.

\section*{Data Availability}
Data and software supporting the results reported in this work, including source code and datasets obtained from QuEra's Aquila system, are publicly available~\cite{https://doi.org/10.5281/zenodo.20520609}.

\appendix
\crefalias{section}{appendix}

\section{Bayesian inference for distribution reconstruction}
\label{sec:bayes}

A central challenge in analyzing coherent state preparation is \emph{measurement error}, where a $0$ is mis-detected as a $1$ or vice versa. This appendix details our use of the expectation maximization (EM) algorithm, a Bayesian inference procedure, to reconstruct the underlying distribution from measured bitstring data. Our approach is related to recent iterative methods for error mitigation \cite{pokharel2024scalable}, which employ histogram-based unfolding. Here, we instead construct a computationally tractable model by parameterizing only the blockaded subspace $\mathcal V$ directly and regularizing contributions from its complement.

\subsection{Bayesian inference}

Bayesian inference provides a framework to address uncertainties arising through noisy measurement~\cite{mackay2003information,blume2010optimal}. Given observed data $Z$ and model parameters $\bm{\phi}$, Bayes' rule states
\begin{equation}
 P(\bm{\phi} \mid Z) \;\propto\; P(Z \mid \bm{\phi})\,P(\bm{\phi}),
\   
\end{equation}
where the \emph{prior} distribution $P(\bm{\phi})$ is the model before data collection, $P(Z \mid \bm{\phi})$ is the \emph{likelihood} of observing $Z$ given parameters $\phi$, and the \emph{posterior} distribution $P(\bm{\phi} \mid Z)$ describes how plausible the values of $\bm{\phi}$ are given $Z$. Obtaining parameters that best explain the observed data then occurs through maximization of the likelihood. 

\subsection{Likelihood under measurement error}

Measurement error on Aquila is well described by an asymmetric bit-flip channel acting independently on each qubit~\cite{wurtz2023aquila}. The probability of recording a bitstring $z$ given pre-measurement string $s$ is
\begin{equation}
\label{eq:readout}
K(z \mid s) \;=\; \prod_{j=1}^N P(z_j \mid s_j),
\end{equation}
where $j$ indexes the bits of $z$ and $s$, which essentially convolves the pre-measurement probabilities with the transition matrix
\[
P(z \mid s) \;=\;
\begin{pmatrix}
P_{00} & P_{01} \\
P_{10} & P_{11}
\end{pmatrix},
\] 
where $P_{ab}$ is the probability for the transition $a\rightarrow b$,  $P_{01} = 1 - P_{00}$ and $P_{10} =  1 - P_{11}$.

As our aim is to construct the pre-measurement distribution over the subspace of graph vertices $\mathcal V$, we consider a model with parameters $\bm{\phi}_{|\mathcal V|} = \{\phi^k\}_{k=1}^{|\mathcal V|}$ for each possible originating bitstring $s^k \in \mathcal{V}$, and a reduced set of parameters $\bm{\phi}^\perp = \{\phi^\perp_j\}_{j=1}^N$ for bitstrings $s \in \mathcal{V}^\perp = \{0,1\}^N \setminus \mathcal{V}$.  
Here each $\phi^\perp_j$ models the probability that bit $j$ of an observed bitstring is $1$ in $\mathcal V^\perp$, using independent Bernoulli parameters,
\begin{equation}
P_{\mathrm{out}}(s;\bm{\phi}^{\perp}) = \prod_{j=1}^N \left(\phi^{\perp}_j\right)^{s_j}\left(1-\phi^{\perp}_j\right)^{1-s_j}, 
\quad s \in \mathcal V^\perp.
\end{equation}
a choice informed by our assumption of independent single-qubit noise channels. The full parameter set is
\begin{equation}
\bm{\phi} =\bm{\phi}_{|\mathcal V|} \oplus \bm{\phi}^\perp = (\phi^1,\dots,\phi^{|\mathcal V|},\,\phi^\perp_1, \dots,\phi^\perp_N).
\end{equation}

Combining the contributions from $\mathcal V$ and $\mathcal V^\perp$, the model probability of measuring bitstring $z^i$ is
\begin{equation}
m_i(\bm{\phi}) \;=\; \sum_{k=1}^{|\mathcal V|} \phi^k L_{k,i} \;+\; L_i^\perp(\bm{\phi}^\perp),
\end{equation}
where $L_{k,i} = K(z^i \mid s^k)$ is the likelihood for $s^k \in \mathcal V$, and the likelihood for $s \in \mathcal V^\perp$ is
\begin{equation}
L_i^\perp(\bm{\phi}^\perp) = \sum_{s\in \mathcal{V}^\perp} K(z^i \mid s)\,P_{\mathrm{out}}(s;\bm{\phi}^\perp).
\end{equation}

\subsection{Expectation Maximization}

The expectation--maximization (EM) algorithm is an iterative method for likelihood maximization. The algorithm alternates between two steps, which are guaranteed to monotonically converge to a local optimum of the likelihood~\cite{dempster1977maximum}. The first is an ``E-step'' that computes \emph{responsibilities} -- the posterior probability that each component of the model generated observation $z^i$: 
\begin{equation}
\rho^{(t)}_{k,i} = \frac{{\phi^k}^{(t)} L_{k,i}}{m_i(\boldsymbol{\phi}^{(t)})} \qquad \rho^{(t)}_{\perp,i} = \frac{L_i^\perp\left({\bm{\phi}^\perp}^{(t)}\right)}{m_i(\boldsymbol{\phi}^{(t)})},
\end{equation} 
with $\sum_{k=1}^{|\mathcal V|}\rho^{(t)}_{k,i} + \rho_{\perp,i}^{(t)} = 1$

The second is the ``M-step'', which updates the parameters using the average of the responsibilities over all observations, which for $s^k \in \mathcal V$ is:
\begin{equation} 
{\phi^k}^{(t+1)} = \frac{1}{|Z|} \sum_{i=1}^{|Z|} \rho^{(t)}_{k,i}.
\end{equation}
where $|Z|$ is the number of observed bitstrings. Updating $\bm{\phi}^\perp$ requires estimation of the expected pre-measurement bit values. For bit $j$ this is given by:
\begin{equation}
\phi^{\perp\,(t+1)}_j = \frac{\sum_{i=1}^{|Z|} \rho^{(t)}_{\perp,i}\,\mathbb{E}[T_j \mid z^i_j] + \alpha}{\sum_{i=1}^{|Z|} \rho^{(t)}_{\perp,i} + \alpha + \beta},
\end{equation}
where $T_j$ denotes the true pre-measurement value of bit $j$, and $\rho^{(t)}_{\perp,i}$ is the responsibility of $\mathcal  V^\perp$  for $z^i$ at iteration $t$. The strength of the Rydberg blockade effect leads to a small contribution from  $\mathcal V^\perp$, so $\alpha$ and $\beta$ are introduced as regularizing terms (a $\mathrm{Beta}(\alpha,\beta)$ prior~\cite{mackay2003information}) on each $\phi^\perp_j$ to prevent the estimate from collapsing to exactly $0$ or $1$. The expectation value $\mathbb{E}[T_j \mid z^i_j]$ is computed by inversion of  the bit-flip channel (\cref{eq:readout}):
\begin{align}
P(T_j=1 \mid z^i_j) \;=\; 
\frac{\phi^{\perp}_j \, P_{1,{z^i_j}}}{\phi^{\perp}_j \, P_{1,{z^i_j}} \;+\; (1-\phi^{\perp}_j)\, P_{0,{z^i_j}}}.
\end{align}

In this work, we choose an ``uninformative'' prior, initializing $\bm\phi_{\mathcal V}$ as a uniform distribution and $\bm{\phi}^{\perp}$ to a uniform baseline with an uniform $\mathrm{Beta}(1,1)$ prior~\cite{blume2010optimal}. Parameters for the bit-flip channel are $(P_{00},P_{11})=(0.99, 0.93)$ under standard operation~\cite{wurtz2023aquila} and $(P_{00},P_{11})=(0.90,0.93)$ when using local detuning. We use the convergence criterion
\begin{equation}
\|\bm{\phi}_{\mathcal V}^{(t+1)}-\bm{\phi}_{\mathcal V}^{(t)}\|_{1}
+ \tfrac{1}{N}\|\bm{\phi}^{\perp\,(t+1)}-\bm{\phi}^{\perp\,(t)}\|_{1}
< \epsilon,
\end{equation}
where we set $\epsilon=10^{-8}$. Uncertainty in the estimated probability of measuring a given target $z^*$, taken from the optimized $\bm{\phi}$, is computed using a nonparametric bootstrap to a $95\%$ confidence interval. We choose this approach as it adapts well to model complexity and finite-sample variability with minimal assumptions about the underlying distribution~\cite{efron1992bootstrap}. 

\section{Blockade subspace guarantees from perturbation theory}
\label{sec:SW}

\begin{figure}
    \centering
    \includegraphics[width=.9\linewidth]{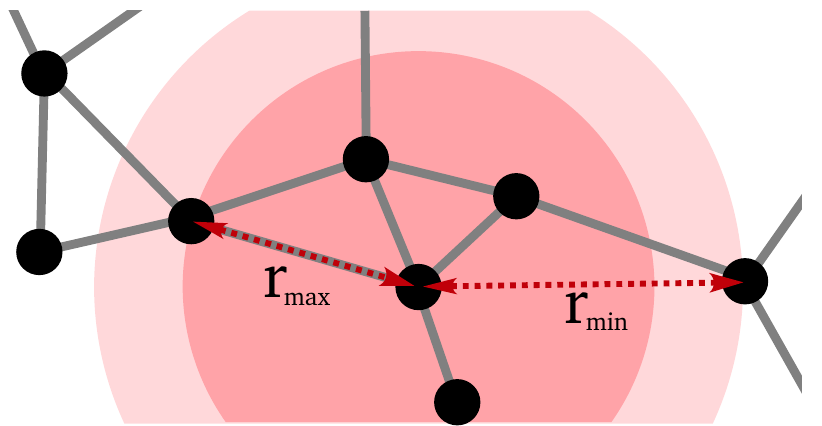}
    \caption{An example unit disk graph to illustrate $r_\text{min}$ and $r_\text{max}$. Given a unit disk graph of vertices (black) at positions $\vec x_i$ and edges (grey) there is a range of unit disk radii $r_\text{max}\leq r_\text{ud}\leq r_\text{min}$ which produce the same graph. $r_\text{max}$ is the maximum distance between any two vertices connected by an edge (equivalently, the minimum unit disk radius, pink). $r_\text{min}$ is the minimum distance between any two vertices not connected by an edge (equivalently, the maximum unit disk radius, light pink).}
    \label{fig:rmin_rmax}
\end{figure}

This appendix details the derivation of the scaling of the positions to best match the blockade radius $r_b$, which may be derived from perturbation theory. In the limit of large interaction strength between atoms within the blockade radius, we may derive the first-order correction and perturbative Hamiltonian from a virtual excitation to the blockaded space. By balancing the contribution from these first-order corrections and long-range $r^6$ interactions outside of the blockade radius, the error Hamiltonian is minimized, maximizing the effective projective Hamiltonian evolution.

For each edge between vertices in the unit disk graph, consider the 2-qubit subspace
\begin{equation}\label{eqn:2qubitsubspace}
    \centering
    \begin{tabular}{cccccccc}
    $\ket{gg}$&\qquad&\qquad&$\ket{gr}$&$\leftrightarrow$&$\ket{rr}$&$\leftrightarrow$&$\ket{rg}$\\
    &&&&\;$\Omega/2$\;&\;$V_{ij}$\;&\;$\Omega/2$\;&
    \end{tabular}
\end{equation}
where the labels represent Hamiltonian terms. Given a definition of Pauli matrices $\sigma_x\equiv \ket{g}\bra{r} + \ket{r}\bra{g}$ and $n\equiv |r\rangle\langle r|$, the effective error Hamiltonian is \cite{Bluvstein2021}
\begin{align}
    H_\text{err} = &\sum_{\text{edges }(i,j)}\frac{-|\Omega|^2}{4V_{ij}}\big(\sigma_x^i\sigma_x^j + \sigma_y^i\sigma_y^j + n_i + n_j\big)\nonumber\\
    +&\sum_{\neg \text{ edges }(i,j)}V_{ij} n_in_j.
\end{align}

The first term is the sum over the perturbative corrections from a virtual excitation to the doubly excited subspace, which results in a hopping term $\ket{gr}\leftrightarrow\ket{rg}$ and Stark shift. The second term is the sum over all van der Waals interactions outside the blockade radius.

Minimizing the contribution of the error Hamiltonian is equivalent to minimizing its norm
\begin{equation}
    ||H_\text{err}|| = \sum_{\text{edges }(i,j)}\frac{|\Omega|^4}{4V_{ij}^2}
    +\sum_{\neg \text{ edges }(i,j)}V_{ij}^2.
\end{equation}

Defining the Blockade radius $|\Omega|\equiv C_6/r_b^6$ and assume that the contribution from blockaded edges comes from $n_b$ edges per vertex at a maximum distance $r_\text{max}$, and the contribution from unblockaded edges comes from $n_u$ edges per vertex at a minimum distance of $r_\text{min}$; $n_b$ and $n_u$ can be scaled to include more weakly interacting edges by rescaling the sum. Then, the norm of the error is
\begin{equation}
    \frac{||H_\text{err}||}{N} = \frac{n_b}{4} \bigg(\frac{C_6}{r_b^6}\bigg)^4\bigg{(\frac{r_\text{max}^6}{C_6}}\bigg)^2 +\, n_u \bigg{(\frac{C_6}{r_\text{min}^6}}\bigg)^2.
\end{equation}

\begin{figure}
    \centering
    \includegraphics[width=\columnwidth]{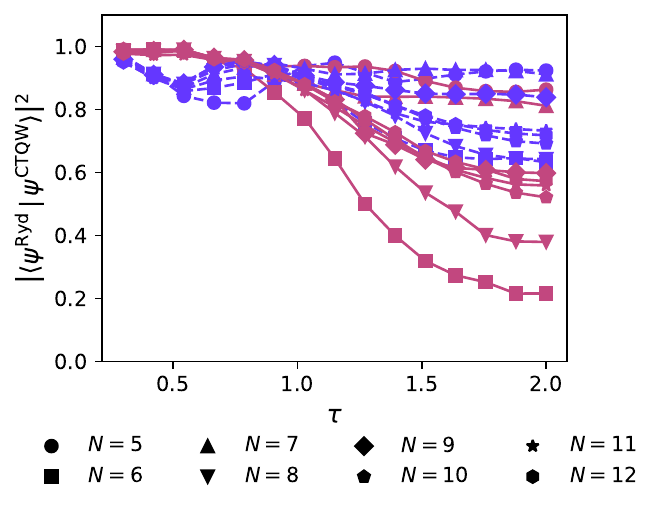}
    \caption{Emulated fidelity for approximation of a CTQW with the $\ket{\psi^{\rm CTQW}}$ with walk time $\tau$ by evolution under the Rydberg atom Hamiltonian $\ket{\psi^{\rm R}}$ (see \cref{sec:rydberg_atoms}), shown with (red, solid lines) and without (purple, dashed lines) inclusion of the perturbative correction factor $\eta$ for blockaded rings of size $N=5,\dots,12$. Over this range the prefactor takes values $\eta=(0.849,\,0.875,\,0.893,\,0.905,\,0.914,\,0.920,\,0.924,\,0.927)$, asymptotically approaching $~0.939$.}
    \label{fig:blockade_radius_prefactor}
\end{figure}

Given a scaling $\lambda$ between virtual coordinates $\overline x$ and physical atom positions $x$ as $x=\lambda \overline x$, the error can be minimized by minimizing with respect to $\lambda$ as $\partial_\lambda ||H|| = 0$ to find
\begin{equation}
    r_b=\eta\,\sqrt{r_\text{min}r_\text{max}}, \qquad  \eta=\bigg(\frac{n_b}{4n_u}\bigg)^{1/24}.
\end{equation}
Although the prefactor scales weakly with the $1/24$th power, as shown in \cref{fig:blockade_radius_prefactor}, we find that its inclusion tangibly improves the CTQW fidelity at small $\tau$. For the 1d chain, $n_b=2$ and $n_u\approx2.220$, so the prefactor shifts the blockade radius by a factor of $0.939$. 

For a 2d King's lattice, $n_b=2.25$ and $n_u\approx7.791$, so the prefactor shifts the blockade radius by a factor of $0.940$. The magnitude of the minimized error term is
\begin{equation}
    \frac{||H_\text{err}||}{N} = \sqrt{n_bn_u}\bigg(\frac{C_6^2}{r_b^{12}}\bigg)\bigg(\frac{r_\text{max}}{r_\text{min}}\bigg)^6.
\end{equation}
Reducing the ratio between the maximum distance between vertices connected by a unit disk edge, and the minimum distance between vertices not connected by a unit disk edge will reduce the error. For the 1d chain, $r_\text{max}/r_\text{min}=1/2$, so $||H_\text{err}||/N\approx (0.181\Omega)^2$. For the 2d king's graph, $r_\text{max}/r_\text{min}=1/\sqrt{2}$, so $||H_\text{err}||/N\approx (0.723\Omega)^2$.

\section{State Preparation Results}
\label{app:table_details}
This appendix collects detailed numerical results for state preparation demonstrations. \Cref{tab:power_law_fits} reports the power-law fit parameters ($c$, $\alpha$, and $R^2$) for the amplification scaling $A = c\,|\mathcal V|^\alpha$ shown in \cref{fig:4_product_state_all_platforms}.~\Cref{tab:product,tab:bracelet} provide per-instance success probabilities for product states and bracelet states, respectively.

In both \cref{tab:product,tab:bracelet}, the \textbf{State} column identifies the target configuration, where $N$ denotes the number of vertices in the constraint graph and $|\mathcal V|$ the number of vertices in the corresponding walk graph $\mathcal G$. The \textbf{Depth ($p$)} column records the number of ansatz layers. For product state preparation (\cref{tab:product}), the variationally optimised parameters are the fiducial and walk times $\tau_0$ and $\tau_1$, and $J_{\rm eff}$ is the effective coupling (see \cref{eq:product_state_effective_time}) computed from the walk times. For bracelet states (\cref{tab:bracelet}), the relevant parameters are the effective evolution time $\tau_\text{eff}=\tau p$ and the phase jumps $\gamma$.

The \textbf{Na\"{i}ve count} column gives the fraction of bitstrings matching the target state directly from raw shot data. The \textbf{Perfect} column reports the corresponding preparation probability under ideal CTQW dynamics, while the \textbf{Emulation} column shows the same quantity obtained from noiseless Rydberg-atom emulation. Hardware performance is reported in the \textbf{EM (H)} column, which shows the success probability obtained from demonstration on Aquila after measurement-error mitigation (see \cref{sec:bayes}), together with the $95\%$ CI. As a check for consistency, the \textbf{EM (E)} column presents values obtained from $1000$ emulated shots convolved with the noisy measurement channel.

\begin{table*}[p]
\begingroup
\tiny
\setlength{\tabcolsep}{4pt}
\renewcommand{\arraystretch}{1.1}
\begin{tabularx}{\textwidth}{S[table-format=3.0]S[table-format=4.0]>{\RaggedRight\arraybackslash}XS[table-format=2.0]S[table-format=1.3]S[table-format=1.3]S[table-format=2.3]>{\centering\arraybackslash}XS[table-format=1.3]S[table-format=1.3]>{\centering\arraybackslash}X>{\centering\arraybackslash}X}
\toprule
{$N$} & {$|\mathcal V|$} & {State} & {Depth ($p$)} & {$\tau_0$} & {$\tau_{1}$} & {$J_{\rm eff}$} & {Na\"{i}ve Count} & {Perfect} & {Emulation} & {EM (H)} & {EM (E)} \\
\midrule
5 & 11 & $00101$ & 1 & 0.532 & 1.146 & 0.936 & 278/954 & 0.964 & 0.481 & $0.32\,[0.28, 0.35]$ & $0.48\,[0.44, 0.51]$ \\
5 & 11 & $00101$ & 2 & 0.330 & 0.642 & 0.973 & 166/948 & 1.000 & 0.554 & $0.15\,[0.12, 0.18]$ & $0.56\,[0.52, 0.59]$ \\
5 & 11 & $00101$ & 3 & 0.225 & 0.456 & 0.987 & 127/955 & 1.000 & 0.503 & $0.08\,[0.05, 0.10]$ & $0.50\,[0.45, 0.54]$ \\
6 & 18 & $000101$ & 1 & 0.606 & 0.960 & 1.003 & 444/1882 & 0.789 & 0.396 & $0.28\,[0.25, 0.30]$ & $0.40\,[0.35, 0.44]$ \\
6 & 18 & $000101$ & 2 & 0.268 & 0.655 & 0.995 & 456/1874 & 0.915 & 0.504 & $0.26\,[0.24, 0.29]$ & $0.51\,[0.46, 0.55]$ \\
6 & 18 & $000101$ & 3 & 0.275 & 0.435 & 0.995 & 314/1870 & 0.963 & 0.620 & $0.15\,[0.12, 0.17]$ & $0.59\,[0.54, 0.64]$ \\
7 & 29 & $0000101$ & 1 & 0.639 & 0.878 & 1.035 & 336/1884 & 0.704 & 0.356 & $0.21\,[0.18, 0.23]$ & $0.38\,[0.34, 0.43]$ \\
7 & 29 & $0000101$ & 2 & 0.226 & 0.663 & 1.012 & 632/1855 & 0.857 & 0.613 & $0.47\,[0.43, 0.50]$ & $0.63\,[0.58, 0.68]$ \\
7 & 29 & $0000101$ & 3 & 0.302 & 0.423 & 0.999 & 454/1862 & 0.947 & 0.669 & $0.28\,[0.25, 0.30]$ & $0.69\,[0.64, 0.73]$ \\
8 & 47 & $00010101$ & 1 & 0.578 & 0.992 & 1.001 & 351/1857 & 0.706 & 0.487 & $0.23\,[0.21, 0.26]$ & $0.45\,[0.40, 0.50]$ \\
8 & 47 & $00010101$ & 2 & 0.285 & 0.649 & 0.992 & 314/1859 & 0.911 & 0.511 & $0.19\,[0.16, 0.21]$ & $0.51\,[0.46, 0.55]$ \\
8 & 47 & $00010101$ & 3 & 0.264 & 0.439 & 0.994 & 255/1902 & 0.957 & 0.614 & $0.12\,[0.10, 0.15]$ & $0.63\,[0.58, 0.68]$ \\
\rowcolor{lightgray}9 & 76 & $000010101$ & 1 & 0.615 & 0.908 & 1.032 & 144/929 & 0.657 & 0.374 & $0.18\,[0.15, 0.22]$ & $0.39\,[0.35, 0.43]$ \\
\rowcolor{lightgray}9 & 76 & $000010101$ & 2 & 0.250 & 0.656 & 1.006 & 361/1841 & 0.849 & 0.596 & $0.26\,[0.23, 0.29]$ & $0.60\,[0.55, 0.65]$ \\
\rowcolor{lightgray}9 & 76 & $000010101$ & 3 & 0.289 & 0.428 & 0.998 & 202/1841 & 0.929 & 0.551 & $0.10\,[0.08, 0.12]$ & $0.54\,[0.49, 0.60]$ \\
10 & 123 & $0000010101$ & 1 & 0.636 & 0.861 & 1.049 & 276/1822 & 0.620 & 0.397 & $0.18\,[0.16, 0.20]$ & $0.40\,[0.34, 0.44]$ \\
10 & 123 & $0000010101$ & 2 & 0.223 & 0.661 & 1.016 & 238/944 & 0.796 & 0.552 & $0.35\,[0.31, 0.39]$ & $0.54\,[0.49, 0.59]$ \\
10 & 123 & $0000010101$ & 3 & 0.305 & 0.422 & 1.000 & 247/1844 & 0.926 & 0.584 & $0.17\,[0.15, 0.19]$ & $0.61\,[0.55, 0.66]$ \\
11 & 199 & $00000010101$ & 1 & 0.650 & 0.831 & 1.061 & 115/914 & 0.564 & 0.326 & $0.14\,[0.11, 0.17]$ & $0.31\,[0.27, 0.36]$ \\
11 & 199 & $00000010101$ & 2 & 0.202 & 0.665 & 1.025 & 333/1828 & 0.748 & 0.474 & $0.27\,[0.24, 0.29]$ & $0.52\,[0.46, 0.57]$ \\
11 & 199 & $00000010101$ & 3 & 0.317 & 0.417 & 1.002 & 135/1812 & 0.913 & 0.442 & $0.08\,[0.06, 0.10]$ & $0.42\,[0.37, 0.47]$ \\
12 & 322 & $000001010101$ & 1 & 0.620 & 0.885 & 1.044 & 172/1821 & 0.525 & 0.298 & $0.13\,[0.11, 0.15]$ & $0.25\,[0.21, 0.30]$ \\
12 & 322 & $000001010101$ & 2 & 0.241 & 0.657 & 1.011 & 252/1792 & 0.761 & 0.514 & $0.21\,[0.18, 0.24]$ & $0.52\,[0.47, 0.58]$ \\
12 & 322 & $000001010101$ & 3 & 0.296 & 0.426 & 0.999 & 123/1800 & 0.919 & 0.512 & $0.07\,[0.05, 0.09]$ & $0.56\,[0.50, 0.60]$ \\
13 & 521 & $0000001010101$ & 1 & 0.635 & 0.852 & 1.056 & 135/1765 & 0.483 & 0.324 & $0.11\,[0.08, 0.13]$ & $0.32\,[0.27, 0.37]$ \\
13 & 521 & $0000001010101$ & 2 & 0.221 & 0.660 & 1.019 & 253/1784 & 0.755 & 0.510 & $0.23\,[0.20, 0.26]$ & $0.51\,[0.46, 0.58]$ \\
13 & 521 & $0000001010101$ & 3 & 0.308 & 0.421 & 1.001 & 122/1802 & 0.892 & 0.562 & $0.09\,[0.07, 0.11]$ & $0.54\,[0.48, 0.59]$ \\
14 & 843 & $00000001010101$ & 1 & 0.646 & 0.829 & 1.065 & 132/1806 & 0.482 & 0.288 & $0.09\,[0.07, 0.11]$ & $0.29\,[0.25, 0.35]$ \\
14 & 843 & $00000001010101$ & 2 & 0.205 & 0.664 & 1.025 & 279/1804 & 0.689 & 0.463 & $0.25\,[0.22, 0.28]$ & $0.43\,[0.38, 0.50]$ \\
14 & 843 & $00000001010101$ & 3 & 0.317 & 0.417 & 1.002 & 120/1814 & 0.882 & 0.500 & $0.09\,[0.07, 0.11]$ & $0.52\,[0.47, 0.59]$ \\
15 & 1364 & $000000001010101$ & 1 & 0.654 & 0.812 & 1.071 & 121/1756 & 0.459 & 0.296 & $0.10\,[0.08, 0.12]$ & $0.31\,[0.26, 0.35]$ \\
15 & 1364 & $000000001010101$ & 2 & 0.191 & 0.667 & 1.031 & 245/1767 & 0.684 & 0.485 & $0.27\,[0.24, 0.29]$ & $0.52\,[0.48, 0.60]$ \\
15 & 1364 & $000000001010101$ & 3 & 0.324 & 0.414 & 1.003 & 126/1776 & 0.884 & 0.544 & $0.11\,[0.09, 0.13]$ & $0.53\,[0.48, 0.60]$ \\
16 & 2207 & $0000000101010101$ & 1 & 0.635 & 0.847 & 1.060 & 78/1754 & 0.399 & 0.221 & $0.07\,[0.05, 0.09]$ & $0.27\,[0.23, 0.32]$ \\
16 & 2207 & $0000000101010101$ & 2 & 0.220 & 0.660 & 1.020 & 153/1749 & 0.664 & 0.439 & $0.15\,[0.12, 0.17]$ & $0.49\,[0.44, 0.56]$ \\
16 & 2207 & $0000000101010101$ & 3 & 0.309 & 0.420 & 1.001 & 40/1727 & 0.896 & 0.412 & $0.03\,[0.02, 0.04]$ & $0.42\,[0.37, 0.49]$ \\
17 & 3571 & $00000000101010101$ & 1 & 0.644 & 0.828 & 1.067 & 74/1740 & 0.385 & 0.206 & $0.06\,[0.04, 0.08]$ & $0.22\,[0.18, 0.26]$ \\
17 & 3571 & $00000000101010101$ & 2 & 0.207 & 0.663 & 1.025 & 106/1700 & 0.662 & 0.440 & $0.10\,[0.08, 0.12]$ & $0.42\,[0.37, 0.49]$ \\
17 & 3571 & $00000000101010101$ & 3 & 0.317 & 0.417 & 1.002 & 60/1741 & 0.856 & 0.446 & $0.06\,[0.04, 0.08]$ & $0.49\,[0.43, 0.53]$ \\
18 & 5778 & $000000000101010101$ & 1 & 0.651 & 0.814 & 1.072 & 66/1706 & 0.370 & 0.246 & $0.05\,[0.04, 0.07]$ & $0.24\,[0.20, 0.28]$ \\
18 & 5778 & $000000000101010101$ & 2 & 0.195 & 0.665 & 1.030 & 166/1711 & 0.637 & 0.419 & $0.22\,[0.19, 0.24]$ & $0.45\,[0.40, 0.51]$ \\
18 & 5778 & $000000000101010101$ & 3 & 0.323 & 0.414 & 1.003 & 52/1685 & 0.857 & 0.442 & $0.06\,[0.04, 0.07]$ & $0.39\,[0.34, 0.45]$ \\
19 & 9349 & $0000000000101010101$ & 1 & 0.657 & 0.802 & 1.076 & 67/1713 & 0.350 & 0.204 & $0.07\,[0.05, 0.08]$ & $0.21\,[0.17, 0.26]$ \\
19 & 9349 & $0000000000101010101$ & 2 & 0.184 & 0.667 & 1.034 & 118/1704 & 0.622 & 0.386 & $0.15\,[0.12, 0.18]$ & $0.47\,[0.41, 0.52]$ \\
19 & 9349 & $0000000000101010101$ & 3 & 0.328 & 0.412 & 1.004 & 56/1678 & 0.856 & 0.446 & $0.07\,[0.05, 0.09]$ & $0.48\,[0.41, 0.51]$ \\
20 & 15127 & $00000000010101010101$ & 1 & 0.643 & 0.828 & 1.068 & 40/1698 & 0.352 & 0.161 & $0.03\,[0.02, 0.05]$ & $0.20\,[0.16, 0.24]$ \\
20 & 15127 & $00000000010101010101$ & 2 & 0.208 & 0.662 & 1.025 & 112/1753 & 0.608 & 0.356 & $0.14\,[0.12, 0.17]$ & $0.35\,[0.31, 0.44]$ \\
20 & 15127 & $00000000010101010101$ & 3 & 0.316 & 0.417 & 1.002 & 23/1721 & 0.837 & 0.331 & $0.02\,[0.01, 0.04]$ & $0.36\,[0.30, 0.41]$ \\
21 & 24476 & $000000000010101010101$ & 1 & 0.649 & 0.815 & 1.073 & 35/1708 & 0.324 & 0.175 & $0.03\,[0.01, 0.04]$ & $0.19\,[0.14, 0.22]$ \\
21 & 24476 & $000000000010101010101$ & 2 & 0.198 & 0.664 & 1.029 & 58/1726 & 0.558 & 0.319 & $0.07\,[0.05, 0.09]$ & $0.36\,[0.31, 0.42]$ \\
21 & 24476 & $000000000010101010101$ & 3 & 0.322 & 0.415 & 1.003 & 30/1722 & 0.851 & 0.407 & $0.02\,[0.01, 0.04]$ & $0.41\,[0.35, 0.46]$ \\
22 & 39603 & $0000000000010101010101$ & 1 & 0.655 & 0.804 & 1.077 & 39/1696 & 0.322 & 0.192 & $0.04\,[0.02, 0.05]$ & $0.21\,[0.16, 0.24]$ \\
22 & 39603 & $0000000000010101010101$ & 2 & 0.189 & 0.666 & 1.033 & 106/1719 & 0.557 & 0.335 & $0.17\,[0.14, 0.20]$ & $0.44\,[0.38, 0.49]$ \\
22 & 39603 & $0000000000010101010101$ & 3 & 0.326 & 0.413 & 1.004 & 13/1691 & 0.827 & 0.368 & $0.010\,[0.000, 0.019]$ & $0.43\,[0.37, 0.48]$ \\
23 & 64079 & $00000000000010101010101$ & 1 & 0.659 & 0.796 & 1.080 & 33/1652 & 0.322 & 0.150 & $0.04\,[0.02, 0.05]$ & $0.17\,[0.13, 0.20]$ \\
23 & 64079 & $00000000000010101010101$ & 2 & 0.180 & 0.668 & 1.036 & 64/1631 & 0.546 & 0.346 & $0.13\,[0.10, 0.15]$ & $0.47\,[0.39, 0.49]$ \\
23 & 64079 & $00000000000010101010101$ & 3 & 0.330 & 0.411 & 1.004 & 19/1667 & 0.838 & 0.382 & $0.02\,[0.01, 0.04]$ & $0.41\,[0.35, 0.46]$ \\

6 & 18 & $010101$ & 1 & 0.439 & 1.202 & 0.957 & 534/1880 & 0.990 & 0.617 & $0.32\,[0.29, 0.35]$ & $0.60\,[0.56, 0.64]$ \\
6 & 18 & $010101$ & 2 & 0.340 & 0.632 & 0.980 & 74/860 & 0.998 & 0.232 & $0.06\,[0.03, 0.08]$ & $0.22\,[0.18, 0.26]$ \\
7 & 29 & $0010101$ & 1 & 0.506 & 1.153 & 0.947 & 475/1889 & 0.954 & 0.633 & $0.29\,[0.26, 0.32]$ & $0.63\,[0.59, 0.66]$ \\
7 & 29 & $0010101$ & 2 & 0.332 & 0.639 & 0.975 & 157/860 & 0.999 & 0.536 & $0.18\,[0.14, 0.21]$ & $0.52\,[0.47, 0.56]$ \\
8 & 47 & $01010101$ & 1 & 0.439 & 1.201 & 0.958 & 353/1862 & 0.981 & 0.690 & $0.22\,[0.19, 0.25]$ & $0.73\,[0.68, 0.77]$ \\
8 & 47 & $01010101$ & 2 & 0.340 & 0.632 & 0.980 & 57/821 & 0.998 & 0.364 & $0.07\,[0.05, 0.10]$ & $0.40\,[0.36, 0.44]$ \\
9 & 76 & $001010101$ & 1 & 0.492 & 1.160 & 0.950 & 342/1821 & 0.930 & 0.675 & $0.23\,[0.20, 0.26]$ & $0.68\,[0.64, 0.72]$ \\
9 & 76 & $001010101$ & 2 & 0.334 & 0.637 & 0.977 & 115/840 & 0.996 & 0.540 & $0.14\,[0.11, 0.17]$ & $0.53\,[0.48, 0.57]$ \\
10 & 123 & $0101010101$ & 1 & 0.439 & 1.201 & 0.958 & 202/933 & 0.991 & 0.722 & $0.29\,[0.25, 0.33]$ & $0.73\,[0.68, 0.77]$ \\
10 & 123 & $0101010101$ & 2 & 0.340 & 0.632 & 0.980 & 54/809 & 0.998 & 0.390 & $0.07\,[0.04, 0.10]$ & $0.40\,[0.35, 0.44]$ \\
11 & 199 & $00101010101$ & 1 & 0.483 & 1.166 & 0.952 & 143/884 & 0.934 & 0.612 & $0.23\,[0.20, 0.27]$ & $0.64\,[0.59, 0.68]$ \\
11 & 199 & $00101010101$ & 2 & 0.335 & 0.636 & 0.977 & 56/819 & 0.995 & 0.432 & $0.07\,[0.04, 0.10]$ & $0.42\,[0.38, 0.47]$ \\
12 & 322 & $010101010101$ & 1 & 0.439 & 1.201 & 0.958 & 262/1789 & 0.990 & 0.731 & $0.20\,[0.17, 0.23]$ & $0.77\,[0.71, 0.80]$ \\
12 & 322 & $010101010101$ & 2 & 0.340 & 0.632 & 0.980 & 48/834 & 0.998 & 0.436 & $0.06\,[0.04, 0.09]$ & $0.42\,[0.37, 0.47]$ \\
13 & 521 & $0010101010101$ & 1 & 0.477 & 1.170 & 0.953 & 100/844 & 0.926 & 0.653 & $0.19\,[0.15, 0.23]$ & $0.66\,[0.61, 0.70]$ \\
13 & 521 & $0010101010101$ & 2 & 0.336 & 0.635 & 0.978 & 77/814 & 0.997 & 0.530 & $0.12\,[0.08, 0.15]$ & $0.54\,[0.49, 0.58]$ \\
14 & 843 & $01010101010101$ & 1 & 0.439 & 1.201 & 0.958 & 98/896 & 0.980 & 0.729 & $0.16\,[0.12, 0.20]$ & $0.70\,[0.65, 0.75]$ \\
14 & 843 & $01010101010101$ & 2 & 0.340 & 0.632 & 0.980 & 34/828 & 0.996 & 0.409 & $0.04\,[0.02, 0.07]$ & $0.40\,[0.35, 0.45]$ \\
15 & 1364 & $001010101010101$ & 1 & 0.472 & 1.174 & 0.954 & 156/1760 & 0.904 & 0.671 & $0.13\,[0.11, 0.15]$ & $0.67\,[0.62, 0.71]$ \\
15 & 1364 & $001010101010101$ & 2 & 0.336 & 0.635 & 0.978 & 46/800 & 0.998 & 0.456 & $0.08\,[0.05, 0.11]$ & $0.49\,[0.44, 0.54]$ \\
16 & 2207 & $0101010101010101$ & 1 & 0.439 & 1.201 & 0.958 & 157/1776 & 0.978 & 0.689 & $0.14\,[0.11, 0.17]$ & $0.71\,[0.65, 0.75]$ \\
16 & 2207 & $0101010101010101$ & 2 & 0.340 & 0.632 & 0.980 & 61/2576 & 0.997 & 0.388 & $0.03\,[0.02, 0.04]$ & $0.42\,[0.37, 0.48]$ \\
17 & 3571 & $00101010101010101$ & 1 & 0.468 & 1.177 & 0.955 & 114/1769 & 0.905 & 0.611 & $0.10\,[0.08, 0.12]$ & $0.64\,[0.58, 0.67]$ \\
17 & 3571 & $00101010101010101$ & 2 & 0.337 & 0.635 & 0.978 & 80/2644 & 0.992 & 0.396 & $0.04\,[0.03, 0.05]$ & $0.40\,[0.35, 0.46]$ \\
18 & 5778 & $010101010101010101$ & 1 & 0.439 & 1.201 & 0.958 & 108/1716 & 0.968 & 0.675 & $0.10\,[0.08, 0.13]$ & $0.67\,[0.61, 0.72]$ \\
18 & 5778 & $010101010101010101$ & 2 & 0.340 & 0.632 & 0.980 & 32/2558 & 0.997 & 0.362 & $0.01\,[0.00, 0.02]$ & $0.39\,[0.34, 0.45]$ \\
19 & 9349 & $0010101010101010101$ & 1 & 0.465 & 1.179 & 0.955 & 84/1708 & 0.907 & 0.621 & $0.08\,[0.06, 0.10]$ & $0.67\,[0.61, 0.70]$ \\
19 & 9349 & $0010101010101010101$ & 2 & 0.337 & 0.634 & 0.978 & 18/1671 & 0.997 & 0.386 & $0.006\,[0.000, 0.017]$ & $0.40\,[0.35, 0.46]$ \\
20 & 15127 & $01010101010101010101$ & 1 & 0.439 & 1.201 & 0.958 & 66/1736 & 0.964 & 0.658 & $0.06\,[0.04, 0.09]$ & $0.67\,[0.62, 0.72]$ \\
20 & 15127 & $01010101010101010101$ & 2 & 0.340 & 0.632 & 0.980 & 22/2576 & 0.995 & 0.317 & $0.008\,[0.000, 0.017]$ & $0.34\,[0.30, 0.40]$ \\
21 & 24476 & $001010101010101010101$ & 1 & 0.463 & 1.181 & 0.955 & 103/1736 & 0.901 & 0.617 & $0.11\,[0.09, 0.14]$ & $0.67\,[0.61, 0.70]$ \\
21 & 24476 & $001010101010101010101$ & 2 & 0.337 & 0.634 & 0.979 & 30/2534 & 0.995 & 0.411 & $0.02\,[0.01, 0.03]$ & $0.41\,[0.36, 0.47]$ \\
22 & 39603 & $0101010101010101010101$ & 1 & 0.439 & 1.201 & 0.958 & 73/1718 & 0.974 & 0.612 & $0.08\,[0.06, 0.11]$ & $0.64\,[0.57, 0.67]$ \\
22 & 39603 & $0101010101010101010101$ & 2 & 0.340 & 0.632 & 0.980 & 25/2537 & 0.993 & 0.292 & $0.015\,[0.006, 0.024]$ & $0.30\,[0.26, 0.38]$ \\
23 & 64079 & $00101010101010101010101$ & 1 & 0.461 & 1.183 & 0.956 & 58/1630 & 0.902 & 0.561 & $0.08\,[0.06, 0.10]$ & $0.60\,[0.54, 0.63]$ \\
23 & 64079 & $00101010101010101010101$ & 2 & 0.337 & 0.634 & 0.979 & 19/2520 & 0.995 & 0.355 & $0.012\,[0.004, 0.021]$ & $0.38\,[0.33, 0.44]$ \\

\bottomrule
\end{tabularx}
\endgroup
\caption{Numerical and hardware results for $\ket{z^\text{half}}$ and $\ket{z^\text{MIS}}$ product states, as detailed at the start of \ref{app:table_details}. Highlighted rows emphasize the target state of Fig.~\ref{tab:stateprep}}\label{tab:product}
\end{table*}
\begin{table*}[p]
\begingroup
\tiny
\setlength{\tabcolsep}{2.5pt}
\renewcommand{\arraystretch}{1.03}
\begin{tabularx}{\textwidth}{S[table-format=3.0]S[table-format=4.0]p{1.8cm}S[table-format=2.0]S[table-format=2.3]>{\RaggedRight\arraybackslash}Xp{1.5cm}S[table-format=1.3]S[table-format=1.4]p{1.5cm}p{1.5cm}}
\toprule
{$N$} & {$|\mathcal V|$} & {State} & {Depth ($p$)} & {$\tau_{\rm eff}$} & \multicolumn{1}{c}{$\bm \gamma$} & \multicolumn{1}{c}{Na\"{i}ve Count} & {Perfect} & {Emulation} & \multicolumn{1}{c}{EM (H)}  & \multicolumn{1}{c}{EM (E)}  \\
\midrule
5 & 11 & $[00101]$ & 38 & 15.708 & \begin{tabular}[c]{@{}l@{}}0.494, 0.074, -0.008, 0.019, -0.023, 0.211, -0.046 \\ 0.009, 0.388, 0.012, 0.056, -0.075, -0.040, 0.019 \\ 0.031, 0.015, -0.283, 0.709, 0.051, -0.011, -0.174 \\ 0.034, 0.034, 0.020, 0.035, -0.043, 0.033, -0.163 \\ 0.110, -0.170, -0.048, -0.122, 0.116, -0.029, 0.008 \\ 0.064, -0.225, 0.107\end{tabular} & 460/947 & 1.000 & 0.994 & \begin{tabular}[c]{@{}l@{}}$0.57$ \\ {}[0.52, 0.61]\end{tabular} & \begin{tabular}[c]{@{}l@{}}$0.97$ \\ {}[0.94, 0.99]\end{tabular} \\
\addlinespace[2pt]
6 & 18 & $[000101]$ & 25 & 10.485 & \begin{tabular}[c]{@{}l@{}}0.109, 0.230, 0.002, 0.087, -0.285, 0.556, 0.141 \\ 0.108, 0.578, -0.244, 0.191, -0.060, -0.106, -0.304 \\ -0.077, -0.108, -0.330, -0.162, -0.014, 0.526, 0.155 \\ 0.591, 0.284, -0.266, 0.104\end{tabular} & 453/940 & 0.996 & 0.994 & \begin{tabular}[c]{@{}l@{}}$0.53$ \\ {}[0.49, 0.57]\end{tabular} & \begin{tabular}[c]{@{}l@{}}$0.98$ \\ {}[0.95, 0.99]\end{tabular} \\
\addlinespace[2pt]
7 & 29 & $[0000101]$ & 19 & 8.234 & \begin{tabular}[c]{@{}l@{}}-0.286, -0.024, 0.156, 0.335, 0.451, 1.000, 0.814 \\ -0.008, -0.623, -0.354, 0.564, 0.230, 0.066, -0.640 \\ -0.225, -0.349, -0.173, -0.221, -0.486\end{tabular} & 529/924 & 0.960 & 0.943 & \begin{tabular}[c]{@{}l@{}}$0.66$ \\ {}[0.62, 0.69]\end{tabular} & \begin{tabular}[c]{@{}l@{}}$0.93$ \\ {}[0.90, 0.95]\end{tabular} \\
\addlinespace[2pt]
8 & 47 & $[00010101]$ & 47 & 19.320 & \begin{tabular}[c]{@{}l@{}}0.123, 0.300, -0.245, 1.259, 0.720, -0.236, -0.194 \\ -0.003, -0.098, 0.006, 1.157, 0.730, -0.447, -0.100 \\ -0.068, 0.092, -0.244, 0.405, 0.095, 0.168, 0.118 \\ -0.369, -0.277, -0.255, -0.318, -0.342, -0.078, -0.090 \\ -0.005, 0.066, -0.451, -0.102, 0.352, 0.252, -0.023 \\ 0.609, 0.361, -0.013, -0.287, -0.389, -0.114, -0.537 \\ 0.009, -0.276, -0.126, 0.509, 0.043\end{tabular} & 190/938 & 0.984 & 0.970 & \begin{tabular}[c]{@{}l@{}}$0.19$ \\ {}[0.15, 0.23]\end{tabular} & \begin{tabular}[c]{@{}l@{}}$0.94$ \\ {}[0.91, 0.96]\end{tabular} \\
\addlinespace[2pt]
9 & 76 & $[000010101]$ & 41 & 16.708 & \begin{tabular}[c]{@{}l@{}}-0.026, -0.205, -0.368, -0.118, -0.337, -0.132, 0.319 \\ -0.064, -0.066, 0.981, -0.275, -0.109, -0.011, -0.037 \\ -0.208, -0.522, 0.924, 0.205, -0.071, 0.078, -0.050 \\ 0.201, 0.255, -0.030, -0.116, 0.037, 0.011, -0.025 \\ 0.016, 0.012, 0.270, 0.203, 0.059, 0.273, -0.044 \\ 1.012, 0.049, 0.288, -0.183, 0.799, 0.206\end{tabular} & 181/934 & 0.970 & 0.797 & \begin{tabular}[c]{@{}l@{}}$0.22$ \\ {}[0.19, 0.25]\end{tabular} & \begin{tabular}[c]{@{}l@{}}$0.78$ \\ {}[0.74, 0.81]\end{tabular} \\
\addlinespace[2pt]
10 & 123 & $[0000010101]$ & 47 & 18.979 & \begin{tabular}[c]{@{}l@{}}1.204, -0.112, 0.011, -0.081, 0.427, 0.420, -0.922 \\ 0.020, 0.077, -0.402, 0.323, 0.834, -0.081, 0.417 \\ -0.013, -0.368, 1.169, 0.916, -0.199, 0.341, 0.284 \\ -0.046, -0.087, -0.264, 0.085, -0.548, 0.349, -0.217 \\ 0.102, 0.125, 0.135, 0.057, -0.022, 0.076, -0.371 \\ -0.178, -0.153, 0.306, 0.758, -0.036, 0.006, 0.024 \\ -0.035, 0.002, -0.039, 0.100, -0.020\end{tabular} & 98/907 & 0.859 & 0.835 & \begin{tabular}[c]{@{}l@{}}$0.12$ \\ {}[0.09, 0.15]\end{tabular} & \begin{tabular}[c]{@{}l@{}}$0.81$ \\ {}[0.76, 0.82]\end{tabular} \\
\addlinespace[2pt]
11 & 199 & $[00000010101]$ & 48 & 19.670 & \begin{tabular}[c]{@{}l@{}}0.002, 0.274, 0.363, 0.147, 0.039, 0.054, -0.247 \\ -0.058, -0.068, 0.066, 0.711, -0.184, 0.104, 0.929 \\ 0.098, 0.051, 0.191, -0.083, -0.227, 0.185, -0.208 \\ -0.173, -0.173, -0.106, 0.004, -0.031, -0.232, -0.285 \\ -0.218, -0.120, -0.197, -0.131, -0.287, -0.149, -0.044 \\ -0.446, -0.271, -0.210, -0.290, -0.067, -0.157, -0.007 \\ -0.069, 0.141, 0.056, -0.234, -0.177, 0.332\end{tabular} & 65/909 & 0.751 & 0.643 & \begin{tabular}[c]{@{}l@{}}$0.07$ \\ {}[0.05, 0.09]\end{tabular} & \begin{tabular}[c]{@{}l@{}}$0.61$ \\ {}[0.56, 0.64]\end{tabular} \\
\addlinespace[2pt]
12 & 322 & $[000001010101]$ & 43 & 17.459 & \begin{tabular}[c]{@{}l@{}}0.983, 0.241, 0.820, 0.094, 0.044, 0.996, 0.136 \\ -0.012, 0.072, 0.547, 0.165, 0.120, -0.004, -0.083 \\ 0.487, -0.094, 0.271, -0.016, 0.020, 0.281, 0.409 \\ 0.319, 0.681, -0.069, -0.061, 0.887, -0.056, -0.238 \\ -0.505, -0.036, 0.073, -0.192, 0.535, -0.234, 0.073 \\ -0.534, -0.750, -0.393, -0.169, -0.522, -0.436, 0.019 \\ 0.472\end{tabular} & 68/909 & 0.580 & 0.568 & \begin{tabular}[c]{@{}l@{}}$0.10$ \\ {}[0.07, 0.12]\end{tabular} & \begin{tabular}[c]{@{}l@{}}$0.56$ \\ {}[0.50, 0.58]\end{tabular} \\
\addlinespace[2pt]
5 & 11 & $[00101]$ & 38 & 15.708 & \begin{tabular}[c]{@{}l@{}}0.494, 0.074, -0.008, 0.019, -0.023, 0.211, -0.046 \\ 0.009, 0.388, 0.012, 0.056, -0.075, -0.040, 0.019 \\ 0.031, 0.015, -0.283, 0.709, 0.051, -0.011, -0.174 \\ 0.034, 0.034, 0.020, 0.035, -0.043, 0.033, -0.163 \\ 0.110, -0.170, -0.048, -0.122, 0.116, -0.029, 0.008 \\ 0.064, -0.225, 0.107\end{tabular} & 460/947 & 1.000 & 0.994 & \begin{tabular}[c]{@{}l@{}}$0.57$ \\ {}[0.52, 0.61]\end{tabular} & \begin{tabular}[c]{@{}l@{}}$0.97$ \\ {}[0.94, 0.99]\end{tabular} \\
\addlinespace[2pt]
6 & 18 & $[010101]$ & 24 & 10.095 & \begin{tabular}[c]{@{}l@{}}0.921, 0.081, -0.386, -0.371, -0.530, 0.084, 0.064 \\ 0.035, -0.321, 0.254, 0.245, 0.035, -0.056, 0.248 \\ 0.065, 0.009, -0.068, 0.102, 0.016, 0.103, 0.353 \\ 0.396, -0.422, -0.587\end{tabular} & 521/934 & 0.996 & 0.986 & \begin{tabular}[c]{@{}l@{}}$0.72$ \\ {}[0.67, 0.75]\end{tabular} & \begin{tabular}[c]{@{}l@{}}$0.98$ \\ {}[0.95, 0.98]\end{tabular} \\
\addlinespace[2pt]
7 & 29 & $[0010101]$ & 24 & 9.925 & \begin{tabular}[c]{@{}l@{}}0.244, -0.004, -0.228, 0.175, 0.096, 0.023, -0.277 \\ -0.143, -0.022, -0.300, 1.014, 0.060, -0.009, -0.176 \\ -0.094, -0.212, 0.061, 0.056, 0.343, 0.190, 0.000 \\ -0.164, -0.121, -0.242\end{tabular} & 456/936 & 0.988 & 0.988 & \begin{tabular}[c]{@{}l@{}}$0.62$ \\ {}[0.58, 0.66]\end{tabular} & \begin{tabular}[c]{@{}l@{}}$0.96$ \\ {}[0.93, 0.98]\end{tabular} \\
\addlinespace[2pt]
8 & 47 & $[01010101]$ & 40 & 16.398 & \begin{tabular}[c]{@{}l@{}}0.826, -0.028, -0.126, -0.179, 0.098, -0.082, -0.030 \\ -0.202, -0.426, -0.028, 0.686, 0.090, -0.212, 0.150 \\ 0.202, -0.386, -0.154, 0.122, 0.044, 0.225, -0.001 \\ 0.182, 0.127, -0.315, -0.191, 0.481, 0.259, -0.473 \\ 0.095, 0.977, 1.219, -0.070, 0.359, -0.027, 0.213 \\ -0.088, -0.018, 0.212, 0.141, -0.035\end{tabular} & 259/920 & 0.990 & 0.957 & \begin{tabular}[c]{@{}l@{}}$0.38$ \\ {}[0.35, 0.42]\end{tabular} & \begin{tabular}[c]{@{}l@{}}$0.95$ \\ {}[0.91, 0.96]\end{tabular} \\
\addlinespace[2pt]
9 & 76 & $[001010101]$ & 18 & 7.884 & \begin{tabular}[c]{@{}l@{}}1.435, 0.430, 0.101, 0.216, -0.042, 0.105, -0.295 \\ 0.352, -0.023, 0.099, -0.002, 0.212, 0.417, -0.052 \\ -0.118, 0.243, -0.634, -1.151\end{tabular} & 542/895 & 0.976 & 0.938 & \begin{tabular}[c]{@{}l@{}}$0.87$ \\ {}[0.81, 0.88]\end{tabular} & \begin{tabular}[c]{@{}l@{}}$0.92$ \\ {}[0.88, 0.93]\end{tabular} \\
\addlinespace[2pt]
10 & 123 & $[0101010101]$ & 34 & 14.027 & \begin{tabular}[c]{@{}l@{}}-0.869, -0.385, -0.024, 0.184, -0.497, 0.415, 0.155 \\ 0.084, 0.025, -0.128, 0.238, 0.162, 0.214, 0.964 \\ 0.872, 0.155, 0.056, 0.115, -0.170, -0.052, -0.037 \\ 0.192, 0.103, 0.148, 0.061, 0.061, 0.124, -0.244 \\ -0.121, 0.063, -0.111, 0.035, 1.110, 1.184\end{tabular} & 272/911 & 0.960 & 0.903 & \begin{tabular}[c]{@{}l@{}}$0.44$ \\ {}[0.40, 0.48]\end{tabular} & \begin{tabular}[c]{@{}l@{}}$0.88$ \\ {}[0.84, 0.90]\end{tabular} \\
\addlinespace[2pt]
11 & 199 & $[00101010101]$ & 10 & 4.572 & \begin{tabular}[c]{@{}l@{}}1.577, 0.240, -0.008, 0.129, -0.432, 0.005, -0.759 \\ -0.388, -0.622, -1.846\end{tabular} & 513/913 & 0.990 & 0.929 & \begin{tabular}[c]{@{}l@{}}$0.85$ \\ {}[0.79, 0.86]\end{tabular} & \begin{tabular}[c]{@{}l@{}}$0.93$ \\ {}[0.88, 0.93]\end{tabular} \\
\addlinespace[2pt]
12 & 322 & $[010101010101]$ & 26 & 10.825 & \begin{tabular}[c]{@{}l@{}}1.178, 0.160, -0.266, -0.031, 0.387, 0.053, 0.220 \\ 0.115, -0.098, -0.315, -0.460, -0.357, -0.380, -0.076 \\ -0.167, -0.070, 0.228, 0.123, 0.036, 0.485, 0.331 \\ 0.258, 0.252, 0.086, -0.358, -1.016\end{tabular} & 213/883 & 0.765 & 0.856 & \begin{tabular}[c]{@{}l@{}}$0.39$ \\ {}[0.35, 0.43]\end{tabular} & \begin{tabular}[c]{@{}l@{}}$0.83$ \\ {}[0.79, 0.85]\end{tabular} \\
\addlinespace[2pt]
13 & 521 & $[0010101010101]$ & 20 & 8.674 & \begin{tabular}[c]{@{}l@{}}-1.181, -0.232, -0.033, 0.031, 0.086, 0.149, -0.183 \\ 0.216, -0.231, 0.038, -0.242, 0.014, -0.107, 0.246 \\ 0.217, 0.356, 0.689, 0.745, 0.809, 1.815\end{tabular} & 369/893 & 0.974 & 0.893 & \begin{tabular}[c]{@{}l@{}}$0.68$ \\ {}[0.62, 0.70]\end{tabular} & \begin{tabular}[c]{@{}l@{}}$0.89$ \\ {}[0.83, 0.89]\end{tabular} \\
\addlinespace[2pt]
14 & 843 & $[01010101010101]$ & 26 & 10.825 & \begin{tabular}[c]{@{}l@{}}-1.159, -0.479, -0.159, -0.025, -0.041, 0.027, 0.270 \\ 0.414, 0.449, 0.621, 0.533, 0.429, 0.264, -0.251 \\ 0.199, 0.416, 0.132, 1.116, 1.318, 0.926, 0.774 \\ 0.951, 0.156, 0.191, 1.008, 1.032\end{tabular} & 218/913 & 0.845 & 0.875 & \begin{tabular}[c]{@{}l@{}}$0.41$ \\ {}[0.36, 0.44]\end{tabular} & \begin{tabular}[c]{@{}l@{}}$0.86$ \\ {}[0.81, 0.87]\end{tabular} \\
\addlinespace[2pt]
15 & 1364 & $[001010101010101]$ & 20 & 8.624 & \begin{tabular}[c]{@{}l@{}}1.492, 0.149, 0.268, 0.127, 0.035, -0.067, -0.073 \\ 0.060, -0.261, -0.003, 0.059, -0.452, 0.259, -0.545 \\ -0.513, 0.329, 1.458, -0.224, 0.320, 0.192\end{tabular} & 183/876 & 0.830 & 0.737 & \begin{tabular}[c]{@{}l@{}}$0.35$ \\ {}[0.29, 0.37]\end{tabular} & \begin{tabular}[c]{@{}l@{}}$0.74$ \\ {}[0.67, 0.75]\end{tabular} \\
\addlinespace[2pt]
17 & 3571 & $[00101010101010101]$ & 15 & 6.383 & \begin{tabular}[c]{@{}l@{}}0.972, 1.184, 0.297, 1.054, -0.300, 1.005, -0.772 \\ -0.410, -0.740, -0.186, 0.058, 0.397, 0.054, 0.015 \\ -0.583\end{tabular} & 182/876 & 0.708 & 0.557 & \begin{tabular}[c]{@{}l@{}}$0.37$ \\ {}[0.30, 0.38]\end{tabular} & \begin{tabular}[c]{@{}l@{}}$0.52$ \\ {}[0.45, 0.53]\end{tabular} \\
\bottomrule
\end{tabularx}
\endgroup
\caption{Numerical and hardware results for bracelet states $\ket{[z^\text{half}]}$ and $\ket{[z^\text{MIS}]}$ product states, as detailed at the start of \cref{app:table_details}.}\label{tab:bracelet}
\end{table*}

\begin{table*}[h]
\centering
\setlength{\tabcolsep}{10pt}
\begin{tabular}{l l c c c c}
\toprule
Target & Platform & $p$ & $c$ & $\alpha$ & $R^2$ \\
\midrule
\multirow{9}{*}{$z^{\rm half}$}
& \multirow{3}{*}{CTQW} & 1 & 1.099 & $0.878 \pm 0.006$ & 0.9870 \\
& & 2 & 1.112 & $0.935 \pm 0.003$ & 0.9997 \\
& & 3 & 1.015 & $0.981 \pm 0.002$ & 0.9997 \\
\cmidrule{2-6}
& \multirow{3}{*}{Emulation} & 1 & 0.632 & $0.877 \pm 0.009$ & 0.9881 \\
& & 2 & 0.719 & $0.934 \pm 0.006$ & 0.9961 \\
& & 3 & 0.785 & $0.930 \pm 0.006$ & 0.9959 \\
\cmidrule{2-6}
& \multirow{3}{*}{Hardware} & 1 & 0.595 & $0.729 \pm 0.041$ & 0.9326 \\
& & 2 & 0.463 & $0.876 \pm 0.025$ & 0.9374 \\
& & 3 & 0.418 & $0.734 \pm 0.051$ & 0.7889 \\
\midrule
\multirow{6}{*}{$z^{\rm MIS}$}
& \multirow{2}{*}{CTQW} & 1 & 0.995 & $0.996 \pm 0.001$ & 0.9972 \\
& & 2 & 1.000 & $1.000 \pm 0.000$ & 1.0000 \\
\cmidrule{2-6}
& \multirow{2}{*}{Emulation} & 1 & 0.641 & $1.000 \pm 0.004$ & 0.9838 \\
& & 2 & 0.507 & $0.967 \pm 0.006$ & 0.9857 \\
\cmidrule{2-6}
& \multirow{2}{*}{Hardware} & 1 & 0.512 & $0.820 \pm 0.029$ & 0.9724 \\
& & 2 & 0.285 & $0.703 \pm 0.067$ & 0.9250 \\
\bottomrule
\end{tabular}
\caption{Power-law fit parameters for amplification scaling $A = c\,|\mathcal V|^\alpha$ (see \cref{eq:amp_power_law}) across CTQW (Perfect), noiseless Rydberg emulation, and Aquila hardware. Values are listed for depths $p=1,2,3$ for $z^{\rm half}$ and $p=1,2$ for $z^{\rm MIS}$).}
\label{tab:power_law_fits}
\end{table*}

\bibliographystyle{apsrev4-2}
\bibliography{qis}

@misc{setonixcomputer,
	author       = {{Pawsey Supercomputing Research Centre}},
	title        = {{Setonix} {Supercomputer}},
        note        = {{Pawsey} {Supercomputing} {Research} {Centre}, {Perth}, {Western} {Australia}},
	year         = 2023,
	doi          = {10.48569/18sb-8s43},
}

@article{boulebnane2024,
  title = {Towards a linear-ramp {QAOA} protocol: evidence of a scaling advantage in solving some combinatorial optimization problems},
  volume = {10},
  ISSN = {2056-6387},
  url = {http://dx.doi.org/10.1038/s41534-024-00972-2},
  DOI = {10.1038/s41534-024-00972-2},
  number = {1},
  journal = {npj Quantum Information},
  publisher = {Springer Science and Business Media LLC},
  author = {Boulebnane, Sami and Montanez-Barrera, J. A. and von Borstel, Mikel Sanz and Biercuk, Michael J. and Gu{\'e}ry-Odelin, David and Meyer-Scott, Evan and Thomas, Emmanuel and Baum, Yuval},
  year = {2024},
  month = dec
}

@article{Fuchs2024,
	title        = {{LX}-mixers for {QAOA}: Optimal mixers restricted to subspaces and the stabilizer formalism},
	author       = {Fuchs, Franz G. and Bassa, Ruben},
	year         = 2024,
	journal      = {Quantum},
	publisher    = {Verein zur F{\"{o}}rderung des Open Access Publizierens in den Quantenwissenschaften},
	volume       = 8,
	pages        = 1535,
	doi          = {10.22331/q-2024-12-04-1535},
	url          = {https://doi.org/10.22331/q-2024-12-04-1535}
}

@article{Gonzales2025,
  title = {Efficient sparse state preparation via quantum walks},
  volume = {11},
  ISSN = {2056-6387},
  url = {http://dx.doi.org/10.1038/s41534-025-01093-y},
  DOI = {10.1038/s41534-025-01093-y},
  number = {1},
  journal = {npj Quantum Information},
  publisher = {Springer Science and Business Media LLC},
  author = {Gonzales,  Alvin and Herrman,  Rebekah and Campbell,  Colin and Gaidai,  Igor and Liu,  Ji and Tomesh,  Teague and Saleem,  Zain H.},
  year = {2025},
  month = aug,
  pages = {143}
}

@article{blekos2023review,
	title        = {A review on Quantum Approximate Optimization Algorithm and its variants},
	author       = {Blekos, Kostas and Brand, Dean and Ceschini, Andrea and Chou, Chiao-Hui and Li, Rui-Hao and Pandya, Komal and Summer, Alessandro},
	year         = 2024,
	month        = jun,
	journal      = {Physics Reports},
	publisher    = {Elsevier BV},
	volume       = 1068,
	pages        = {1–66},
	doi          = {10.1016/j.physrep.2024.03.002},
	issn         = {0370-1573},
	url          = {http://dx.doi.org/10.1016/j.physrep.2024.03.002}
}

@misc{farhi2014quantum,
  doi = {10.48550/ARXIV.1411.4028},
  url = {https://arxiv.org/abs/1411.4028},
  author = {Farhi,  Edward and Goldstone,  Jeffrey and Gutmann,  Sam},
  title = {A Quantum Approximate Optimization Algorithm},
  publisher = {arXiv},
  year = {2014},
  copyright = {arXiv.org perpetual,  non-exclusive license},
  eprint        = {1411.4028},
  archivePrefix = {arXiv},
  primaryClass  = {quant-ph}
}

@misc{wurtz2023aquila,
  doi = {10.48550/ARXIV.2306.11727},
  url = {https://arxiv.org/abs/2306.11727},
  author = {Wurtz,  Jonathan and Bylinskii,  Alexei and Braverman,  Boris and Amato-Grill,  Jesse and Cantu,  Sergio H. and Huber,  Florian and Lukin,  Alexander and Liu,  Fangli and Weinberg,  Phillip and Long,  John and Wang,  Sheng-Tao and Gemelke,  Nathan and Keesling,  Alexander},
  title = {Aquila: {QuEra}'s 256-qubit neutral-atom quantum computer},
  publisher = {arXiv},
  year = {2023},
  copyright = {arXiv.org perpetual,  non-exclusive license}
}

@article{Ebadi_2022,
	title        = {Quantum optimization of maximum independent set using {Rydberg} atom arrays},
	allauthor       = {Ebadi, S. and Keesling, A. and Cain, M. and Wang, T. T. and Levine, H. and Bluvstein, D. and Semeghini, G. and Omran, A. and Liu, J.-G. and Samajdar, R. and Luo, X.-Z. and Nash, B. and Gao, X. and Barak, B. and Farhi, E. and Sachdev, S. and Gemelke, N. and Zhou, L. and Choi, S. and Pichler, H. and Wang, S.-T. and Greiner, M. and Vuleti\'{c}, V. and Lukin, M. D.},
        author = {Ebadi, S. and Keesling, A. and Cain, M. and Wang, T. T. and Levine, H. and Bluvstein, D. and Semeghini, G. and Omran, A. and Liu, J.-G. and Samajdar, R. and Luo, X.-Z. and Nash, B. and Gao, X. and Barak, B. and Farhi, E. and Sachdev, S. and Gemelke, N. and Zhou, L. and Choi, S. and Pichler, H. and et al.},
	year         = 2022,
	month        = jun,
	journal      = {Science},
	publisher    = {American Association for the Advancement of Science (AAAS)},
	volume       = 376,
	number       = 6598,
	pages        = {1209–1215},
	doi          = {10.1126/science.abo6587},
	issn         = {1095-9203},
	url          = {http://dx.doi.org/10.1126/science.abo6587}
}

@article{Hadfield_2019,
	title        = {From the Quantum Approximate Optimization Algorithm to a Quantum Alternating Operator Ansatz},
	author       = {Hadfield, Stuart and Wang, Zhihui and O'Gorman, Bryan and Rieffel, Eleanor and Venturelli, Davide and Biswas, Rupak},
	year         = 2019,
	month        = feb,
	journal      = {Algorithms},
	publisher    = {MDPI AG},
	volume       = 12,
	number       = 2,
	pages        = 34,
	doi          = {10.3390/a12020034},
	issn         = {1999-4893},
	url          = {http://dx.doi.org/10.3390/a12020034}
}

@article{Albash_2018,
	title        = {Adiabatic quantum computation},
	author       = {Albash, Tameem and Lidar, Daniel A.},
	year         = 2018,
	month        = jan,
	journal      = {Reviews of Modern Physics},
	publisher    = {American Physical Society (APS)},
	volume       = 90,
	number       = 1,
	pages        = {015002},
	doi          = {10.1103/revmodphys.90.015002},
	issn         = {1539-0756},
	url          = {http://dx.doi.org/10.1103/RevModPhys.90.015002}
}

@misc{Lukin2024shortcut,
	title        = {Quantum quench dynamics as a shortcut to adiabaticity},
	author       = {Lukin, Alexander and Schiffer, Benjamin F. and Braverman, Boris and Cantu, Sergio H. and Huber, Florian and Bylinskii, Alexei and Amato-Grill, Jesse and Maskara, Nishad and Cain, Madelyn and Wild, Dominik S. and Samajdar, Rhine and Lukin, Mikhail D.},
	year         = 2024,
	month        = may,
	eprint       = {2405.21019},
	archiveprefix = {arXiv},
	primaryclass = {quant-ph}
}

@inproceedings{Bartschi2019,
	title        = {Deterministic Preparation of {Dicke} States},
	author       = {B\"{a}rtschi, Andreas and Eidenbenz, Stephan},
	booktitle    = {Fundamentals of Computation Theory (FCT 2019)},
	year         = 2019,
	pages        = {126--139},
	doi          = {10.1007/978-3-030-25027-0_9},
	eprint       = {1904.07358},
	archiveprefix = {arXiv},
	primaryclass = {quant-ph}
}

@inproceedings{Bartschi2020,
	title        = {Grover Mixers for {QAOA}: Shifting Complexity from Mixer Design to State Preparation},
	author       = {B\"{a}rtschi, Andreas and Eidenbenz, Stephan},
	booktitle    = {IEEE International Conference on Quantum Computing and Engineering (QCE 2020)},
	year         = 2020,
	pages        = {72--82},
	doi          = {10.1109/QCE49297.2020.00020},
	eprint       = {2006.00354},
	archiveprefix = {arXiv},
	primaryclass = {quant-ph}
}

@article{Bluvstein2021,
	title        = {Controlling quantum many-body dynamics in driven {Rydberg} atom arrays},
	author       = {Bluvstein,  D. and Omran,  A. and Levine,  H. and Keesling,  A. and Semeghini,  G. and Ebadi,  S. and Wang,  T. T. and Michailidis,  A. A. and Maskara,  N. and Ho,  W. W. and Choi,  S. and Serbyn,  M. and Greiner,  M. and Vuleti\'{c},  V. and Lukin,  M. D.},
	year         = 2021,
	month        = mar,
	journal      = {Science},
	publisher    = {American Association for the Advancement of Science (AAAS)},
	volume       = 371,
	number       = 6536,
	pages        = {1355–1359},
	doi          = {10.1126/science.abg2530},
	issn         = {1095-9203},
	url          = {http://dx.doi.org/10.1126/science.abg2530}
}

@misc{reichardt2025,
  doi = {10.48550/ARXIV.2411.11822},
  url = {https://arxiv.org/abs/2411.11822},
  allauthor = {Reichardt,  Ben W. and Paetznick,  Adam and Aasen,  David and Basov,  Ivan and Bello-Rivas,  Juan M. and Bonderson,  Parsa and Chao,  Rui and van Dam,  Wim and Hastings,  Matthew B. and Mishmash,  Ryan V. and Paz,  Andres and da Silva,  Marcus P. and Sundaram,  Aarthi and Svore,  Krysta M. and Vaschillo,  Alexander and Wang,  Zhenghan and Zanner,  Matt and Cairncross,  William B. and Chen,  Cheng-An and Crow,  Daniel and Kim,  Hyosub and Kindem,  Jonathan M. and King,  Jonathan and McDonald,  Michael and Norcia,  Matthew A. and Ryou,  Albert and Stone,  Mark and Wadleigh,  Laura and Barnes,  Katrina and Battaglino,  Peter and Bohdanowicz,  Thomas C. and Booth,  Graham and Brown,  Andrew and Brown,  Mark O. and Cassella,  Kayleigh and Coxe,  Robin and Epstein,  Jeffrey M. and Feldkamp,  Max and Griger,  Christopher and Halperin,  Eli and Heinz,  Andre and Hummel,  Frederic and Jaffe,  Matthew and Jones,  Antonia M. W. and Kapit,  Eliot and Kotru,  Krish and Lauigan,  Joseph and Li,  Ming and Marjanovic,  Jan and Megidish,  Eli and Meredith,  Matthew and Morshead,  Ryan and Muniz,  Juan A. and Narayanaswami,  Sandeep and Nishiguchi,  Ciro and Paule,  Timothy and Pawlak,  Kelly A. and Pudenz,  Kristen L. and Pérez,  David Rodríguez and Simon,  Jon and Smull,  Aaron and Stack,  Daniel and Urbanek,  Miroslav and van de Veerdonk,  René J. M. and Vendeiro,  Zachary and Weverka,  Robert T. and Wilkason,  Thomas and Wu,  Tsung-Yao and Xie,  Xin and Zalys-Geller,  Evan and Zhang,  Xiaogang and Bloom,  Benjamin J.},
  author = {Reichardt, Ben W. and Paetznick, Adam and Aasen, David and Basov, Ivan and Bello-Rivas, Juan M. and Bonderson, Parsa and Chao, Rui and van Dam, Wim and Hastings, Matthew B. and Mishmash, Ryan V. and Paz, Andres and da Silva, Marcus P. and Sundaram, Aarthi and Svore, Krysta M. and Vaschillo, Alexander and Wang, Zhenghan and Zanner, Matt and Cairncross, William B. and Chen, Cheng-An and Crow, Daniel and et al.},
  title = {Fault-tolerant quantum computation with a neutral atom processor},
  publisher = {arXiv},
  year = {2024},
  copyright = {arXiv.org perpetual,  non-exclusive license},
  eprint        = {2411.11822},
  archivePrefix = {arXiv},
  primaryClass  = {quant-ph}
}

@article{Shaw2024,
	title        = {Benchmarking highly entangled states on a 60-atom analogue quantum simulator},
	author       = {Shaw, Adam L. and Chen, Zhuo and Choi, Joonhee and Mark, Daniel K. and Scholl, Pascal and Finkelstein, Ran and Elben, Andreas and Choi, Soonwon and Endres, Manuel},
	year         = 2024,
	month        = mar,
	journal      = {Nature},
	publisher    = {Springer Science and Business Media LLC},
	volume       = 628,
	number       = 8006,
	pages        = {71–77},
	doi          = {10.1038/s41586-024-07173-x},
	issn         = {1476-4687},
	url          = {http://dx.doi.org/10.1038/s41586-024-07173-x}
}

@article{Evered2023,
	title        = {High-fidelity parallel entangling gates on a neutral-atom quantum computer},
	author       = {Evered,  Simon J. and Bluvstein,  Dolev and Kalinowski,  Marcin and Ebadi,  Sepehr and Manovitz,  Tom and Zhou,  Hengyun and Li,  Sophie H. and Geim,  Alexandra A. and Wang,  Tout T. and Maskara,  Nishad and Levine,  Harry and Semeghini,  Giulia and Greiner,  Markus and Vuleti\'{c},  Vladan and Lukin,  Mikhail D.},
	year         = 2023,
	month        = oct,
	journal      = {Nature},
	publisher    = {Springer Science and Business Media LLC},
	volume       = 622,
	number       = 7982,
	pages        = {268–272},
	doi          = {10.1038/s41586-023-06481-y},
	issn         = {1476-4687},
	url          = {http://dx.doi.org/10.1038/s41586-023-06481-y}
}

@misc{evered2025,
	title        = {Probing topological matter and fermion dynamics on a neutral-atom quantum computer},
	author       = {Simon J. Evered and Marcin Kalinowski and Alexandra A. Geim and Tom Manovitz and Dolev Bluvstein and Sophie H. Li and Nishad Maskara and Hengyun Zhou and Sepehr Ebadi and Muqing Xu and Joseph Campo and Madelyn Cain and Stefan Ostermann and Susanne F. Yelin and Subir Sachdev and Markus Greiner and Vladan Vuleti\'{c} and Mikhail D. Lukin},
	year         = 2025,
	url          = {https://arxiv.org/abs/2501.18554},
	eprint       = {2501.18554},
	archiveprefix = {arXiv},
	primaryclass = {quant-ph}
}

@article{Gonz_lez_Cuadra_2025,
	title        = {Observation of string breaking on a (2 + 1){D} {Rydberg} quantum simulator},
	author       = {Gonz\'{a}lez-Cuadra, Daniel and Hamdan, Majd and Zache, Torsten V. and Braverman, Boris and Kornja\v{c}a, Milan and Lukin, Alexander and Cant\'{u}, Sergio H. and Liu, Fangli and Wang, Sheng-Tao and Keesling, Alexander and Lukin, Mikhail D. and Zoller, Peter and Bylinskii, Alexei},
	year         = 2025,
	month        = jun,
	journal      = {Nature},
	publisher    = {Springer Science and Business Media LLC},
	volume       = 642,
	number       = 8067,
	pages        = {321–326},
	doi          = {10.1038/s41586-025-09051-6},
	issn         = {1476-4687},
	url          = {http://dx.doi.org/10.1038/s41586-025-09051-6}
}

@misc{bluvstein2025,
	title        = {Architectural mechanisms of a universal fault-tolerant quantum computer},
	author       = {Dolev Bluvstein and Alexandra A. Geim and Sophie H. Li and Simon J. Evered and J. Pablo Bonilla Ataides and Gefen Baranes and Andi Gu and Tom Manovitz and Muqing Xu and Marcin Kalinowski and Shayan Majidy and Christian Kokail and Nishad Maskara and Elias C. Trapp and Luke M. Stewart and Simon Hollerith and Hengyun Zhou and Michael J. Gullans and Susanne F. Yelin and Markus Greiner and Vladan Vuletic and Madelyn Cain and Mikhail D. Lukin},
	year         = 2025,
	url          = {https://arxiv.org/abs/2506.20661},
	eprint       = {2506.20661},
	archiveprefix = {arXiv},
	primaryclass = {quant-ph}
}

@article{Morgado2021,
  title = {Quantum simulation and computing with {Rydberg}-interacting qubits},
  volume = {3},
  ISSN = {2639-0213},
  url = {http://dx.doi.org/10.1116/5.0036562},
  DOI = {10.1116/5.0036562},
  number = {2},
  journal = {AVS Quantum Science},
  publisher = {American Vacuum Society},
  author = {Morgado,  M. and Whitlock,  S.},
  year = {2021},
  month = may,
  pages = {023501},
}

@article{Saffman2016,
	title        = {Quantum computing with atomic qubits and {Rydberg} interactions: progress and challenges},
	author       = {Saffman, M},
	year         = 2016,
	month        = oct,
	journal      = {Journal of Physics B: Atomic, Molecular and Optical Physics},
	publisher    = {IOP Publishing},
	volume       = 49,
	number       = 20,
	pages        = 202001,
	doi          = {10.1088/0953-4075/49/20/202001},
	issn         = {1361-6455},
	url          = {http://dx.doi.org/10.1088/0953-4075/49/20/202001}
}

@article{rodriguez2024,
  title = {Experimental demonstration of logical magic state distillation},
  ISSN = {1476-4687},
  url = {http://dx.doi.org/10.1038/s41586-025-09367-3},
  DOI = {10.1038/s41586-025-09367-3},
  journal = {Nature},
  volume = {645},
  number = {8081},
  pages = {620-625},
  allauthor = {Rodriguez,  Pedro Sales and Robinson,  John M. and Jepsen,  Paul Niklas and He,  Zhiyang and Duckering,  Casey and Zhao,  Chen and Wu,  Kai-Hsin and Campo,  Joseph and Bagnall,  Kevin and Kwon,  Minho and Karolyshyn,  Thomas and Weinberg,  Phillip and Cain,  Madelyn and Evered,  Simon J. and Geim,  Alexandra A. and Kalinowski,  Marcin and Li,  Sophie H. and Manovitz,  Tom and Amato-Grill,  Jesse and Basham,  James I. and Bernstein,  Liane and Braverman,  Boris and Bylinskii,  Alexei and Choukri,  Adam and DeAngelo,  Robert and Fang,  Fang and Fieweger,  Connor and Frederick,  Paige and Haines,  David and Hamdan,  Majd and Hammett,  Julian and Hsu,  Ning and Hu,  Ming-Guang and Huber,  Florian and Jia,  Ningyuan and Kedar,  Dhruv and Kornjača,  Milan and Liu,  Fangli and Long,  John and Lopatin,  Jonathan and Lopes,  Pedro L. S. and Luo,  Xiu-Zhe and Macrì,  Tommaso and Marković,  Ognjen and Martínez-Martínez,  Luis A. and Meng,  Xianmei and Ostermann,  Stefan and Ostroumov,  Evgeny and Paquette,  David and Qiang,  Zexuan and Shofman,  Vadim and Singh,  Anshuman and Singh,  Manuj and Sinha,  Nandan and Thoreen,  Henry and Wan,  Noel and Wang,  Yiping and Waxman-Lenz,  Daniel and Wong,  Tak and Wurtz,  Jonathan and Zhdanov,  Andrii and Zheng,  Laurent and Greiner,  Markus and Keesling,  Alexander and Gemelke,  Nathan and Vuletić,  Vladan and Kitagawa,  Takuya and Wang,  Sheng-Tao and Bluvstein,  Dolev and Lukin,  Mikhail D. and Lukin,  Alexander and Zhou,  Hengyun and Cantú,  Sergio H.},
  author = {Rodriguez, Pedro Sales and Robinson, John M. and Jepsen, Paul Niklas and He, Zhiyang and Duckering, Casey and Zhao, Chen and Wu, Kai-Hsin and Campo, Joseph and Bagnall, Kevin and Kwon, Minho and Karolyshyn, Thomas and Weinberg, Phillip and Cain, Madelyn and Evered, Simon J. and Geim, Alexandra A. and Kalinowski, Marcin and Li, Sophie H. and Manovitz, Tom and Amato-Grill, Jesse and Basham, James I. and et al.},
  publisher = {Springer Science and Business Media LLC},
  year = {2025},
  month = jul 
}

@article{Bluvstein2023,
	title        = {Logical quantum processor based on reconfigurable atom arrays},
	author       = {Bluvstein, Dolev and Evered, Simon J. and Geim, Alexandra A. and Li, Sophie H. and Zhou, Hengyun and Manovitz, Tom and Ebadi, Sepehr and Cain, Madelyn and Kalinowski, Marcin and Hangleiter, Dominik and Bonilla Ataides, J. Pablo and Maskara, Nishad and Cong, Iris and Gao, Xun and Sales Rodriguez, Pedro and Karolyshyn, Thomas and Semeghini, Giulia and Gullans, Michael J. and Greiner, Markus and Vuleti\'{c}, Vladan and Lukin, Mikhail D.},
	year         = 2023,
	month        = dec,
	journal      = {Nature},
	publisher    = {Springer Science and Business Media LLC},
	volume       = 626,
	number       = 7997,
	pages        = {58–65},
	doi          = {10.1038/s41586-023-06927-3},
	issn         = {1476-4687},
	url          = {http://dx.doi.org/10.1038/s41586-023-06927-3}
}

@article{Bernien2017,
	title        = {Probing many-body dynamics on a 51-atom quantum simulator},
	author       = {Bernien, Hannes and Schwartz, Sylvain and Keesling, Alexander and Levine, Harry and Omran, Ahmed and Pichler, Hannes and Choi, Soonwon and Zibrov, Alexander S. and Endres, Manuel and Greiner, Markus and Vuleti\'{c}, Vladan and Lukin, Mikhail D.},
	year         = 2017,
	month        = nov,
	journal      = {Nature},
	publisher    = {Springer Science and Business Media LLC},
	volume       = 551,
	number       = 7682,
	pages        = {579–584},
	doi          = {10.1038/nature24622},
	issn         = {1476-4687},
	url          = {http://dx.doi.org/10.1038/nature24622}
}

@article{Choi2020,
	title        = {Robust Dynamic {Hamiltonian} Engineering of Many-Body Spin Systems},
	author       = {Choi,  Joonhee and Zhou,  Hengyun and Knowles,  Helena S. and Landig,  Renate and Choi,  Soonwon and Lukin,  Mikhail D.},
	year         = 2020,
	month        = jul,
	journal      = {Physical Review X},
	publisher    = {American Physical Society (APS)},
	volume       = 10,
	number       = 3,
	doi          = {10.1103/physrevx.10.031002},
	issn         = {2160-3308},
	url          = {http://dx.doi.org/10.1103/PhysRevX.10.031002},
        pages = {031002},
}

@article{Choi2019,
	title        = {Emergent {SU}(2) Dynamics and Perfect Quantum Many-Body Scars},
	author       = {Choi,  Soonwon and Turner,  Christopher J. and Pichler,  Hannes and Ho,  Wen Wei and Michailidis,  Alexios A. and Papi\'{c},  Zlatko and Serbyn,  Maksym and Lukin,  Mikhail D. and Abanin,  Dmitry A.},
	year         = 2019,
	month        = jun,
	journal      = {Physical Review Letters},
	publisher    = {American Physical Society (APS)},
	volume       = 122,
	number       = 22,
	doi          = {10.1103/physrevlett.122.220603},
	issn         = {1079-7114},
	url          = {http://dx.doi.org/10.1103/PhysRevLett.122.220603},
        pages        = {220603},
}

@article{Cerezo2022,
	title        = {Challenges and opportunities in quantum machine learning},
	author       = {Cerezo,  M. and Verdon,  Guillaume and Huang,  Hsin-Yuan and Cincio,  Lukasz and Coles,  Patrick J.},
	year         = 2022,
	month        = sep,
	journal      = {Nature Computational Science},
	publisher    = {Springer Science and Business Media LLC},
	volume       = 2,
	number       = 9,
	pages        = {567–576},
	doi          = {10.1038/s43588-022-00311-3},
	issn         = {2662-8457},
	url          = {http://dx.doi.org/10.1038/s43588-022-00311-3}
}

@article{Wurtz2024,
	title        = {Solving Nonnative Combinatorial Optimization Problems Using Hybrid Quantum–Classical Algorithms},
	author       = {Wurtz,  Jonathan and Sack,  Stefan H. and Wang,  Sheng-Tao},
	year         = 2024,
	journal      = {IEEE Transactions on Quantum Engineering},
	publisher    = {Institute of Electrical and Electronics Engineers (IEEE)},
	volume       = 5,
	pages        = {1–14},
	doi          = {10.1109/tqe.2024.3443660},
	issn         = {2689-1808},
	url          = {http://dx.doi.org/10.1109/TQE.2024.3443660}
}

@article{Wurtz2022,
	title        = {Counterdiabaticity and the quantum approximate optimization algorithm},
	author       = {Wurtz,  Jonathan and Love,  Peter J.},
	year         = 2022,
	month        = jan,
	journal      = {Quantum},
	publisher    = {Verein zur Forderung des Open Access Publizierens in den Quantenwissenschaften},
	volume       = 6,
	pages        = 635,
	doi          = {10.22331/q-2022-01-27-635},
	issn         = {2521-327X},
	url          = {http://dx.doi.org/10.22331/q-2022-01-27-635}
}

@article{GuryOdelin2019,
  title = {Shortcuts to adiabaticity: Concepts,  methods,  and applications},
  volume = {91},
  ISSN = {1539-0756},
  url = {http://dx.doi.org/10.1103/RevModPhys.91.045001},
  DOI = {10.1103/revmodphys.91.045001},
  number = {4},
  journal = {Reviews of Modern Physics},
  publisher = {American Physical Society (APS)},
  author = {Guéry-Odelin,  D. and Ruschhaupt,  A. and Kiely,  A. and Torrontegui,  E. and Martínez-Garaot,  S. and Muga,  J. G.},
  year = {2019},
  month = oct,
  pages = {045001},
}

@article{sawada_generating_2001,
  title = {Generating Bracelets in Constant Amortized Time},
  volume = {31},
  ISSN = {1095-7111},
  url = {http://dx.doi.org/10.1137/S0097539700377037},
  DOI = {10.1137/s0097539700377037},
  number = {1},
  journal = {SIAM Journal on Computing},
  publisher = {Society for Industrial & Applied Mathematics (SIAM)},
  author = {Sawada,  Joe},
  year = {2001},
  month = jan,
  pages = {259–268}
}

@article{Preskill18,
  title = {Quantum Computing in the {NISQ} era and beyond},
  volume = {2},
  ISSN = {2521-327X},
  url = {http://dx.doi.org/10.22331/q-2018-08-06-79},
  DOI = {10.22331/q-2018-08-06-79},
  journal = {Quantum},
  publisher = {Verein zur Forderung des Open Access Publizierens in den Quantenwissenschaften},
  author = {Preskill,  John},
  year = {2018},
  month = aug,
  pages = {79}
}

@article{marsh_combinatorial_2020,
  title = {Combinatorial optimization via highly efficient quantum walks},
  volume = {2},
  ISSN = {2643-1564},
  url = {http://dx.doi.org/10.1103/PhysRevResearch.2.023302},
  DOI = {10.1103/physrevresearch.2.023302},
  number = {2},
  journal = {Physical Review Research},
  publisher = {American Physical Society (APS)},
  author = {Marsh,  S. and Wang,  J. B.},
  year = {2020},
  month = jun,
  pages = {023302}
}

@article{matwiejew_quantum_2023,
  title = {Quantum optimisation for continuous multivariable functions by a structured search},
  volume = {8},
  ISSN = {2058-9565},
  url = {http://dx.doi.org/10.1088/2058-9565/ace6cc},
  DOI = {10.1088/2058-9565/ace6cc},
  number = {4},
  journal = {Quantum Science and Technology},
  publisher = {IOP Publishing},
  author = {Matwiejew,  Edric and Pye,  Jason and Wang,  Jingbo B},
  year = {2023},
  month = jul,
  pages = {045013}
}

@inproceedings{childs_exponential_2003,
	title        = {Exponential algorithmic speedup by a quantum walk},
	author       = {Childs, Andrew M. and Cleve, Richard and Deotto, Enrico and Farhi, Edward and Gutmann, Sam and Spielman, Daniel A.},
	year         = 2003,
	month        = jun,
	booktitle    = {Proceedings of the thirty-fifth annual {ACM} symposium on Theory of computing},
	publisher    = {Association for Computing Machinery},
	address      = {San Diego, CA, USA},
	series       = {{STOC} '03},
	pages        = {59--68},
	doi          = {10.1145/780542.780552},
	isbn         = {978-1-58113-674-6},
	url          = {https://doi.org/10.1145/780542.780552},
	urldate      = {2020-06-14},
}

@article{Ambainis2007,
	title        = {Quantum Walk Algorithm for Element Distinctness},
	author       = {Ambainis, Andris},
	year         = 2007,
	journal      = {SIAM Journal on Computing},
	volume       = 37,
	number       = 1,
	pages        = {210--239},
	doi          = {10.1137/S0097539705447311}
}

@article{young_tweezer-programmable_2022,
	title        = {Tweezer-programmable {2D} quantum walks in a Hubbard-regime lattice},
	author       = {Young, Aaron W. and Eckner, William J. and Schine, Nathan and Childs, Andrew M. and Kaufman, Adam M.},
	volume       = 377,
	number       = 6608,
	pages        = {885--889},
	doi          = {10.1126/science.abo0608},
	url          = {https://www.science.org/doi/10.1126/science.abo0608},
	urldate      = {2025-07-30},
	journal = {Science},
	date         = {2022-08-19},
	year = 2022
}

@inproceedings{bennett2024non,
  title = {Non-Variational Quantum Combinatorial Optimisation},
  url = {http://dx.doi.org/10.1109/QCE60285.2024.00014},
  DOI = {10.1109/qce60285.2024.00014},
  booktitle = {2024 IEEE International Conference on Quantum Computing and Engineering (QCE)},
  publisher = {IEEE},
  author = {Bennett,  Tavis and Noakes,  Lyle and Wang,  Jingbo},
  year = {2024},
  month = sep,
  pages = {31–41}
}

@article{peruzzo2010quantum,
  title = {Quantum Walks of Correlated Photons},
  volume = {329},
  ISSN = {1095-9203},
  url = {http://dx.doi.org/10.1126/science.1193515},
  DOI = {10.1126/science.1193515},
  number = {5998},
  journal = {Science},
  publisher = {American Association for the Advancement of Science (AAAS)},
  author = {Peruzzo,  Alberto and Lobino,  Mirko and Matthews,  Jonathan C. F. and Matsuda,  Nobuyuki and Politi,  Alberto and Poulios,  Konstantinos and Zhou,  Xiao-Qi and Lahini,  Yoav and Ismail,  Nur and W\"{o}rhoff,  Kerstin and Bromberg,  Yaron and Silberberg,  Yaron and Thompson,  Mark G. and O'Brien,  Jeremy L.},
  year = {2010},
  month = sep,
  pages = {1500–1503}
}

@article{yan2019strongly,
  title = {Strongly correlated quantum walks with a 12-qubit superconducting processor},
  volume = {364},
  ISSN = {1095-9203},
  url = {http://dx.doi.org/10.1126/science.aaw1611},
  DOI = {10.1126/science.aaw1611},
  number = {6442},
  journal = {Science},
  publisher = {American Association for the Advancement of Science (AAAS)},
  author = {Yan,  Zhiguang and Zhang,  Yu-Ran and Gong,  Ming and Wu,  Yulin and Zheng,  Yarui and Li,  Shaowei and Wang,  Can and Liang,  Futian and Lin,  Jin and Xu,  Yu and Guo,  Cheng and Sun,  Lihua and Peng,  Cheng-Zhi and Xia,  Keyu and Deng,  Hui and Rong,  Hao and You,  J. Q. and Nori,  Franco and Fan,  Heng and Zhu,  Xiaobo and Pan,  Jian-Wei},
  year = {2019},
  month = may,
  pages = {753–756}
}

@article{gong2021quantum,
  title = {Quantum walks on a programmable two-dimensional 62-qubit superconducting processor},
  volume = {372},
  ISSN = {1095-9203},
  url = {http://dx.doi.org/10.1126/science.abg7812},
  DOI = {10.1126/science.abg7812},
  number = {6545},
  journal = {Science},
  publisher = {American Association for the Advancement of Science (AAAS)},
  allauthor = {Gong,  Ming and Wang,  Shiyu and Zha,  Chen and Chen,  Ming-Cheng and Huang,  He-Liang and Wu,  Yulin and Zhu,  Qingling and Zhao,  Youwei and Li,  Shaowei and Guo,  Shaojun and Qian,  Haoran and Ye,  Yangsen and Chen,  Fusheng and Ying,  Chong and Yu,  Jiale and Fan,  Daojin and Wu,  Dachao and Su,  Hong and Deng,  Hui and Rong,  Hao and Zhang,  Kaili and Cao,  Sirui and Lin,  Jin and Xu,  Yu and Sun,  Lihua and Guo,  Cheng and Li,  Na and Liang,  Futian and Bastidas,  V. M. and Nemoto,  Kae and Munro,  W. J. and Huo,  Yong-Heng and Lu,  Chao-Yang and Peng,  Cheng-Zhi and Zhu,  Xiaobo and Pan,  Jian-Wei},
  author = {Gong, Ming and Wang, Shiyu and Zha, Chen and Chen, Ming-Cheng and Huang, He-Liang and Wu, Yulin and Zhu, Qingling and Zhao, Youwei and Li, Shaowei and Guo, Shaojun and Qian, Haoran and Ye, Yangsen and Chen, Fusheng and Ying, Chong and Yu, Jiale and Fan, Daojin and Wu, Dachao and Su, Hong and Deng, Hui and Rong, Hao and et al.},
  year = {2021},
  month = may,
  pages = {948–952}
}

@article{preiss2015strongly,
  title = {Strongly correlated quantum walks in optical lattices},
  volume = {347},
  ISSN = {1095-9203},
  url = {http://dx.doi.org/10.1126/science.1260364},
  DOI = {10.1126/science.1260364},
  number = {6227},
  journal = {Science},
  publisher = {American Association for the Advancement of Science (AAAS)},
  author = {Preiss,  Philipp M. and Ma,  Ruichao and Tai,  M. Eric and Lukin,  Alexander and Rispoli,  Matthew and Zupancic,  Philip and Lahini,  Yoav and Islam,  Rajibul and Greiner,  Markus},
  year = {2015},
  month = mar,
  pages = {1229–1233}
}

@article{desaules2022hypergrid,
  title = {Hypergrid subgraphs and the origin of scarred quantum walks in many-body Hilbert space},
  volume = {105},
  ISSN = {2469-9969},
  url = {http://dx.doi.org/10.1103/PhysRevB.105.245137},
  DOI = {10.1103/physrevb.105.245137},
  number = {24},
  journal = {Physical Review B},
  publisher = {American Physical Society (APS)},
  author = {Desaules,  Jean-Yves and Bull,  Kieran and Daniel,  Aiden and Papić,  Zlatko},
  year = {2022},
  month = jun,
  pages = {245137}
}

@book{mackay2003information,
	title        = {Information theory, inference and learning algorithms},
	author       = {MacKay, David JC},
	year         = 2003,
	publisher    = {Cambridge University Press}
}

@article{pokharel2024scalable,
  title = {Scalable measurement error mitigation via iterative bayesian unfolding},
  volume = {6},
  ISSN = {2643-1564},
  url = {http://dx.doi.org/10.1103/PhysRevResearch.6.013187},
  DOI = {10.1103/physrevresearch.6.013187},
  number = {1},
  journal = {Physical Review Research},
  publisher = {American Physical Society (APS)},
  author = {Pokharel,  Bibek and Srinivasan,  Siddarth and Quiroz,  Gregory and Boots,  Byron},
  year = {2024},
  month = feb,
  pages = {013187},
}

@inbook{efron1992bootstrap,
  title = {Bootstrap Methods: Another Look at the Jackknife},
  ISBN = {9781461243809},
  ISSN = {0172-7397},
  url = {http://dx.doi.org/10.1007/978-1-4612-4380-9_41},
  DOI = {10.1007/978-1-4612-4380-9_41},
  booktitle = {Breakthroughs in Statistics},
  publisher = {Springer New York},
  author = {Efron,  Bradley},
  year = {1992},
  pages = {569–593}
}

@article{dempster1977maximum,
  title = {Maximum Likelihood from Incomplete Data Via the {EM} Algorithm},
  volume = {39},
  ISSN = {1467-9868},
  url = {http://dx.doi.org/10.1111/j.2517-6161.1977.tb01600.x},
  DOI = {10.1111/j.2517-6161.1977.tb01600.x},
  number = {1},
  journal = {Journal of the Royal Statistical Society Series B: Statistical Methodology},
  publisher = {Oxford University Press (OUP)},
  author = {Dempster,  A. P. and Laird,  N. M. and Rubin,  D. B.},
  year = {1977},
  month = sep,
  pages = {1–22}
}

@article{blume2010optimal,
  title = {Optimal,  reliable estimation of quantum states},
  volume = {12},
  ISSN = {1367-2630},
  url = {http://dx.doi.org/10.1088/1367-2630/12/4/043034},
  DOI = {10.1088/1367-2630/12/4/043034},
  number = {4},
  journal = {New Journal of Physics},
  publisher = {IOP Publishing},
  author = {Blume-Kohout,  Robin},
  year = {2010},
  month = apr,
  pages = {043034}
}

@article{childsuniversal2009,
	title        = {Universal Computation by Quantum Walk},
	author       = {Childs, Andrew M.},
	volume       = 102,
	number       = 18,
	pages        = {180501},
	url          = {https://link.aps.org/doi/10.1103/PhysRevLett.102.180501},
	urldate      = {2023-02-19},
	doi          = {10.1103/PhysRevLett.102.180501},
	journal = {Physical Review Letters},
	shortjournal = {Phys. Rev. Lett.},
	date         = {2009-05-04},
	year = 2009
}

@article{tamuraquantum2020,
	title        = {Quantum Walks of a Phonon in Trapped Ions},
	author       = {Tamura, Masaya and Mukaiyama, Takashi and Toyoda, Kenji},
	volume       = 124,
	number       = 20,
	pages        = {200501},
	doi          = {10.1103/PhysRevLett.124.200501},
	journal = {Physical Review Letters},
	shortjournal = {Phys. Rev. Lett.},
	date         = {2020-05-19},
	year = 2020
}

@article{venegasandracaquantum2012,
	title        = {Quantum walks: a comprehensive review},
	shorttitle   = {Quantum walks},
	author       = {Venegas-Andraca, Salvador Elias},
	volume       = 11,
	number       = 5,
	pages        = {1015--1106},
	doi          = {10.1007/s11128-012-0432-5},
	issn         = {1573-1332},
	url          = {https://doi.org/10.1007/s11128-012-0432-5},
	urldate      = {2019-10-16},
	journal = {Quantum Information Processing},
	shortjournal = {Quantum Inf Process},
	date         = {2012-10-01},
	year = 2012
}

@article{quexperimental2024,
	title        = {Experimental implementation of quantum-walk-based portfolio optimization},
	author       = {Qu, Dengke and Matwiejew, Edric and Wang, Kunkun and Wang, Jingbo and Xue, Peng},
	volume       = 9,
	number       = 2,
	pages        = {025014},
	doi          = {10.1088/2058-9565/ad27e9},
	issn         = {2058-9565},
	url          = {https://iopscience.iop.org/article/10.1088/2058-9565/ad27e9},
	urldate      = {2024-04-05},
	journal = {Quantum Science and Technology},
	shortjournal = {Quantum Sci. Technol.},
	date         = {2024-04-01},
	year = 2024
}

@article{munarinilucas2001,
	title        = {On the Lucas Cubes},
	author       = {Munarini, Emanuele and Cippo, Claudio Perelli and Salvi, Norma Zagaglia},
	volume       = 39,
	number       = 1,
	pages        = {12--21},
	doi          = {10.1080/00150517.2001.12428753},
	issn         = {0015-0517, 2641-340X},
	url          = {https://www.tandfonline.com/doi/full/10.1080/00150517.2001.12428753},
	urldate      = {2024-12-07},
	journal = {The Fibonacci Quarterly},
	shortjournal = {The Fibonacci Quarterly},
	date         = {2001-02},
	year = 2001
}

@article{ashrafi2016vertex,
  title = {Vertex and Edge Orbits of Fibonacci and Lucas Cubes},
  volume = {20},
  ISSN = {0219-3094},
  url = {http://dx.doi.org/10.1007/s00026-016-0318-9},
  DOI = {10.1007/s00026-016-0318-9},
  number = {2},
  journal = {Annals of Combinatorics},
  publisher = {Springer Science and Business Media LLC},
  author = {Ashrafi,  Ali Reza and Azarija,  Jernej and Fathalikhani,  Khadijeh and Klavžar,  Sandi and Petkovšek,  Marko},
  year = {2016},
  month = may,
  pages = {209–229}
}

@article{wurtzmaxcut2021,
	title        = {{MaxCut} quantum approximate optimization algorithm performance guarantees for $p>1$},
	author       = {Wurtz, Jonathan and Love, Peter},
	year         = 2021,
	month        = apr,
	journal      = {Physical Review A},
	volume       = 103,
	number       = 4,
	pages        = {042612},
	doi          = {10.1103/PhysRevA.103.042612},
	url          = {https://link.aps.org/doi/10.1103/PhysRevA.103.042612},
	urldate      = {2022-10-08}
}

@article{brassard2000quantum,
  title = {Quantum amplitude amplification and estimation},
  ISSN = {0271-4132},
  url = {http://dx.doi.org/10.1090/conm/305/05215},
  DOI = {10.1090/conm/305/05215},
  journal = {Quantum Computation and Information},
  publisher = {American Mathematical Society},
  author = {Brassard,  Gilles and Høyer,  Peter and Mosca,  Michele and Tapp,  Alain},
  year = {2002},
  pages = {53–74}
}

@article{ambainis2003quantum,
  title = {Quantum walks and their algorithmic applications},
  volume = {01},
  ISSN = {1793-6918},
  url = {http://dx.doi.org/10.1142/S0219749903000383},
  DOI = {10.1142/s0219749903000383},
  number = {04},
  journal = {International Journal of Quantum Information},
  publisher = {World Scientific Pub Co Pte Lt},
  author = {Ambainis,  ANDRIS},
  year = {2003},
  month = dec,
  pages = {507–518}
}

@article{zalkagrovers1999,
  title = {Grover’s quantum searching algorithm is optimal},
  volume = {60},
  ISSN = {1094-1622},
  url = {http://dx.doi.org/10.1103/PhysRevA.60.2746},
  DOI = {10.1103/physreva.60.2746},
  number = {4},
  journal = {Physical Review A},
  publisher = {American Physical Society (APS)},
  author = {Zalka,  Christof},
  year = {1999},
  month = oct,
  pages = {2746–2751}
}

@article{yosi2021,
	title        = {Improved upper bounds for the hitting times of quantum walks},
	author       = {Atia, Yosi and Chakraborty, Shantanav},
	year         = 2021,
	month        = {Sep},
	journal      = {Phys. Rev. A},
	publisher    = {American Physical Society},
	volume       = 104,
	pages        = {032215},
	doi          = {10.1103/PhysRevA.104.032215},
	url          = {https://link.aps.org/doi/10.1103/PhysRevA.104.032215},
	issue        = 3,
	numpages     = 13
}

@article{christandl2004perfect,
  title = {Perfect State Transfer in Quantum Spin Networks},
  volume = {92},
  ISSN = {1079-7114},
  url = {http://dx.doi.org/10.1103/PhysRevLett.92.187902},
  DOI = {10.1103/physrevlett.92.187902},
  number = {18},
  journal = {Physical Review Letters},
  publisher = {American Physical Society (APS)},
  author = {Christandl,  Matthias and Datta,  Nilanjana and Ekert,  Artur and Landahl,  Andrew J.},
  year = {2004},
  month = may,
  pages = {187902},
}

@article{Deffner2017,
  title = {Quantum speed limits: from Heisenberg’s uncertainty principle to optimal quantum control},
  volume = {50},
  ISSN = {1751-8121},
  url = {http://dx.doi.org/10.1088/1751-8121/aa86c6},
  DOI = {10.1088/1751-8121/aa86c6},
  number = {45},
  journal = {Journal of Physics A: Mathematical and Theoretical},
  publisher = {IOP Publishing},
  author = {Deffner,  Sebastian and Campbell,  Steve},
  year = {2017},
  month = oct,
  pages = {453001}
}

@article{Valiant1979,
  title = {The Complexity of Enumeration and Reliability Problems},
  volume = {8},
  ISSN = {1095-7111},
  url = {http://dx.doi.org/10.1137/0208032},
  DOI = {10.1137/0208032},
  number = {3},
  journal = {SIAM Journal on Computing},
  publisher = {Society for Industrial & Applied Mathematics (SIAM)},
  author = {Valiant,  Leslie G.},
  year = {1979},
  month = aug,
  pages = {410–421}
}

@article{Krovi2007,
  title = {Quantum walks on quotient graphs},
  volume = {75},
  ISSN = {1094-1622},
  url = {http://dx.doi.org/10.1103/PHYSREVA.75.062332},
  DOI = {10.1103/physreva.75.062332},
  number = {6},
  journal = {Physical Review A},
  publisher = {American Physical Society (APS)},
  author = {Krovi,  Hari and Brun,  Todd A.},
  year = {2007},
  month = jun ,
  pages = {062332}
}

@article{Harrigan2021,
  title = {Quantum approximate optimization of non-planar graph problems on a planar superconducting processor},
  volume = {17},
  ISSN = {1745-2481},
  url = {http://dx.doi.org/10.1038/s41567-020-01105-y},
  DOI = {10.1038/s41567-020-01105-y},
  number = {3},
  journal = {Nature Physics},
  publisher = {Springer Science and Business Media LLC},
  allauthor = {Harrigan,  Matthew P. and Sung,  Kevin J. and Neeley,  Matthew and Satzinger,  Kevin J. and Arute,  Frank and Arya,  Kunal and Atalaya,  Juan and Bardin,  Joseph C. and Barends,  Rami and Boixo,  Sergio and Broughton,  Michael and Buckley,  Bob B. and Buell,  David A. and Burkett,  Brian and Bushnell,  Nicholas and Chen,  Yu and Chen,  Zijun and Ben Chiaro and Collins,  Roberto and Courtney,  William and Demura,  Sean and Dunsworth,  Andrew and Eppens,  Daniel and Fowler,  Austin and Foxen,  Brooks and Gidney,  Craig and Giustina,  Marissa and Graff,  Rob and Habegger,  Steve and Ho,  Alan and Hong,  Sabrina and Huang,  Trent and Ioffe,  L. B. and Isakov,  Sergei V. and Jeffrey,  Evan and Jiang,  Zhang and Jones,  Cody and Kafri,  Dvir and Kechedzhi,  Kostyantyn and Kelly,  Julian and Kim,  Seon and Klimov,  Paul V. and Korotkov,  Alexander N. and Kostritsa,  Fedor and Landhuis,  David and Laptev,  Pavel and Lindmark,  Mike and Leib,  Martin and Martin,  Orion and Martinis,  John M. and McClean,  Jarrod R. and McEwen,  Matt and Megrant,  Anthony and Mi,  Xiao and Mohseni,  Masoud and Mruczkiewicz,  Wojciech and Mutus,  Josh and Naaman,  Ofer and Neill,  Charles and Neukart,  Florian and Niu,  Murphy Yuezhen and O’Brien,  Thomas E. and O’Gorman,  Bryan and Ostby,  Eric and Petukhov,  Andre and Putterman,  Harald and Quintana,  Chris and Roushan,  Pedram and Rubin,  Nicholas C. and Sank,  Daniel and Skolik,  Andrea and Smelyanskiy,  Vadim and Strain,  Doug and Streif,  Michael and Szalay,  Marco and Vainsencher,  Amit and White,  Theodore and Yao,  Z. Jamie and Yeh,  Ping and Zalcman,  Adam and Zhou,  Leo and Neven,  Hartmut and Bacon,  Dave and Lucero,  Erik and Farhi,  Edward and Babbush,  Ryan},
  author = {Harrigan, Matthew P. and Sung, Kevin J. and Neeley, Matthew and Satzinger, Kevin J. and Arute, Frank and Arya, Kunal and Atalaya, Juan and Bardin, Joseph C. and Barends, Rami and Boixo, Sergio and Broughton, Michael and Buckley, Bob B. and Buell, David A. and Burkett, Brian and Bushnell, Nicholas and Chen, Yu and Chen, Zijun and Ben Chiaro and Collins, Roberto and Courtney, William and et al.},
  year = {2021},
  month = feb,
  pages = {332–336}
}

@article{Shaydulin2024,
  title = {Evidence of scaling advantage for the quantum approximate optimization algorithm on a classically intractable problem},
  volume = {10},
  ISSN = {2375-2548},
  url = {http://dx.doi.org/10.1126/sciadv.adm6761},
  DOI = {10.1126/sciadv.adm6761},
  number = {22},
  pages = {eadm6761},
  journal = {Science Advances},
  publisher = {American Association for the Advancement of Science (AAAS)},
  allauthor = {Shaydulin,  Ruslan and Li,  Changhao and Chakrabarti,  Shouvanik and DeCross,  Matthew and Herman,  Dylan and Kumar,  Niraj and Larson,  Jeffrey and Lykov,  Danylo and Minssen,  Pierre and Sun,  Yue and Alexeev,  Yuri and Dreiling,  Joan M. and Gaebler,  John P. and Gatterman,  Thomas M. and Gerber,  Justin A. and Gilmore,  Kevin and Gresh,  Dan and Hewitt,  Nathan and Horst,  Chandler V. and Hu,  Shaohan and Johansen,  Jacob and Matheny,  Mitchell and Mengle,  Tanner and Mills,  Michael and Moses,  Steven A. and Neyenhuis,  Brian and Siegfried,  Peter and Yalovetzky,  Romina and Pistoia,  Marco},
  author = {Shaydulin, Ruslan and Li, Changhao and Chakrabarti, Shouvanik and DeCross, Matthew and Herman, Dylan and Kumar, Niraj and Larson, Jeffrey and Lykov, Danylo and Minssen, Pierre and Sun, Yue and Alexeev, Yuri and Dreiling, Joan M. and Gaebler, John P. and Gatterman, Thomas M. and Gerber, Justin A. and Gilmore, Kevin and Gresh, Dan and Hewitt, Nathan and Horst, Chandler V. and Hu, Shaohan and et al.},
  year = {2024},
  month = may 
}

@misc{https://doi.org/10.5281/zenodo.20520609,
  doi = {10.5281/ZENODO.20520609},
  url = {https://zenodo.org/doi/10.5281/zenodo.20520609},
  author = {Matwiejew,  Edric and Wurtz,  Jonathan and Chen,  Jing and Elahi,  Pascal Jahan and Macrì,  Tommaso and Varetto,  Ugo},
  title = {Supplementary data and software for ``{C}ontinuous-time quantum-walk-based ans\"{a}tze on neutral-atom hardware''},
  publisher = {Zenodo},
  year = {2026},
  copyright = {MIT License}
}

\end{document}